\documentclass[preprint]{aastex}

\received{}
\accepted{}
\journalid{}{}
\articleid{}{}

\shortauthors{Carpenter, Hillenbrand, \& Skrutskie}
\shorttitle{Variable Stars in Orion~A}

\newcommand{\ts}{\thinspace}
\newcommand{\simless}{\mathbin{\lower 3pt\hbox
     {$\rlap{\raise 5pt\hbox{$\char'074$}}\mathchar"7218$}}}
\newcommand{\simgreat}{\mathbin{\lower 3pt\hbox
     {$\rlap{\raise 5pt\hbox{$\char'076$}}\mathchar"7218$}}}

\newcommand{\about}    {$\sim$\ts}
\newcommand{\aboutless}{$\simless$\ts}
\newcommand{\aboutmore}{$\simgreat$\ts}

\newcommand{\KB}{$K_s$}
\newcommand{\HK}{$H-K_s$}
\newcommand{\JK}{$J-K_s$}

\newcommand{\M}{$^{\rm m}$}

\newcommand{\msun}{\ts M$_\odot$}

\newcommand{\rsun}{\ts R$_\odot$}
\newcommand{\kkms}{\ts K\ts km\ts s$^{-1}$}
\newcommand{\kms}{\ts km\ts s$^{-1}$}
\newcommand{\sqamin}{\ts arcmin$^{-2}$}

\newcommand{\etal}{et~al.}
\newcommand{\mdot}{$M{\raise 1.5ex\hbox{\hskip-6pt$\mathchar"201$}\kern0.2em}_{acc}$ }
\newcommand{\myear}{\msun~year$^{-1}$}

\newcommand{\thcoj}{$^{13}$CO(1--0)}
\newcommand{\NGC}{NGC\ts}

\def\insertplot#1#2#3#4#5#6#7{
\vskip 10pt\nobreak\hbox to \hsize{\hss\dimen0=#3in\hbox to #6\dimen0{%
\dimen0=#2in\vbox to #6\dimen0{\vss
\special{ps: plotfile #1}
\special{ps::[end]
  PGPLOT restore
}
}\hss}\hss}\vskip 10pt}

\begin{document}

\title{Near-Infrared Photometric Variability of\\
       Stars Toward the Orion~A Molecular Cloud}

\author{John M. Carpenter}
\affil{California Institute of Technology, 
       Department of Astronomy, MS 105-24, \\ Pasadena, CA 91125; 
       email: jmc@astro.caltech.edu}

\author{Lynne A. Hillenbrand}
\affil{California Institute of Technology, 
       Department of Astronomy, MS 105-24, \\ Pasadena, CA 91125; 
       email: lah@astro.caltech.edu}

\and

\author{M. F. Skrutskie}
\affil{University of Massachusetts, Department of Astronomy, \\
       Amherst, MA 01003; email: skrutski@north.astro.umass.edu}

\begin{abstract}

We present an analysis of $J$, $H$, and \KB\ time series photometry obtained 
with the southern 2MASS telescope over a $0.84^\circ\times6^\circ$ 
region centered near the Trapezium region of the Orion Nebula Cluster. These 
data are used to establish the 
near-infrared variability properties of pre-main-sequence stars in Orion on 
time scales of \about 1-36 days, \about 2 months, and \about 2 years.
A total of 1235 near-infrared variable stars are identified, \about 93\% of 
which are likely associated with the Orion~A molecular cloud. The variable 
stars exhibit a diversity of photometric behavior with time, including cyclic
fluctuations with periods up to 15 days, aperiodic day-to-day fluctuations, 
eclipses, slow drifts in brightness over one month or longer, colorless 
variability (within the noise limits of the data), stars that become redder as 
they fade, and stars that become bluer as they fade. The mean 
peak-to-peak amplitudes of 
the photometric fluctuations are \about 0.2\M\ in each band and 77\% of the 
variable stars have color variations less than 0.05\M. The more extreme stars 
in our sample have amplitudes as large as \about 2\M\ and change in color by
as much as \about 1\M. The typical time scale of the photometric fluctuations 
is less than a few days, indicating that near-infrared variability results 
primarily from short term processes. We examine rotational modulation of cool 
and hot star spots, variable obscuration from an inner circumstellar disk, 
and changes in the mass accretion rate and other physical properties in a 
circumstellar disk as possible physical origins of the near-infrared 
variability. Cool spots alone can explain the observed variability 
characteristics in \about 56-77\% of the stars, while the properties of the 
photometric fluctuations are more consistent with hot spots or extinction 
changes in at 
least 23\% of the stars, and with variations in the disk mass accretion rate 
or inner disk radius in \about 1\% of our sample. However, differences between 
the details of the observations and the details of variability predicted by 
hot spot, extinction, and accretion disk models suggest either that another 
variability mechanism not considered here may be operative, or that the 
observed variability represents the net results of several of these phenomena. 
Analysis of the star count data indicates that the Orion Nebula Cluster is 
part of a larger area of enhanced stellar surface density which extends over a 
$0.4^\circ\times2.4^\circ$ ($3.4\:\rm{pc}\times20\:\rm{pc}$) region containing
\about 2700 stars brighter than $K_s=14$\M.

\end{abstract}

\keywords{infrared: stars --- 
          stars:pre-main-sequence --- 
          stars:variables ---\hfill\\
          stars:individual (YY Ori, BM Ori) --- 
          open clusters and associations}

\section{Introduction}

Optical photometric variability is one of the original defining 
characteristics of pre-main-sequence stars \citep{Joy45,Herbig62}. Although 
pre-main sequence objects are usually identified by other, less biased 
photometric and spectroscopic survey techniques now, modern variability 
observations remain a valuable probe of the stellar and circumstellar activity 
(e.g. Bouvier \etal~1993,1995).  Such monitoring studies have shown that 
photometric variability is a diverse phenomenon in that the observed flux can 
change by milli-magnitudes to magnitudes on time scales of minutes to years, 
often with periodic as well as aperiodic components. For periodic stars, the 
variability is thought to originate mainly from cool magnetic or hot accretion 
spots on the stellar surface that are hundreds to thousands of degrees 
different in temperature from the photosphere and rotate with the star.
Aperiodic variability may arise from mechanisms such as coronal flares, 
irregular accretion of new material onto the star, and temporal variations in 
circumstellar extinction \citep[see also review by M\'enard \& 
Bertout~1998]{Herbst94}.

Young stellar objects are also variable at x-ray, ultraviolet, infrared, and 
radio wavelengths. Each of these wavelength regimes probes a different aspect 
of the young star and its circumstellar material, and they can be used together
to establish a more comprehensive picture of the temporal properties and 
physical characteristics of young stellar objects (see, e.g., 
Kenyon \etal~1994, Guenther \etal~2000, Yudin~2000).  Near-infrared monitoring 
observations, the focus of this study, are expected to probe relatively 
cool phenomena compared to optical, ultraviolet, and x-ray studies, and
may probe phenomena not accessible to shorter wavelength studies, such as 
temperature, opacity, and geometry changes in the dust- and gas-rich 
near-circumstellar environment.

Variability at near-infrared wavelengths (1.2-2.2\micron) has long been 
established from aperture photometry of young stars 
\citep{Cohen76,Rydgren83,Skrutskie96} and extensive campaigns on individual 
targets (e.g. RU Lup and RY Tau -- Hutchinson \etal~1989; SVS13 -- Liseau, 
Lorenzetti, \& Molinari 1992, Aspin \& Sandell 1994; DR~Tau -- Kenyon 
\etal~1994). More recent monitoring observations have used near-infrared 
arrays to study photometric variability of entire clusters of stars (Serpens 
-- Horrobin, Casali, \& Eiroa~1997; Kaas~1999, Hodapp~1999). In addition to 
variability best modeled by cool 
spots, hot spots, and extinction variations, these studies have detected 
variable near-infrared emission that has been interpreted as originating from 
circumstellar material \citep{Skrutskie96}. However, near-infrared variability 
is still not as well characterized as optical variability, and many questions 
remain as to the fraction of stars that exhibit such variability, the 
amplitudes and time scales of the photometric fluctuations, and the dominant 
physical mechanisms contributing to it. Establishing these properties is of 
interest in its own right, but is also critical for understanding how to 
interpret single-epoch near-infrared photometry as representative of the mean 
flux levels in assessing the extinction, stellar mass and age, and accretion 
characteristics of individual stars.

In this contribution, we present an analysis of $J$, $H$, and \KB\ time 
series photometry over a \about $0.84^\circ\times6^\circ$ area centered on the 
Trapezium region of the Orion Nebula Cluster (ONC) that encompasses the
northern portion of the Orion~A molecular cloud. Section~\ref{data} 
describes the full data set obtained for this study and the data reduction 
procedures. In Section~\ref{variable_stars}, we identify the variable stars 
from the time series photometry and present a sampling of the different types
of observed variability. Section~\ref{properties} and 
Section~\ref{characteristics} analyze the near-infrared variability 
characteristics, and possible origins of the variability are discussed in 
Section~\ref{origin}. The implication of these results are discussed in 
Section~\ref{discussion}, and our conclusions are summarized in 
Section~\ref{summary}.

\section{Data}
\label{data}

\subsection{Observations and Data Processing}
\label{observations}

The $J$, $H$, and \KB\ observations of the Orion region were obtained 
with the 2MASS 1.3 meter telescope at Cerro Tololo, Chile near the 
completion of the southern survey operations when auxiliary projects were 
scheduled in otherwise idle telescope time. All data were collected in 
standard 2MASS observing mode by scanning the telescope in declination to 
cover tiles of size ($\Delta\alpha\times\Delta\delta$) \about 
($8.5'\times6^\circ$) in the three bands simultaneously. A 2MASS frame at each 
position within the tile consists of a doubly-correlated difference of two 
NICMOS readouts separated by the 1.3 second frame integration time. The first 
readout occurs 51 millisecond after reset and independently provides a short 
integration to recover unsaturated images of bright (\about 5-9\M) stars. Each 
position on the sky is observed 6 times in this manner for a total integration
time of 7.8 seconds. The nominal region surveyed for this project consists of 
seven contiguous tiles in right ascension as listed in Table~\ref{tbl:coords}, 
with each tile centered on a declination of $-6$\arcdeg. Adjacent tiles 
overlap by \about 40-50\arcsec\ in right ascension to 
provide a total sky coverage of \about $0.84^\circ \times 6^\circ$. The seven 
tiles were observed on nearly a nightly basis for 16 nights in March 2000. By 
the middle of March, Orion set early enough that it became necessary to 
observe fewer tiles per night, and by early April, just a single tile was 
observed. In total, data were obtained on 29 days in March/April 2000 over
a 36 day time period, with the longest time gap in the observations being 4 
days. The observing log for each tile is provided in Table~\ref{tbl:log}. In
addition to these specially scheduled observations of the Orion region, the 
analysis conducted here incorporates data from normal 2MASS survey operations 
in March 1998 and February 2000. With the above time sampling,
the near-infrared photometric variability characteristics can be assessed on 
time scales of \about 1-36 days, \about 2 months, and \about 2 years. 

The data were reduced using a development version of the 2MASS data processing 
pipeline at the Infrared Processing and Analysis Center (IPAC). This 
development pipeline is the same one that generated the data products for the 
first and second incremental 2MASS release catalogs. The data discussed in 
this paper though were not part of these incremental releases and will 
ultimately be replaced by the results of the final 2MASS processing. The 2MASS 
Explanatory Supplement \citep{Cutri00} contains complete details of the data 
reduction procedures and only a brief summary is provided here. The output 
from the data reduction pipeline for each observed tile includes calibrated 
astronomical coordinates, $J$, $H$, and \KB\ magnitudes, photometric 
uncertainties, and various photometric quality flags. 
Stars were identified in either the individual 51~ms images or 
a coadd of the six 1.3~s images using an algorithm operationally equivalent to 
DAOFIND \citep{Stetson87}. For bright stars (\aboutless 9\M), photometry is 
obtained with aperture photometry on the individual 51 millisecond images 
using an aperture radius of 4\arcsec\ and a sky annulus that extends radially 
from 24\arcsec-30\arcsec. The final magnitude is computed by averaging the six 
aperture measurements and adopting the standard deviation of the mean as the 
photometric uncertainty. Stars brighter than \about 5\M\ are saturated in
the short exposure images. Photometry for stars fainter than \about 9\M\ are 
obtained by fitting a Point Spread Function (PSF) to the six 1.3 second images
simultaneously. The PSF photometric uncertainty is computed from the Poisson 
noise within the PSF fitting radius and the observed fluctuations in the sky 
background. On rare occasions, the PSF fit did not converge in crowded regions 
and in areas with bright backgrounds such as the Orion Nebula. In such cases, 
an aperture magnitude is obtained on the 1.3 second integration images. 

The instrumental magnitudes were calibrated by observation of a 2MASS standard 
field every hour as part of normal 2MASS operations, with typical 1$\sigma$ 
calibration uncertainties in the nightly photometric zero points of
\aboutless 0.015\M\ in each band. 
For our time series data, the internal accuracy of the photometry can 
be improved by defining a grid of bright, isolated stars within the survey 
region as secondary standards and assuming that their average magnitudes as an 
ensemble do not change in time. These secondary standards were selected as
isolated stars more than 20\arcsec\ from the nearest object that are brighter 
than 14\M\ in each band with photometric root-mean-squared (RMS) from the 
repeated observations less than 0.1 magnitudes. A total of 3649 stars 
satisfied these criteria. The fiducial magnitude for each secondary standard 
was defined as the average magnitude computed using all available 
observations. The average photometric offset for each tile relative to this 
secondary grid was then computed with a statistical accuracy of \about 
0.002\M. The applied photometric offsets were typically less than 0.015 
magnitudes per band per tile, except for the 1998 data, where the offsets were 
as large as 0.038\M. The March 1998 observations were some of the first 
observations conducted with the southern 2MASS telescope and were obtained 
and calibrated before the complete 2MASS standard calibration grid for the 
southern survey had been established.

Time series photometry for each star was produced by matching 
point sources with positions coincident to within 1\arcsec. Sources were 
merged in this manner from night-to-night and also within a night to account 
for those stars in the overlap regions between adjacent tiles. The Julian date 
for each observation was estimated from the starting Julian date of the tile 
and assuming that the scan rate (\about 1\arcdeg\ min$^{-1}$) along the 
6\arcdeg\ long tile is constant.  

\subsection{Point Source List}
\label{sample}

As mentioned in the previous section, not all tiles in our image data were 
observed on all nights. To establish as uniform, reliable, and complete a 
source list as possible over the entire region surveyed, the 16 nights in 
March 2000 common to all 7 right ascension tiles were used to define the areal 
completeness limits, the magnitude completeness limits, and the final source 
list. 

First, the maximum angular 
area over which the observations are spatially complete was determined by 
examining the spatial distribution of sources detected on all 16 nights. In 
J2000 equatorial coordinates, this area encompasses the right ascension range 
83.405 to 84.250 degrees and the declination range $-8$.98\arcdeg\ to 
$-2$.88\arcdeg\ for a total area of 5.12~deg$^2$. Second, 
the magnitude completeness limits at $J$,$H$,\KB\ were assessed using all 
stars with at least 8 detections in the appropriate band and no
artifact or confusion flags from the IPAC pipeline 
processing. The requirement that the star be detected at least 8 times in the 
time series is arbitrarily set to remove transient detections (e.g. meteor 
trails) and to reduce the number of double stars that are resolved on only a 
fraction of the nights due to differences in seeing conditions. Stars 
between $-6$\arcdeg\ and $-5$\arcdeg\ declination were excluded from the 
completeness analysis to avoid complications in identifying point sources in 
the Orion Nebula and associated dense cluster. For our data, the magnitude 
above which a star is detected on at least 15 of the 16 nights in the absence 
of source confusion occurs at $J$=16.0\M, $H$=15.4\M, and \KB=14.8\M. The 
5.12~deg$^2$ area of spatial completeness contains 18,552 stars with at least 
8 detections brighter than or equal to these defined magnitude completeness 
limits in one or more bands. 

The preliminary source list thus consisted of 18,552 stars, \about 93\% of 
which have no artifact or confusion flags from the processing pipeline in any 
of the observations. After removing those sources flagged as persistence or 
filter glints, potential lingering artifacts were identified as objects that 
had either unusual stellar colors stars, detected less than 15 times, or had 
flags indicating contaminated or confused photometry from a nearby star. Many, 
but not all, of these \about 1300 sources were visually inspected in the 
images. Most of the objects have processing flags indicating the source is 
either an artifact, or is real but near a diffraction spike or potentially 
confused with a nearby bright star. In 2MASS public data releases, many of 
these sources have been omitted to ensure reliability of the output catalogs. 
For this project, however, we used the repeated nature of our observations and 
visual inspection of the images to formulate criteria to distinguish real 
sources from the artifacts among the objects with processing flags to enhance 
the spatial completeness of the observations while retaining a reliable source 
list. From inspection of the images for many sources, it became apparent that 
real sources are generally detected on every night and usually in all three 
bands. Sources detected less than 15 times but without any flags set are 
generally double sources resolved on only a fraction of the nights, or sources 
near the completeness limits but not detected on every apparition. Artifacts, 
if detected on all 16 nights, are usually present in only one band even though 
the upper limits to the stellar flux reported in the IPAC processing indicate 
the source should have been detected in the other two bands. Artifacts 
detected in all three bands can also be distinguished by their unusual colors. 
These criteria then were used to identify and remove artifacts from the source 
list. 

Finally, point source detection in regions with bright, extended backgrounds 
such as the Orion Nebula are notoriously difficult. Therefore, in the core 
of the Orion Nebula, the 2MASS source list was compared with the deep ($K$ 
\about 17.5\M), high resolution Keck image of the inner $5'\times5'$ of the 
ONC presented by \citet{Lynne00}. While the 2MASS and Keck images were obtained 
\about 1 year apart and stars could in principle be present in the 2MASS 
images but not in the Keck data due to variability, in practice, none of the 
2MASS-only sources looked to be a convincing point source from visual 
inspection of the images. We therefore assumed that any source present in the
2MASS data but not detected in the Keck images is likely a knot in the 
nebulosity and removed it from the source list.

In summary, of the 18,552 stars meeting the spatial and magnitude completeness 
criteria, 744 were deemed artifacts around bright stars or knots of nebulosity 
in the Orion Nebula and were removed from the source list. The final source 
list for our variability analysis contains 17,808 stars brighter than 
the defined completeness limits ($J$=16.0\M, $H$=15.4\M, \KB=14.8\M) in at
least one band and within the angular area $\alpha$(J2000) = 
[83.405\arcdeg - 84.250\arcdeg] 
and $\delta$(J2000) = [$-8.98$\arcdeg - $-2.88$\arcdeg].

\subsection{Photometric Integrity}
\label{photometry}

With the point source list established, our next step was to assure photometric
integrity by removing all photometry suspected of being unreliable so that 
true photometric fluctuations can be distinguished from spurious individual 
measurements. Unreliable photometry can result either from characteristics 
that prevent good photometry from ever being obtained for a particular real 
point source (e.g. a neighboring bright star, a stellar companion, bright 
nebulosity, etc...) or conditions that may effect a single measurement (e.g. a 
cosmic ray hit, meteor trail, a PSF fit that did not converge, etc...). Based 
on visual inspection of many sources, the following criteria were adopted to 
formulate a list of reliable photometric measurements. First, only photometric 
measurements obtained using aperture photometry in the 51 millisecond frames 
or PSF fitting photometry in the 1.3 second images (see 
Section~\ref{observations}) were used. Second, 1.8\% of all objects in the 
source list were repeatedly identified as extended at the resolution of 2MASS,
as the $\chi^2$ from the PSF fit averaged over all apparitions of the object 
exceeded 1.5 in each band. As described in Section~\ref{id}, these sources 
were not considered for inclusion in the uniformly selected list of variable 
stars, but a few of these extended sources were identified as variable objects 
using subjective criteria. Finally, to remove potentially unreliable, 
individual photometric measurements from the time series data for a given 
star, any individual photometric measurement based on PSF fitting that had a 
resulting $\chi^2 > 3.0$ was not used in judging the photometric variability.

\begin{figure}
\epsscale{0.9}
\plotone{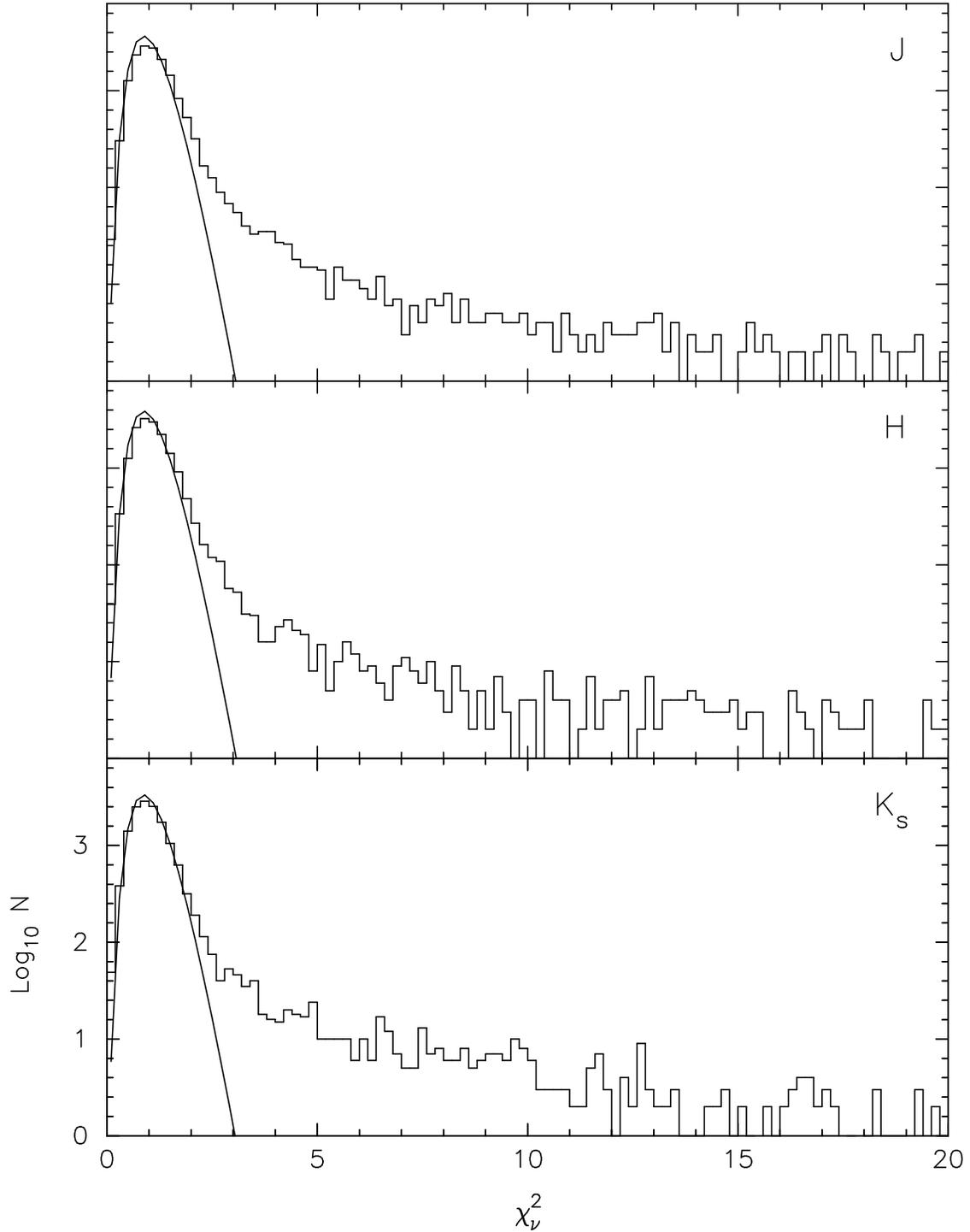}
\epsscale{1.0}
\caption{
  Histogram of the reduced chi-squared ($\chi_\nu^2$) for the observed 
  photometric deviations about the mean for the $J$ (top), $H$ (middle), 
  and \KB-band photometry from the 16 nights in which all 7 tiles 
  were observed. The solid curve in each panel is the 
  expected $\chi_\nu^2$ distribution for 15 degrees of freedom, where the 
  curves have been normalized by the total number of stars. 
  This figure demonstrates that the observed photometric scatter for the 
  majority of the stars is consistent with random noise. Stars with large 
  values of $\chi_\nu^2$ are candidate variable stars.
  \label{fig:chi2}
}
\end{figure}

We tested that the estimated photometric uncertainties (see 
Section~\ref{observations}) accurately reflect the real noise characteristics 
of the 2MASS photometry using the reduced chi-squared ($\chi^2_\nu$), computed 
for each band and each star as
\begin{equation}
\label{eq:chi2}
     \chi^2_\nu = {1\over\nu} \sum_{i=1}^{n}{(m_i - \bar{m})^2\over\sigma_i^2},
\end{equation}
where $n$ is the number of reliable photometric measurements as defined above,
$\nu = n-1$ is the number of degrees of freedom, and $\sigma_i$ is the 
estimated photometric uncertainty. Histograms of $\chi^2_\nu$ are shown in 
Figure~\ref{fig:chi2} for the $J$, $H$, and \KB\ photometry on the 16 
nights common to the entire survey area. The solid curves show the expected 
$\chi^2_\nu$ distribution for 15 degrees of freedom normalized to the total 
number of stars. (In practice, \about 20\% of the stars may have more than or
less than 16 measurements depending on if the star is in the overlap region 
of adjacent tiles or if some of the individual photometric measurements were 
discarded as just described.)
Figure~\ref{fig:chi2} shows that the observed and expected $\chi^2_\nu$ 
distribution agree rather well for the majority of the stars, demonstrating 
that the estimated photometric uncertainties accurately reflect the expected 
photometric scatter due to random noise. A number of stars have significantly 
larger values of $\chi^2_\nu$ than expected for random noise. As discussed in 
Section~\ref{id}, the majority of these objects are true variable stars.

\begin{figure}
\insertplot{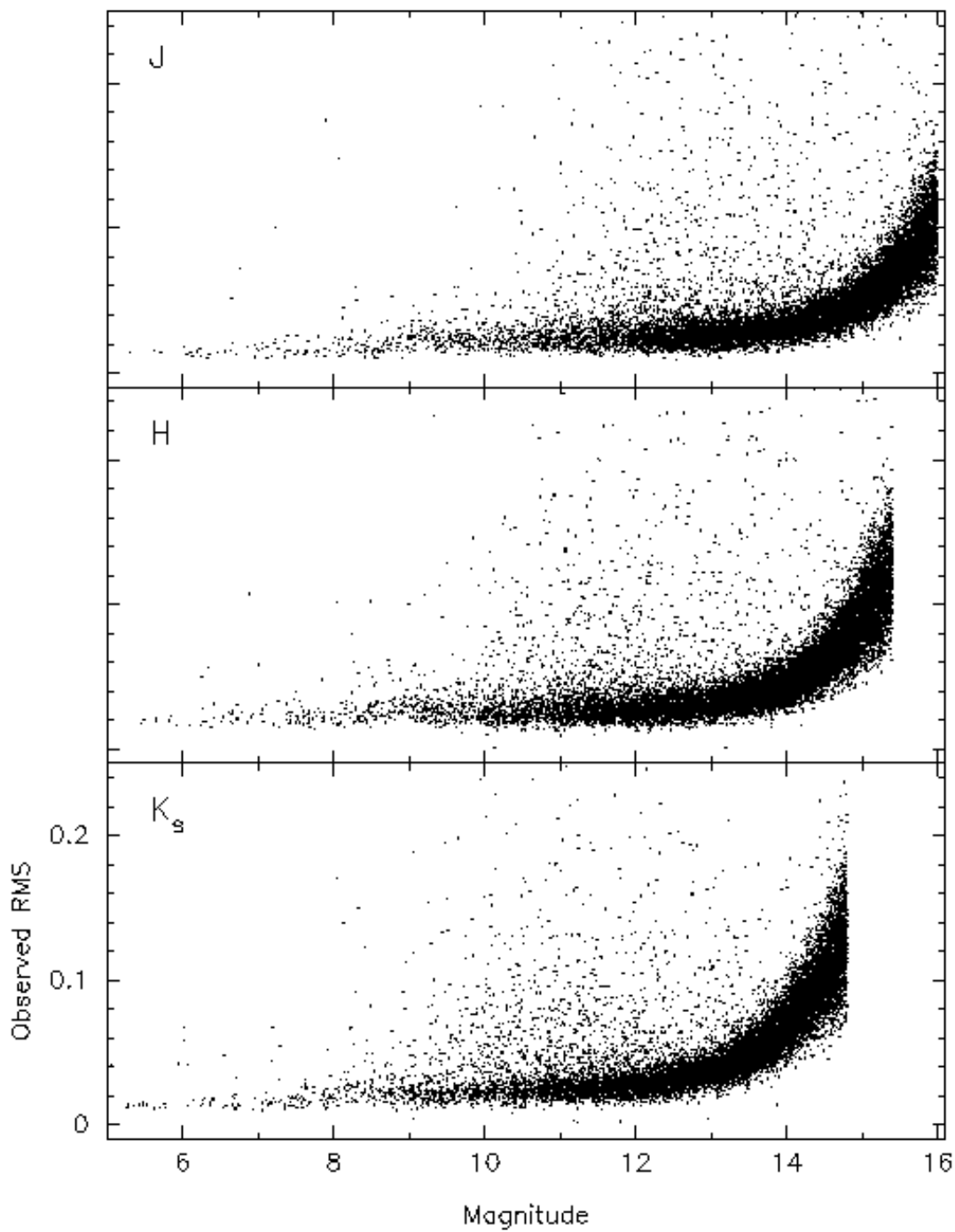}{6.8}{8.4}{0.0}{1.4}{0.90}{0}
\caption{
  The observed photometric RMS in the time series data as a function of 
  magnitude for stars brighter than the defined completeness limits. The 
  observed RMS ranges from \about 0.015\M\ for the brightest stars to 
  \aboutless 0.15 magnitudes (i.e. signal to noise ratio $\ge$ 7) for stars 
  at the completeness limit. 
  \label{fig:rms}
}
\end{figure}

To estimate the signal to noise ratio of the photometry, Figure~\ref{fig:rms} 
plots the observed photometric RMS ($\sigma_{\rm obs}$) in the time series for 
each star as a function of magnitude for objects brighter than the 
completeness limits. The observed RMS in the time series photometry was 
computed from the individual magnitudes ($m_i$) and photometric uncertainties
($\sigma_i$) using
\begin{equation}
\label{eq:sigma_obs}
   \sigma_{\rm obs}^2 = {n\;\sum_{i=1}^{n} w_i(m_i - \bar{m})^2 \over
                         (n-1)\;\sum_{i=1}^{n}w_i},
\end{equation}
where $w_i = 1.0/\sigma_i^2$ is the weight assigned to each observation.
We also define the expected photometric RMS ($\sigma_{\rm noise}$) in the time 
series data due to random noise using
\begin{equation}
\label{eq:sigma_noise}
   \sigma_{\rm noise}^2 = {n\over\sum_{i=1}^{n}(1.0/\sigma_i^2)}.
\end{equation}
Figure~\ref{fig:rms} shows a correlation with magnitude as expected if the 
observed RMS in the time series is mostly due to photometric noise and not due 
to intrinsic variability. The observed RMS values range from a minimum of 
\about 0.015\M\ for the bright stars to \aboutless 0.15\M\ for stars near the 
completeness limit. The observed RMS floor of \about 0.015\M\ for the brighter 
stars is interpreted as the minimum photometric repeatability for these data, 
and consequently, a minimum photometric uncertainty of 0.015\M\ has been 
imposed on all of the photometric measurements. Based upon the estimated 
photometric uncertainties produced by the IPAC data reduction pipeline, we 
find that 97\%, 89\%, 86\% of the stars at $J$, $H$, and \KB\ respectively 
have a signal to noise ratio per measurement $\ge$ 10, and 99\% have a signal 
to noise ratio $\ge$ 7.

\section{Variable Stars in the Orion~A Molecular Cloud}
\label{variable_stars}

\subsection{Identification}
\label{id}

Operationally, a variable star is a star that exhibits larger photometric 
variations over the course of a time series than expected based upon the 
photometric uncertainties. Several techniques have been utilized in the 
literature to identify variable stars, each with its own merits and 
limitations. Historically, variable stars were identified through visual 
inspection of image data or light curves. This approach has the advantage that 
different types of variability can be identified visually that may be 
difficult to pick out in an automated fashion, and suspicious photometric 
measurements can be eliminated by inspection of images. However, the 
subjective nature of this process and the large number of stars in our survey 
makes this an impractical option (although we did look at several thousand 
light curves!), and more quantitative measures of the variability were 
examined. 

One quantitative estimator of photometric variability is the reduced 
chi-squared ($\chi^2_\nu$) of the observed magnitudes (see Eq.~\ref{eq:chi2}), 
which directly translates into a probability that the observed variations can 
result from gaussian random noise \citep{Bevington69}. The $\chi^2_\nu$ 
technique is amenable to multi-band photometry in that the probability that a 
high $\chi^2_\nu$ is observed in 2 or more bands can be readily computed. The 
disadvantage to this approach though, is that noise is often non-gaussian in 
confused regions, and a single ``bad'' measurement can cause a high 
$\chi^2_\nu$ despite our efforts to remove such observations. Further, 
the $\chi^2_\nu$ statistic does not take advantage of correlated changes in 
multi-band magnitudes as a function of time that can be used to identify 
variables with a low amplitude. These considerations motivated \citet[see also 
Stetson~1996]{WS93} to propose an alternate statistic that correlates the 
photometric fluctuations in multi-band photometry. Although the advantages are 
exactly the limitations of the $\chi^2_\nu$ technique, the main difficulties 
with this alternate variability index are that its mathematical properties are 
not as well understood as $\chi^2_\nu$ (except through Monte Carlo 
simulations), plus it dilutes the signatures of variability occurring 
only in a single band.

After examining the results from each of the three techniques described above, 
it became apparent that a single criterion could not be implemented to identify
all variable stars, and a combination of these methods was therefore used.
First, a uniformly selected list of variable stars was created using the 
Stetson variability index, $J$, as defined in \citet{Stetson96}, but re-naming 
the index $S$ to avoid confusion with the 2MASS $J$-band magnitude. Much of 
the analysis of variable star properties that follows uses this uniformly 
selected list. The Stetson variability index was computed for each star from 
the observed $J$, $H$, and \KB\ magnitudes and associated photometric 
uncertainties as 
\begin{equation}
    S = {\sum_{i=1}^p g_i\ {\rm sgn}(P_i)\sqrt{|P_i|}\over\sum_{i=1}^n g_i},
\end{equation}
where $p$ is the number of pairs of observations for a star taken at the same
time, $P_i = \delta_{j(i)}\delta_{k(i)}$ is the product of the normalized 
residuals of two observations, and $g_i$ is the weight assigned to each 
normalized residual (see Stetson~1996). The normalized residual for a given 
band, $\delta_i$, is computed as 
\begin{equation}
    \delta_i = {\sqrt{n\over{n-1}}\ {m_i - \bar{m}\over\sigma_i}},
\end{equation}
where $n$ is the number of measurements used to determine the mean magnitude
$\bar{m}$ and $\sigma_i$ is the photometric uncertainty (see 
Section~\ref{photometry}). In the case of a star with a measurement in only 
one time sample, the product of the normalized residuals was set to $P_i = 
\delta_{j(i)}^2 - 1$. The formula for $P_i$ is different for stars with only a 
single measurement since the expectation value for $\delta_{j(i)}^2$ is 1 for 
random, gaussian noise. The weights, $g_i$, are set based upon the number of 
detections in a given time sample. Time samples with three band detections 
were assigned a weight of 2/3 (implying a total weight of 2.0 for the 3 
possible pair combinations), while 1 or 2 band detections were assigned a 
weight of 1.

\begin{figure}
\insertplot{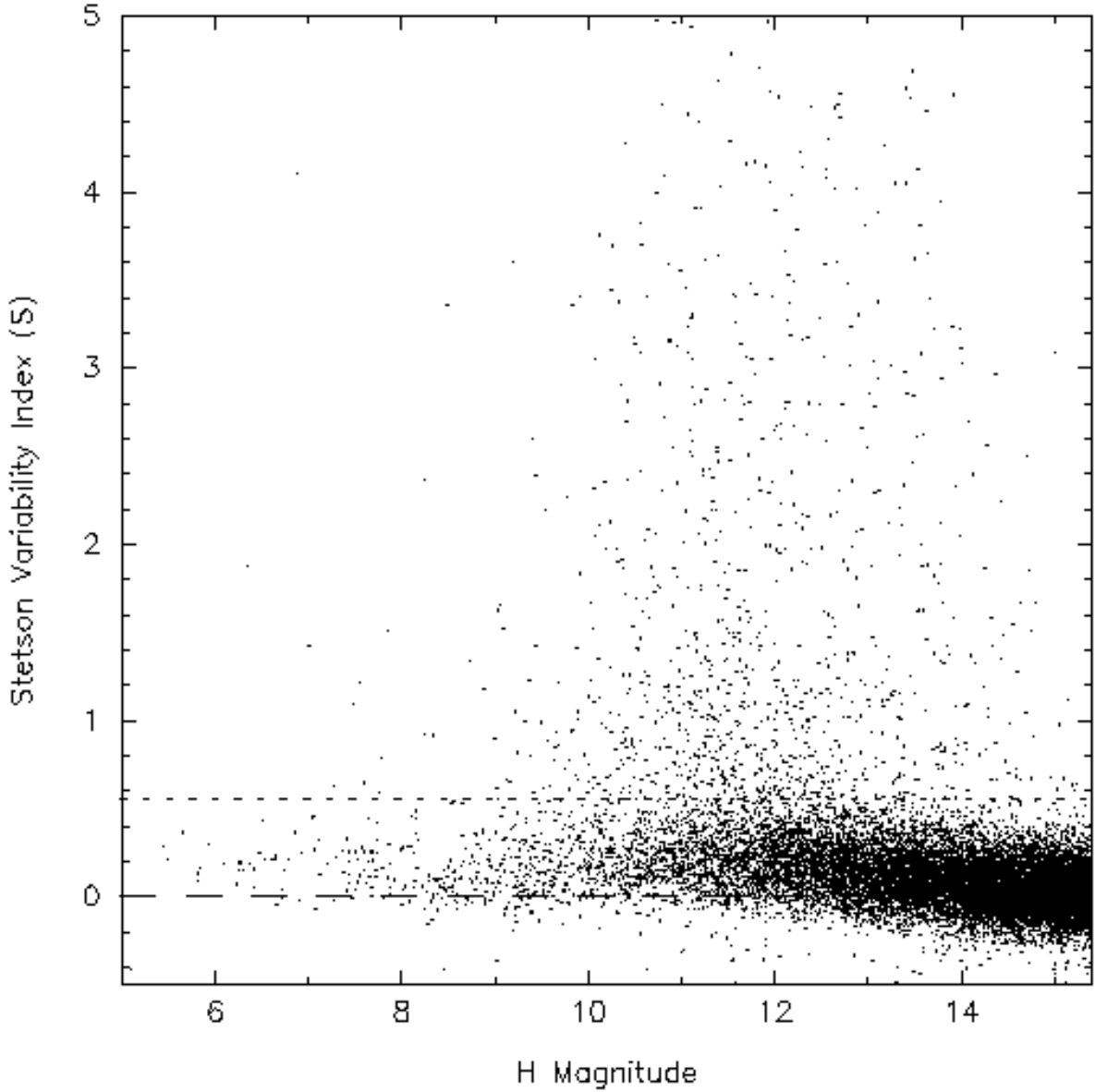}{6.8}{8.4}{0.0}{1.8}{1.00}{0}
\caption{
  The Stetson variable index (S) plotted as a function of the $H$ magnitude 
  for stars brighter than $H$=15.4\M. Only data for stars in the 16 nights 
  common to all seven tiles were used to compute the variability index.
  The dashed line at $S=0$ shows the expected value of the variability index
  for non-variable stars, and the dotted line at $S=0.55$ represents the 
  minimum adopted value used to identify variable stars in this study.
  The origin of the positive bias in the computed index values is unknown, 
  and suggests that a weak correlation exists between the $J$, $H$, and \KB\ 
  photometry, possibly from the fact that the three bands were observed at the 
  same time. Note that 53 stars with $S > 5.0$ are not shown for clarity.
  \label{fig:stetson}
}
\end{figure}

Figure~\ref{fig:stetson} shows the Stetson statistic as a function of the 
$H$-band magnitude for photometry obtained on the 16 nights common to the 
entire survey area. For random noise, the Stetson variability index should be 
scattered around zero, and have higher, positive, values for stars with 
correlated variability. For this data set, the Stetson variability index has a 
positive value on average for the brighter stars. The origin of this offset is 
unclear, but suggests that a weak correlation exists between the $J$, $H$, and 
\KB\ photometry, possibly from the fact that the three bands were observed 
simultaneously. Nonetheless, the Stetson variability index is skewed toward 
large positive values around the nominal value, indicating that a number of 
stars exhibit real variability that is correlated between the three 2MASS 
bands. The minimum value of the Stetson variability index which likely 
represents real variability, as opposed to random noise, was estimated through 
visual examination of the light curves as a function of the Stetson index and 
plotting the Stetson index versus the observed $\chi^2_\nu$. Based on this 
analysis, variable stars were defined as objects having a Stetson index 
$S \ge 0.55$. 

The Stetson index was computed as described above for all stars in the sample 
for two time periods. In Sample~1, the 16 nights in common to the entire 
survey area were used to select a spatially complete set of variable stars
based on identical temporal sampling. In Sample~2, all of the March/April 2000 
data were used to obtain a broader assessment of the variability 
characteristics on time scales up to \about 1 month. (The \about 2 month and 2 
year variability characteristics derived from inclusion of the February~2000
and March~1998 data are considered later.) All light curves and images for 
each candidate variable star were visually examined, and 47 stars were removed 
since the photometry was deemed suspect due to a nearby bright star or bright 
nebulosity from the Orion Nebula. As summarized in Table~\ref{tbl:samples}, a 
total of 1006 variable stars were identified in Sample~1, and 1054 stars in 
Sample~2. These two samples largely overlap, as just 72 variables are 
identified only in Sample~1, and 120 only in Sample~2. The light curves for 
stars identified as variable in only one of the time samples were examined, 
and indeed many appear variable for only a portion of the light curve or have 
Stetson indices close to the adopted threshold, but exceeding or falling short 
of the threshold depending on which data were included. While the observed 
photometric $\chi^2_\nu$ did not formally enter into the selection criteria, 
\about 90\% of the stars in both samples have $\chi^2_\nu$ values indicating 
that there is less than a 0.1\% chance that the observed fluctuations could be 
due to random, gaussian noise. The observed $\chi^2_\nu$ distributions 
suggest indicate that less than 1\% of the identified variables are
likely to be spurious due to random noise, although the contamination rate in 
practice is likely somewhat higher due to non-gaussian noise (e.g. nebulosity,
source confusion, etc...) in some regions.

To evaluate whether our selection criteria for photometric variability
were too stringent or too liberal, we estimated the number of variable stars 
we should detect based on the statistics of the observed and expected
$\chi^2_\nu$ distributions shown in Figure~\ref{fig:chi2}. 
Assuming for the moment that only stars with 
$\chi^2_\nu \ge 2$ are variables, and subtracting the expected $\chi^2_\nu$
distribution from the observed distribution, the predicted number of variable 
stars is \about 900 at \KB-band and \about 1300 at $J$-band. Compared to 
\about 1100 variables actually identified in our analysis, we conclude that 
our adopted criteria with regard to the Stetson index have identified 
approximately the number of variables expected given the noise characteristics 
of the data and the observed photometric fluctuations.

In addition to the variable star sample selected strictly according to the
Stetson index, we also include stars in our variable star list that have 
sufficiently impressive variability as judged from visual inspection of light 
curves but which did not meet the imposed Stetson index and PSF thresholds.  
Light curves were identified for visual inspection 
based on one or more of the following quantitative criteria: 
(1) large $\chi^2_\nu$ in one or more bands but otherwise a small Stetson 
    index; 
(2) long term photometric variations, as defined in Section~\ref{timescales};
(3) significant periods in a Lomb-Scargle analysis despite small Stetson index
    (see Section~\ref{periodicity});
and
(4) sources extended at the 2MASS resolution, but having a large Stetson index.
These somewhat arbitrary additions comprise a small percentage (\about 10\%)
of the final variable star list.  

Table~\ref{tbl:variables} summarizes the photometric properties of the 1235
variable stars identified from our data. Included in the table are an ID 
number, the equatorial J2000 coordinates, the average $J$, $H$, and \KB\ 
magnitudes, the observed photometric RMS in each, the number of high quality 
photometric measurements used to assess the variability, the Stetson 
variability index, and a 6 digit flag that indicates the variability exhibited 
by the star. The photometric information and Stetson index reported in this
table were computed using all available photometry meeting the criteria in
Section~\ref{photometry}.  Each digit in the 
variability flag represents a different variability indicator, and is set to 
`1' if the star exhibits that type of variability, and is set to `0' 
otherwise. In order starting from the leftmost digit, the flags represent
(1) variability in the 16 nights common to the survey area as indicated by the
    Stetson index;
(2) variability identified in the entire March/April 2000 time series as
    indicated by the Stetson index;
(3) long term variability relative to either the March 1998 or February 2000 
    data (see Section~\ref{timescales});
(4) periodicity as determined from the Lomb-Scargle periodogram analysis
    (see Section~\ref{periodicity})
(5) candidate eclipsing system (see Section~\ref{eclipses})
and 
(6) variability identified from subjective, visual inspection of the light 
    curves, usually among sources that appear extended at the 2MASS resolution.
Note that one or more of the first 5 digits in the variability flag may be
set for any individual star.

\subsection{Light Curves}
\label{lightcurves}

For each variable star listed in Table~\ref{tbl:variables}, a figure has been
generated showing the $J$, $H$, and \KB\ light curves, and the \KB\ vs. \HK\ 
color-magnitude diagram and $J-H$ vs. \HK\ color-color diagram for the 
time series data. It is not feasible to present figures for all 1235 variable 
stars here, but figures for all variable stars are available in the electronic 
version of this article. (The electronic version of the figures also include
the $J-H$ and $H-K$ light curves, the $J$ vs. $J-H$ color-magnitude diagram, 
postage stamps of the $J$, $H$, and $K_s$ images, and a tabular summary of the 
photometric data.) Further, {\tt .gif} images for each star, along with search 
capabilities by object name or coordinates, links to tabular data, and cross 
references to existing optical and near-infrared catalogs are also currently 
available at the web site 
http://www.astro.caltech.edu/$\sim$jmc/variables/orion. 
Many stars display unique variability characteristics that can only be 
appreciated from inspection of these figures. 

\begin{figure}
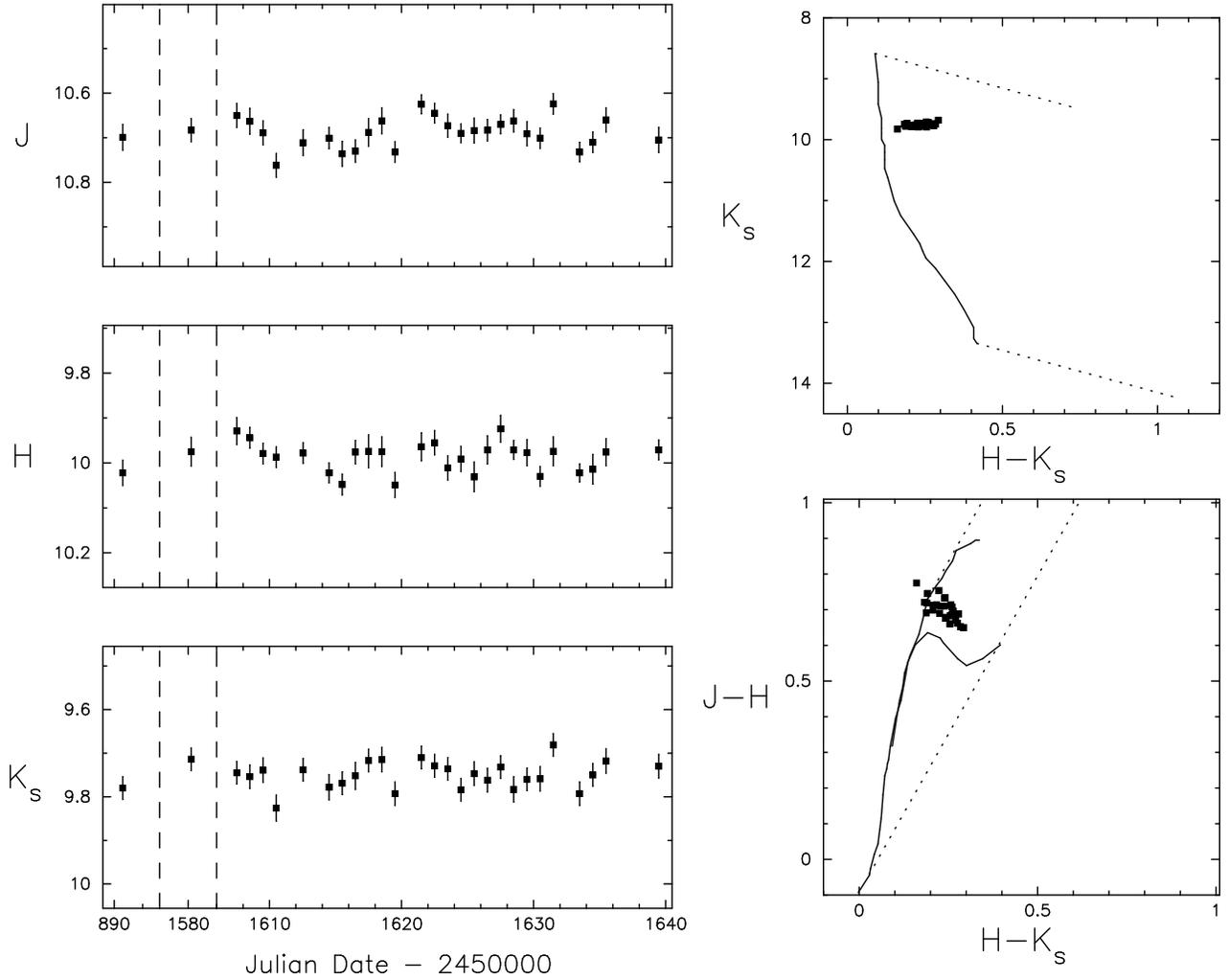

\insertplot{figure04.ps}{7.8}{8.4}{0.0}{0.8}{0.75}{1}
\caption{
  Photometric data for star 5123, which has a Stetson variability index 
  ($S=0.58$) that is just above the limit (0.55) to be classified as a 
  variable. The left panels show the $J$, $H$, and \KB\ light curves.
  The first data point in each light curve is from March 1998, the second data 
  point from February 2000, and the remaining photometry from March/April 
  2000. The vertical bars through the data points represent the $\pm 1\sigma$
  photometric uncertainties. The right panels show the \KB\ vs. 
  \HK\ color-magnitude diagram and the $J-H$ vs. \HK\ color-color diagrams for 
  each data point in the time series, where the dotted line represents the 
  interstellar reddening vector from \citet{Cohen81} transformed into the 
  2MASS photometric system \citep{Carp01}. The uncertainties in the 
  photometric measurements have been omitted for clarity. The solid line in 
  the color-magnitude diagram is the 1~Myr pre-main-sequence isochrone from 
  \citet{DM97} for stellar masses between 0.08\msun\ and 3\msun. The solid 
  curves in the color-color diagram are the loci of red giant and 
  main-sequence stars from \citet{BB88} in the 2MASS color system. 
  \label{fig:ex_limit}
}
\end{figure}

\begin{figure}
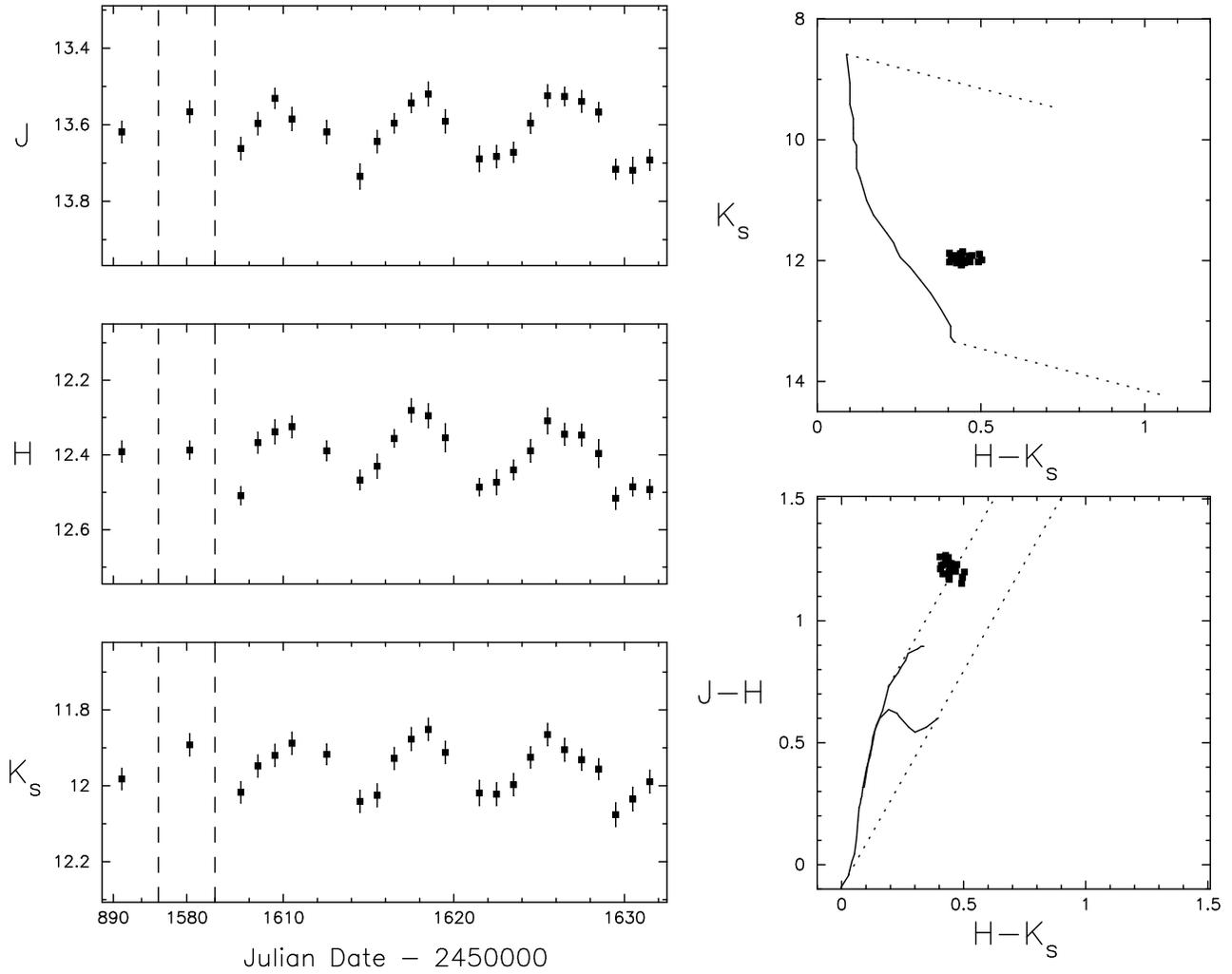

\insertplot{figure05.ps}{7.8}{8.4}{0.0}{0.8}{0.75}{1}
\caption{
  Photometric data for star 3730, classified as a periodic star based upon 
  the Lomb-Scargle periodogram analysis (see Section~\ref{periodicity}).
  The derived period in each band is \about 8 days.
  \label{fig:ex_periodic}
}
\end{figure}

\begin{figure}
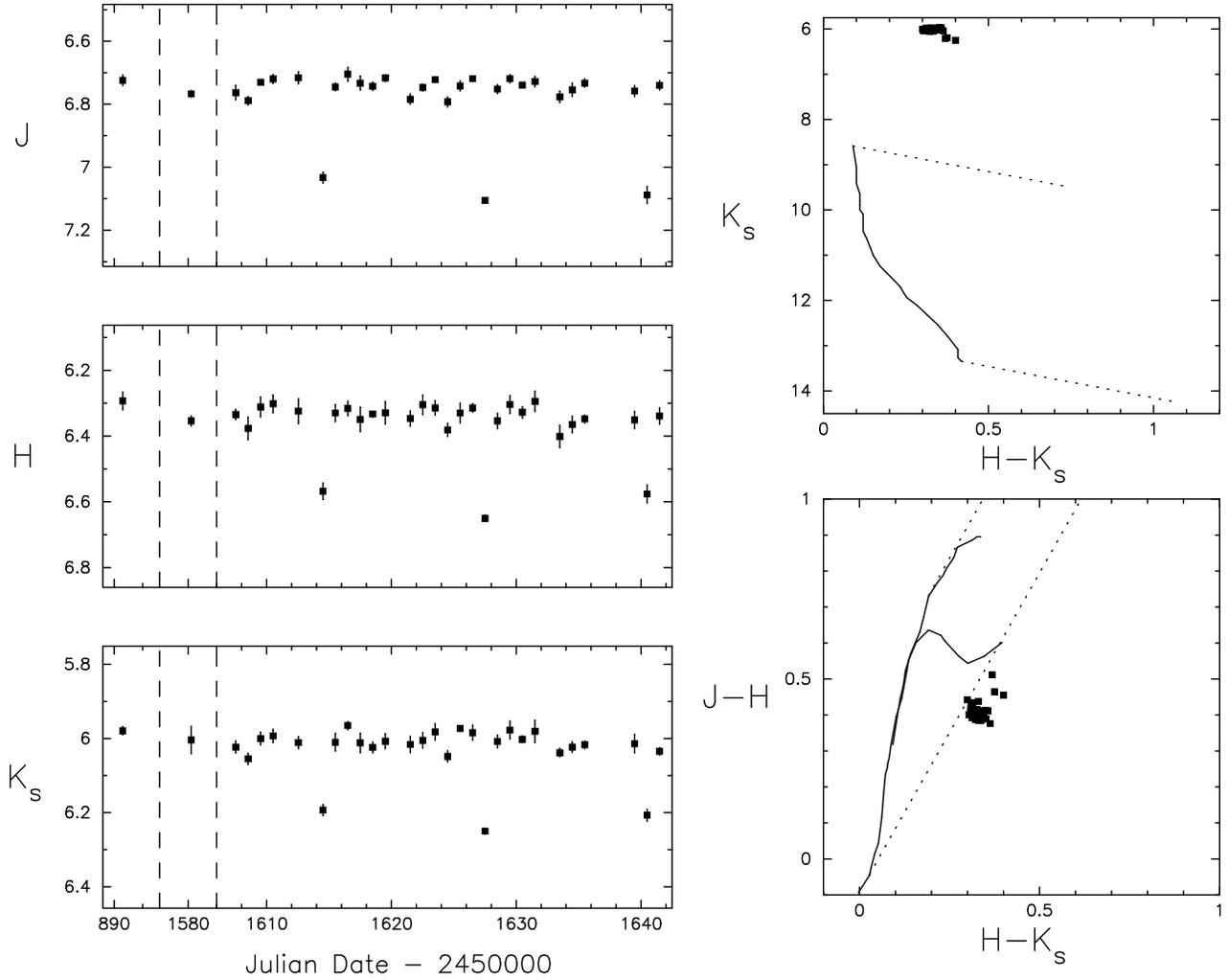

\insertplot{figure06.ps}{7.8}{8.4}{0.0}{0.8}{0.75}{1}
\caption{
  Photometric data for star 8783 (also known as BM Ori), classified as an 
  eclipsing system based on the simultaneous drops in the $J$, $H$, and 
  \KB\ magnitudes on 3 discrete days. This eclipsing system was 
  identified previously from optical observations (see, e.g., Antokhina, 
  Ismailov, \& Cherepashchuk 1989 and references therein).
  \label{fig:ex_eclipse}
}
\end{figure}

\begin{figure}
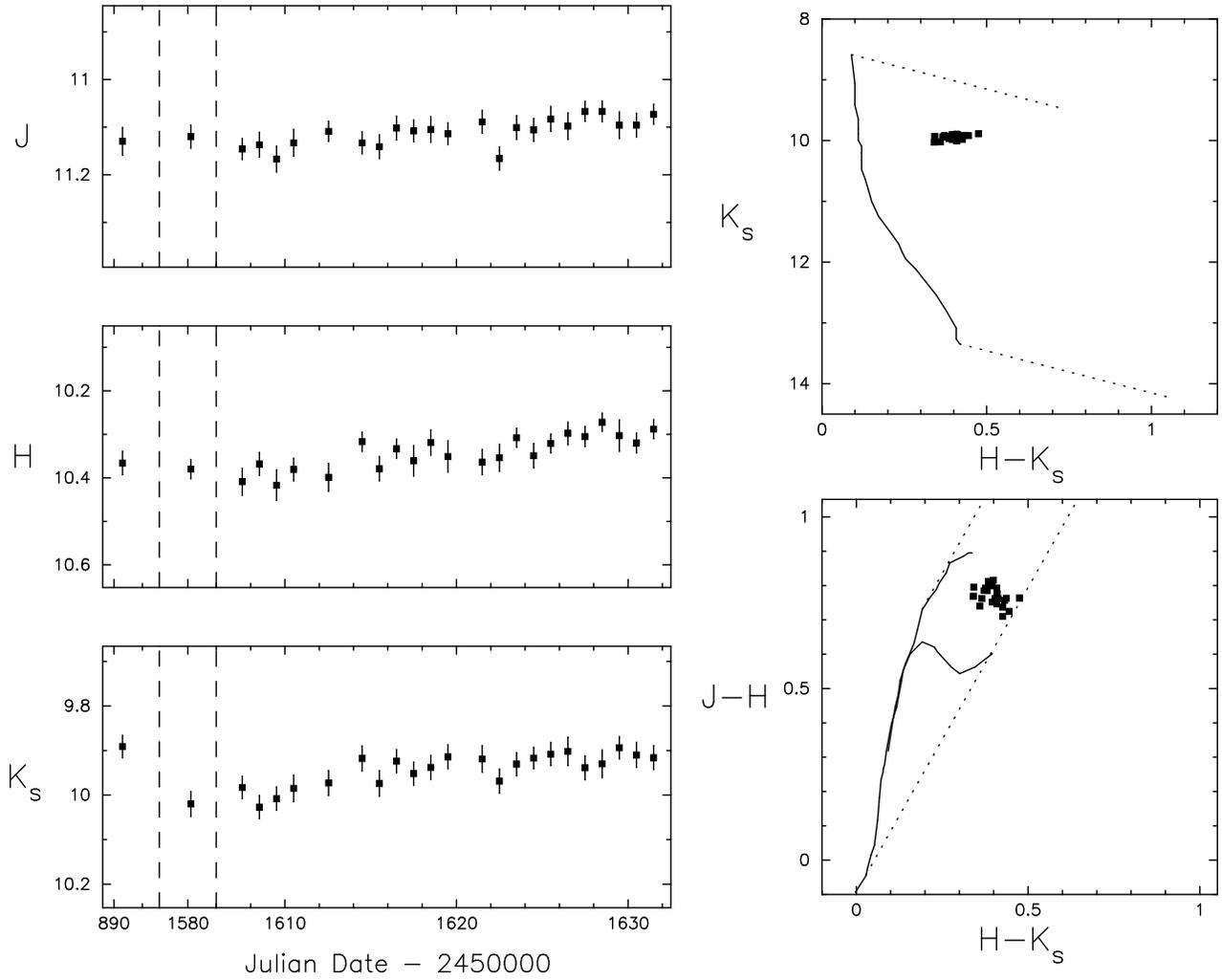

\insertplot{figure07.ps}{7.8}{8.4}{0.0}{0.8}{0.75}{1}
\caption{
  Photometric data for star 4067, an example of a star that steadily
  increased in brightness in the March/April~2000 time period. The March~1998
  photometry though indicates that this has not been a long term trend.
  \label{fig:ex_drift}
}
\end{figure}

\begin{figure}
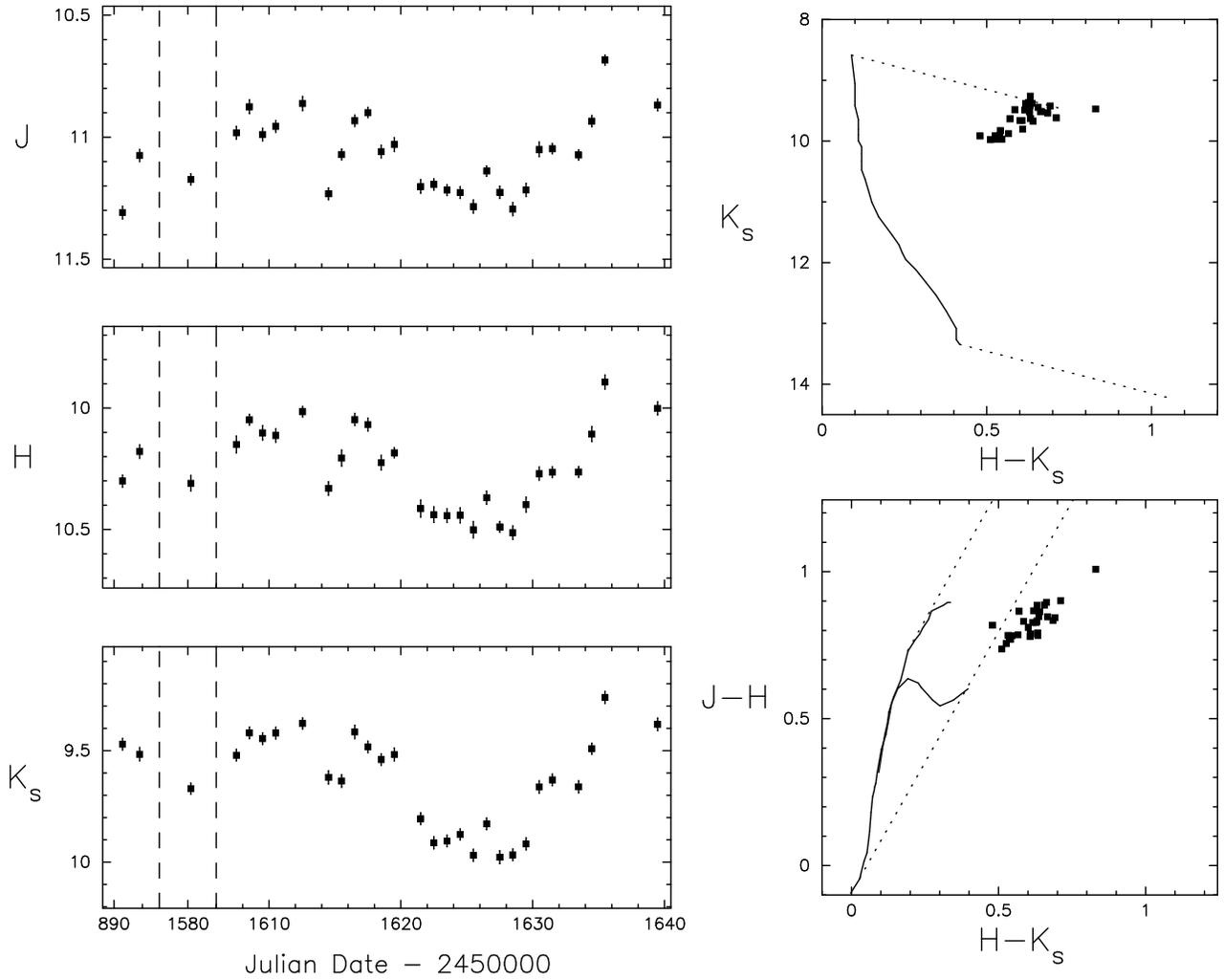

\insertplot{figure08.ps}{7.8}{8.4}{0.0}{0.8}{0.75}{1}
\caption{
  Photometric data for star 5707 (also known as YY~Ori), an example of a star
  in which the stellar colors get bluer as the star gets fainter. The sense
  of the color-magnitude changes are opposite of that expected from either
  rotational modulation by hot spots or extinction variations, but are 
  consistent with a model in which the geometry or mass accretion rate in
  a circumstellar disk changes in time.
  \label{fig:ex_blue}
}
\end{figure}

\begin{figure}
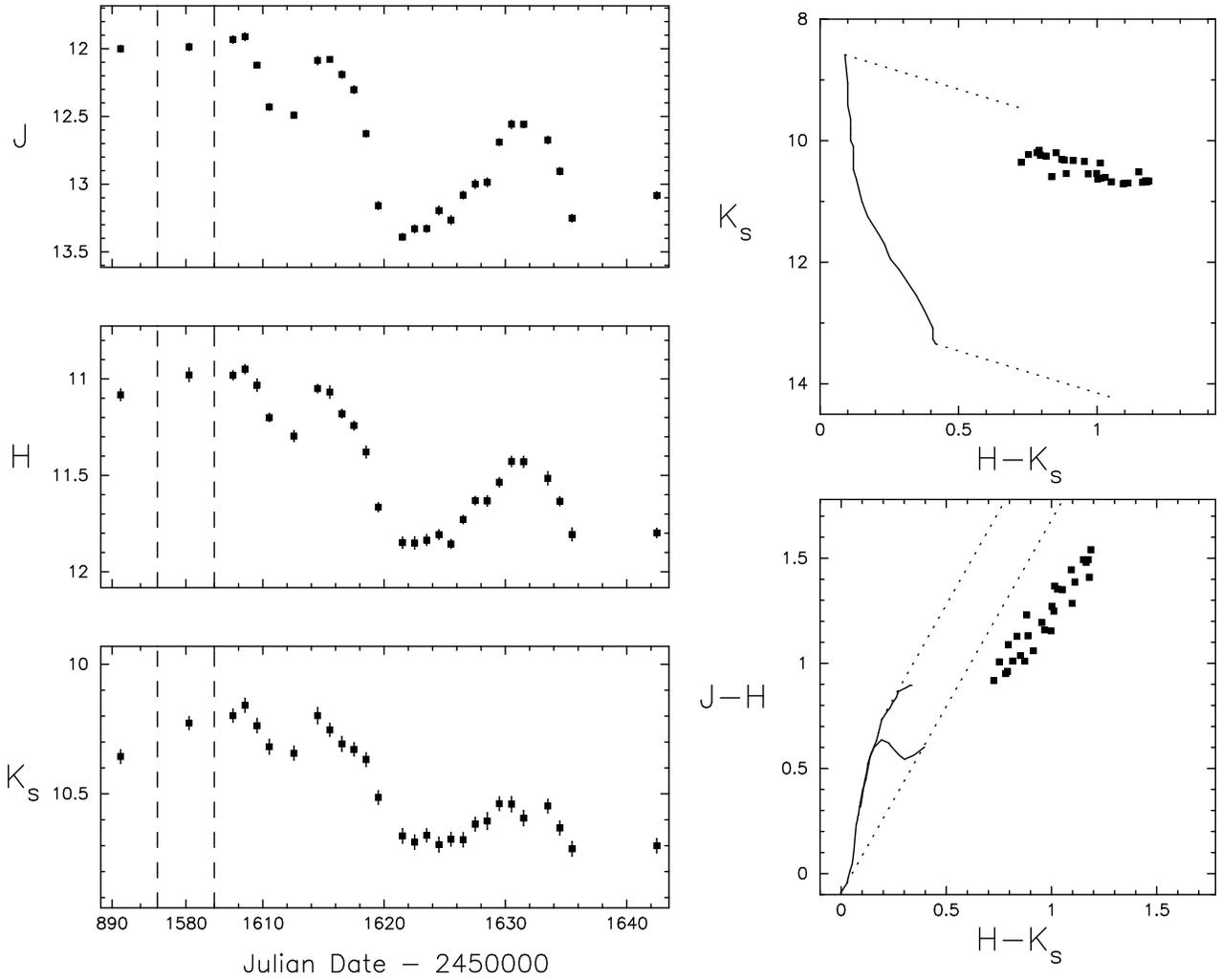

\insertplot{figure09.ps}{7.8}{8.4}{0.0}{0.8}{0.75}{1}
\caption{
  Photometric data for star 11926 (also known as AO~Ori), an example of
  a star where the stellar colors get redder as the star gets fainter.
  Qualitatively the photometric fluctuations are consistent with either
  the presence of time variable hot spots or variations in the amount of
  extinction.
  \label{fig:ex_red1}
}
\end{figure}

\begin{figure}
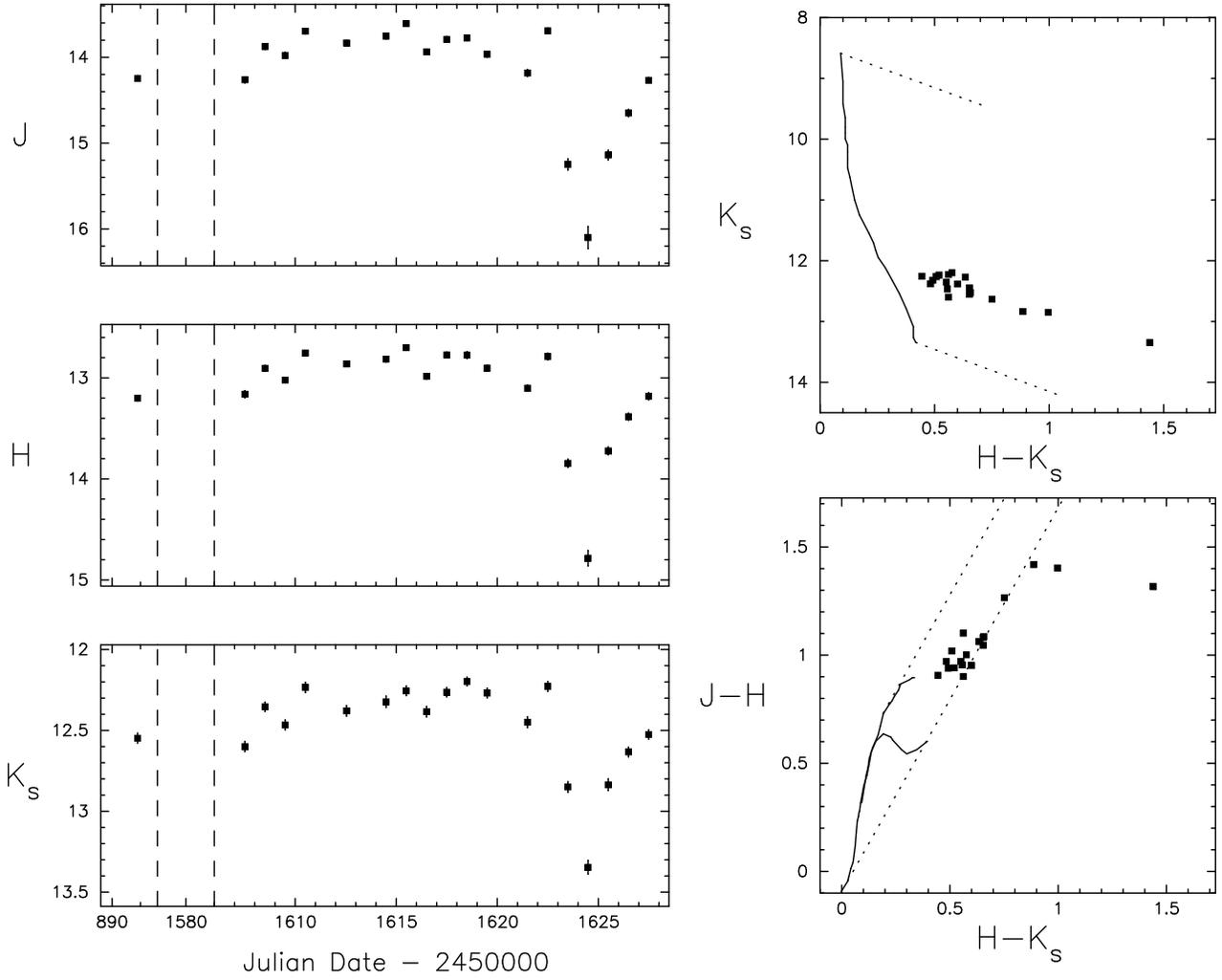

\insertplot{figure10.ps}{7.8}{8.4}{0.0}{0.8}{0.75}{1}
\caption{
  Photometric data for star 1048, another example of a star in which the 
  stellar colors get redder as the star gets fainter (see also 
  Fig.~\ref{fig:ex_red1}). In this instance, the stellar magnitudes are
  relatively constant for the first two weeks of the time series before the
  star becomes fainter by over 1\M\ in each band with progressively redder
  colors over a period of a couple of days. As the star faded in brightness, a
  near-infrared excess become apparent for 2-3 days.
  \label{fig:ex_red2}
}
\end{figure}

\begin{figure}
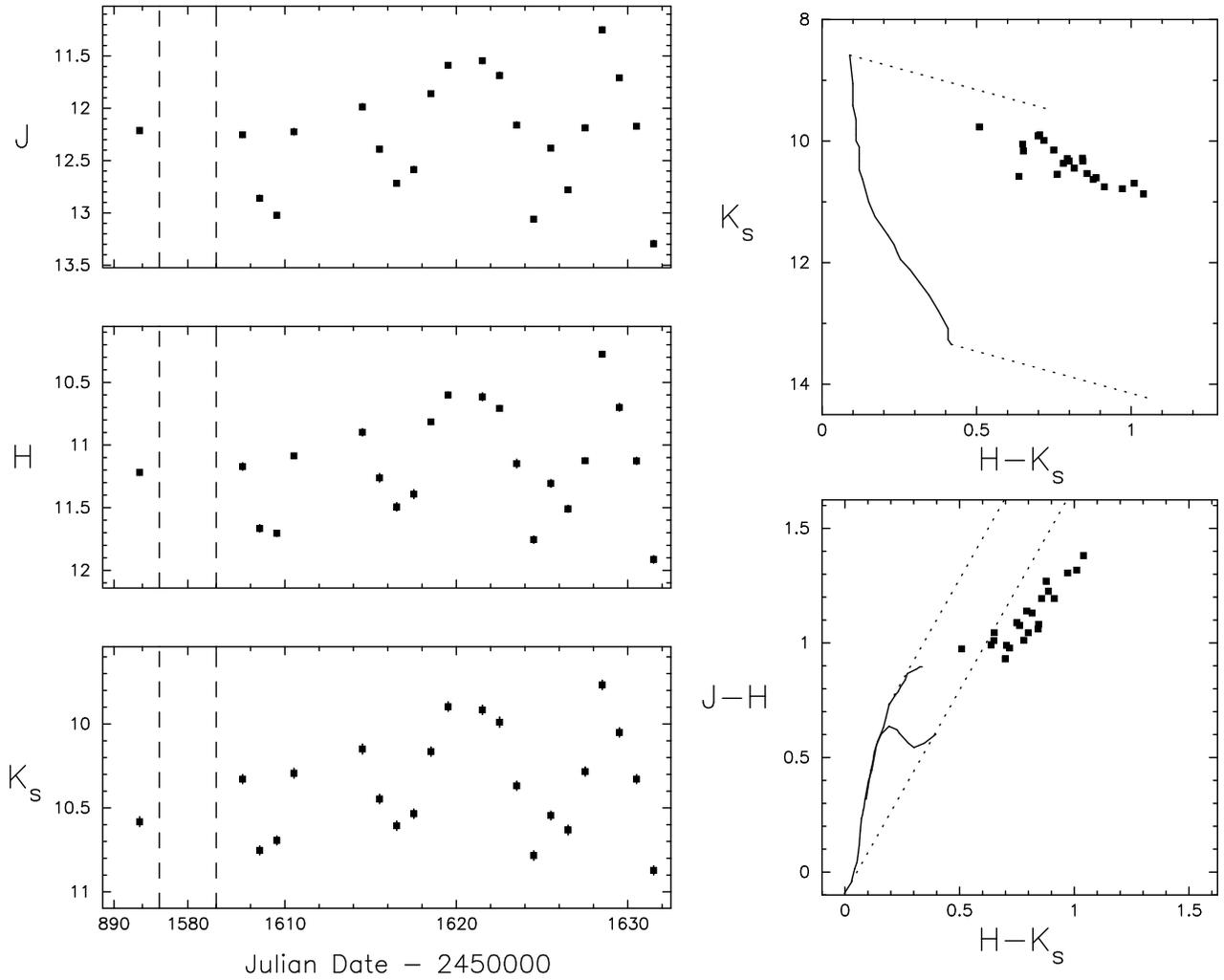

\insertplot{figure11.ps}{7.8}{8.4}{0.0}{0.8}{0.75}{1}
\caption{
  Photometric data for star 13688 (also known as AW~Ori), another example of 
  a star in which the stellar colors get redder as the star gets fainter (see 
  also Figs.~\ref{fig:ex_red1} and \ref{fig:ex_red2}). In this instance, 
  the star exhibits quasi-periodic fluctuations, and the colors and 
  magnitudes vary along a vector that is shallower in the color-color diagram 
  than prior examples and steeper in the color-magnitude diagrams.
  \label{fig:ex_red3}
}
\end{figure}

\begin{figure}
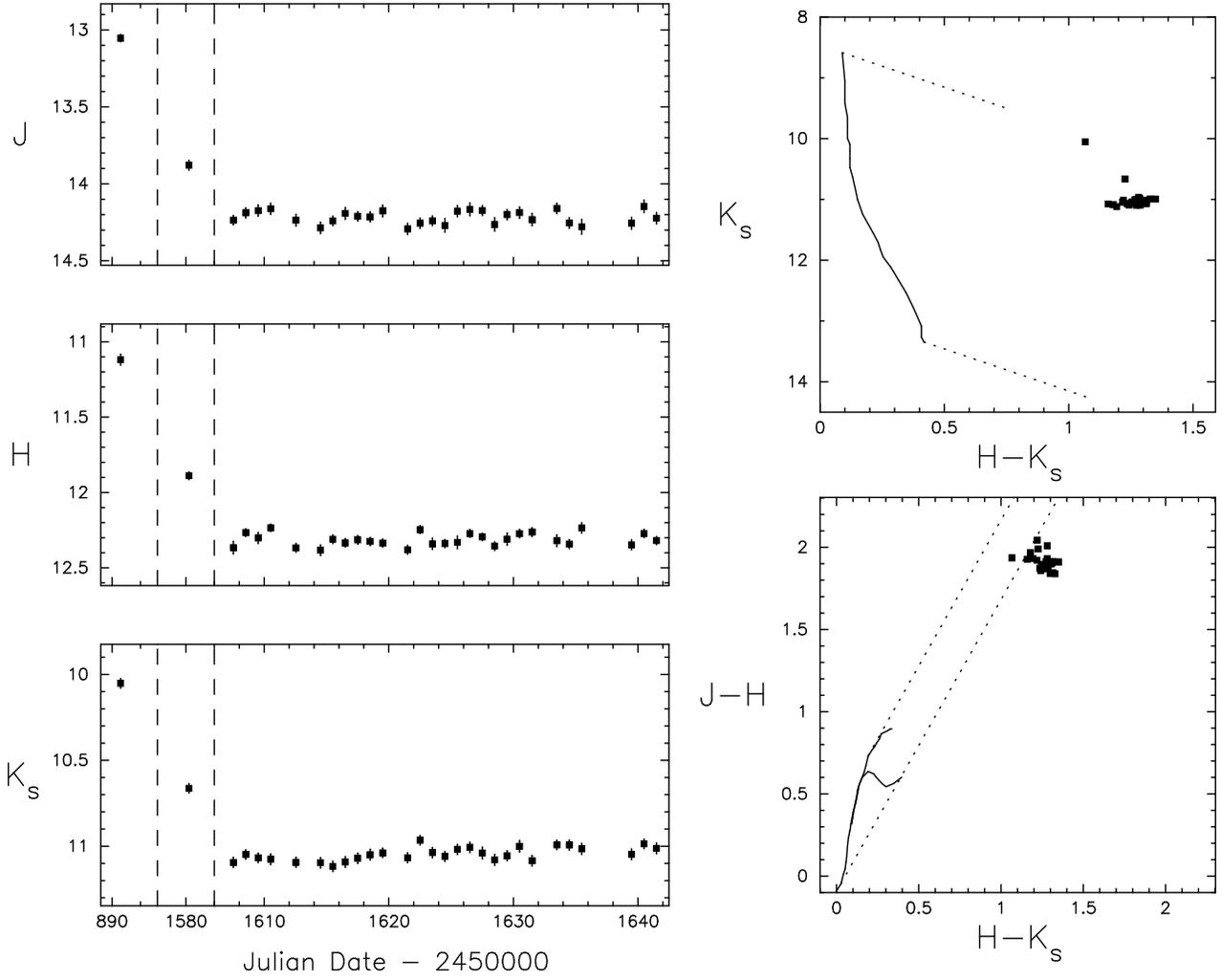

\insertplot{figure12.ps}{7.8}{8.4}{0.0}{0.8}{0.75}{1}
\caption{
  Photometric data for star 10527, an example of a star that is not variable 
  in the March/April 2000 time frame, but exhibits longer term photometric 
  fluctuations in both the March 1998 and February 2000 observations.
  \label{fig:ex_long1}
}
\end{figure}

Figures~\ref{fig:ex_limit}-\ref{fig:ex_long2} present a sampling of the 
observed near-infrared variability characteristics. Figure~\ref{fig:ex_limit} 
shows a star with a Stetson index of 0.58 (just above the adopted threshold of 
0.55 to identify variable stars) that exhibits correlated, low amplitude 
magnitude changes in all three bands. Figure~\ref{fig:ex_periodic} illustrates 
a star with periodic, nearly sinusoidal, colorless variations with a 
peak-to-peak amplitude of \about 0.25\M\ in each band. 
Figure~\ref{fig:ex_eclipse} also shows a possible periodic system where the 
photometric fluctuations do not occur smoothly in time, but on discrete days, 
as expected for an eclipsing binary system. As opposed to the relatively rapid 
fluctuations illustrated thus far, Figure~\ref{fig:ex_drift} shows a star that 
continuously brightened over the March/April~2000 time period. The brightness 
of the star, however, has not changed significantly between March~1998 and 
March~2000, suggesting these variations are not a long term trend.

Some of the larger amplitude variables also display significant color 
variations. Figure~\ref{fig:ex_blue} shows a star (YY~Ori) where 
the photometric fluctuations are largest at \KB-band, and the stellar colors 
become bluer as the star gets fainter. Figure~\ref{fig:ex_red1}, on the other 
hand, illustrates one of the more dramatic instances where the amplitude 
fluctuations are largest at $J$-band and the stellar colors become redder as 
the star gets fainter, with the fluctuations occurring on a day-to-day basis. 
Not all stars with color changes vary continuously in time, as 
Figure~\ref{fig:ex_red2} presents a star in which the photometry was 
relatively constant for the first two weeks of the March/April 2000 
observations before the star became fainter by 1-2\M\ in each band with 
progressively redder colors. Further, as the star faded in brightness, a 
near-infrared excess become apparent for 2-3 days. By the end of the time 
series observations, the magnitudes and colors were nearly back to the values 
at the start of the observations. Figure~\ref{fig:ex_red3} shows another star 
with large color variations, but in which the slope of the photometric 
fluctuations is less steep in the color-color diagram and more steep in the 
color-magnitude diagrams than in the prior examples. The photometric
fluctuations in this star are also suggestive of quasi-periodic variations. 

Finally, we illustrate two examples of stars that display long term photometric 
variability. The star in Figure~\ref{fig:ex_long1} was not identified as a 
variable in the March/April 2000 time series data, but decreased in brightness
by \about 1\M\ relative to the March 1998 observations. The star in 
Figure~\ref{fig:ex_long2} is variable in the March/April 2000 observations, 
and shows even larger photometric fluctuations when the March~1998 data are
included. The long term variability characteristics are such that while the 
star became fainter at $J$-band over a two year period, it simultaneously 
brightened at \KB-band.

\section{Properties of the Variable Stars}
\label{properties}

In this section we discuss the properties of the 1006 uniformly selected 
variables stars from Sample~1 (see Table~\ref{tbl:samples}), including 
the spatial distribution of near-infrared variables compared with other
tracers of the Orion star-forming region, the fraction of the Orion stars 
exhibiting near-infrared variability, and the distribution of the variable 
stars in color-magnitude and color-color diagrams. In the following section we 
analyze the amplitude, color, and time scale characteristics of the variable 
stellar population.

\subsection{Spatial Distribution}
\label{spatial}

The spatial distribution of the variable stars from Sample~1 is presented in 
Figure~\ref{fig:radec}. Also shown for comparison are
(1) the spatial distribution of all stars brighter than \KB=14.8\M\
    displayed both as discrete sources and a surface density map;
(2) H$\alpha$ emitting stars from the Kiso surveys with a Kiso class of
    3, 4, or 5 (Wiramihardja \etal~1991,1993; see also Parsamian \& 
    Chavira~1982);
(3) the 305 variable stars from Sample 1 that have a near-infrared excess in 
    the $J-H$ vs. \HK\ color-color diagram (see Section~\ref{mean}).
and 
(4) a map of the Orion molecular cloud as traced by \thcoj\ emission 
    \citep{Bally87}. 

Figure~\ref{fig:radec} shows that near-infrared variable stars are found in
substantial numbers between declinations of \about $-7$\arcdeg\ and 
$-4.5$\arcdeg\ which closely reflects the distribution of enhanced \KB-band
star counts. The densest concentration of variables is located toward the 
Trapezium region of the ONC at $\delta$ \about $-5.5$\arcdeg\ with a secondary 
density peak near \NGC1977 at ($\alpha,\delta$) \about 
(83.8\arcdeg,$-$4.8\arcdeg). Variable stars with a near-infrared excess are
distributed over a more restricted declination range than the complete 
variable star sample and share a similar spatial distribution with the 
H$\alpha$ stars, although the Orion Nebula prevents identifying H$\alpha$ 
emitting stars in the Trapezium region. The OB stars, not shown in
Figure~\ref{fig:radec}, are a sparser population, but nonetheless are 
concentrated within the same declination range as the variable and H$\alpha$ 
emitting stars \citep{Brown94}. Further, the spatial distribution of each of 
these stellar populations closely resembles that of the large scale molecular 
cloud structure. These properties suggest that most of the near-infrared 
variables are associated with the Orion~A molecular cloud.

\subsection{Fraction of Stars Exhibiting Near-Infrared Variability}
\label{fraction}

We have taken advantage of the large areal coverage of our observations to
estimate the spatial extent and statistical membership of the stellar 
population associated with the northern portion of the Orion~A molecular cloud 
using the \KB-band star counts as outlined in the Appendix. The results of 
this analysis are summarized in Table~\ref{tbl:cluster}, including various 
surface densities at which the stellar density enhancement identified in
Figure~\ref{fig:radec} can be defined, the corresponding angular 
extent, stellar census counts, and the number and fraction of 
variable stars at these surface density limits. At the lowest defined surface 
density level (1$\sigma$ above the mean field star surface density), the 
density enhancement contains \about 2700 stars with \KB$\le$14\M\ distributed 
over a $0.4^\circ\times2.4^\circ$ ($3.4\:\rm{pc}\times20\:\rm{pc}$) region.
The 1$\sigma$ boundary encompasses 786 variable stars, or 78\% of 
the total variable star population. The 220 variable stars not within this
boundary tend to be located on the periphery of the density enhancement and may 
also be cloud members since the 1$\sigma$ boundary is artificially limited by 
the field star contamination. If not true cloud members, these variables may 
be part of the larger scale Orion star forming region which extends
many tens of parsec in all directions, although some may be field stars as 
discussed in the Appendix. 

With the total number of stars in the Orion~A molecular cloud detectable to 
the sensitivity limits of our observations quantified, we can now establish 
the fraction of the stellar population that is variable.
At the lowest surface density at which we can define the density enhancement, 
\about 29\% of the stellar members are variable 
within the photometric noise limits of the data (see 
Table~\ref{tbl:cluster}). This fraction does not change appreciably with the 
surface density used to define the density enhancement. While there is some 
suggestion that the variable fraction decreases in the inner $5'\times5'$
near the Trapezium, it is more difficult to identify variable stars in this 
region due to the bright sky background from the Orion Nebula which increases 
the photometric noise. 

To investigate the variable star fraction as a function of magnitude, 
Figure~\ref{fig:khist} shows \KB-band histograms for the cloud population and 
for the variable stars within the 1$\sigma$ boundary. The field stars
have been subtracted from the cloud population using a procedure similar 
to that described in the Appendix, but applied to the differential magnitude 
intervals. The \KB-band histogram for the cloud population as a whole is 
broadly peaked at 
\about 11-13\M\ with a median value of 11.7\M\ that is significantly brighter
than the completeness limit of the observations (\KB=14.8\M). Similarly, the 
\KB-band histogram for the variable star population peaks well above the 
completeness limit, but at a median magnitude (11.1\M) that is brighter than 
the cloud population. Further, \about 45\% of the cloud members 
brighter than \KB=11\M\ show near-infrared variability, but only 14\% of stars 
fainter than \KB=12\M\ are near-infrared variable within the noise limits of 
the data.

To determine whether the fainter cloud members are intrinsically less 
variable or whether the increased photometric noise (see Fig.~\ref{fig:rms}) 
is masking low amplitude fluctuations, a Monte Carlo simulation was run in 
which simulated variability common to all 3 bands was added to random noise. 
The amplitude of the noise in each band was set based on the photometric noise 
in the actual data. The variability was simulated with a gaussian random
number generated with a dispersion ranging from 0.01 to 0.2 mag. Variable 
stars in the simulated photometry were then identified using a Stetson index 
threshold of 0.55. For \KB=12\M, \about 1/2 of the simulated stars are 
identified as variable when the dispersion in the photometric fluctuations is 
0.03\M. Since the median \KB-band amplitude dispersion for identified 
variables over all magnitudes is \about 0.05\M\ (see 
Section~\ref{amplitudes}), these results suggest that the lack of variable 
stars at faint magnitudes is indeed a result of increased photometric noise.

\subsection{Mean Colors and Magnitudes}
\label{mean}

Basic physical properties of the variable star population can be constrained
by the observed stellar magnitudes and colors. As already shown in
Figure~\ref{fig:khist}, the histogram of the \KB-band magnitudes for the
variable star population peaks slightly brighter than the peak for the cloud
population. The overall peak has been seen in several previous studies of the 
Trapezium region of the ONC \citep{Z93,M96,Lada96,Simon99,Lynne00}, and can
be accounted for by a combination of pre-main-sequence evolution for a \about
1~Myr cluster and a Miller-Scalo Initial Mass Function \citep{Z93}. Assuming
this age and neglecting extinction and near-infrared excesses, the peak in the
variable star \KB-band histogram corresponds to a stellar mass of \about
0.2-0.6\msun\ according to the evolutionary theory of \citet{DM97} for a 
cluster distance of 480~pc \citep{Genzel81}. Extensive photometric and 
spectroscopic observations have shown that the ONC stellar population near the
Trapezium does in fact consist predominantly of young 
(\aboutless 1~Myr) low mass stars \citep{Herbig86,Lynne97}. Since the peak in 
the \KB-band histogram is a feature of the entire stellar density enhancement
and not just near the Trapezium region, this result suggests that the majority 
of the cloud population and hence variable stars is relatively young and of 
low mass. 

\begin{figure}
\epsscale{1.0}
\plotone{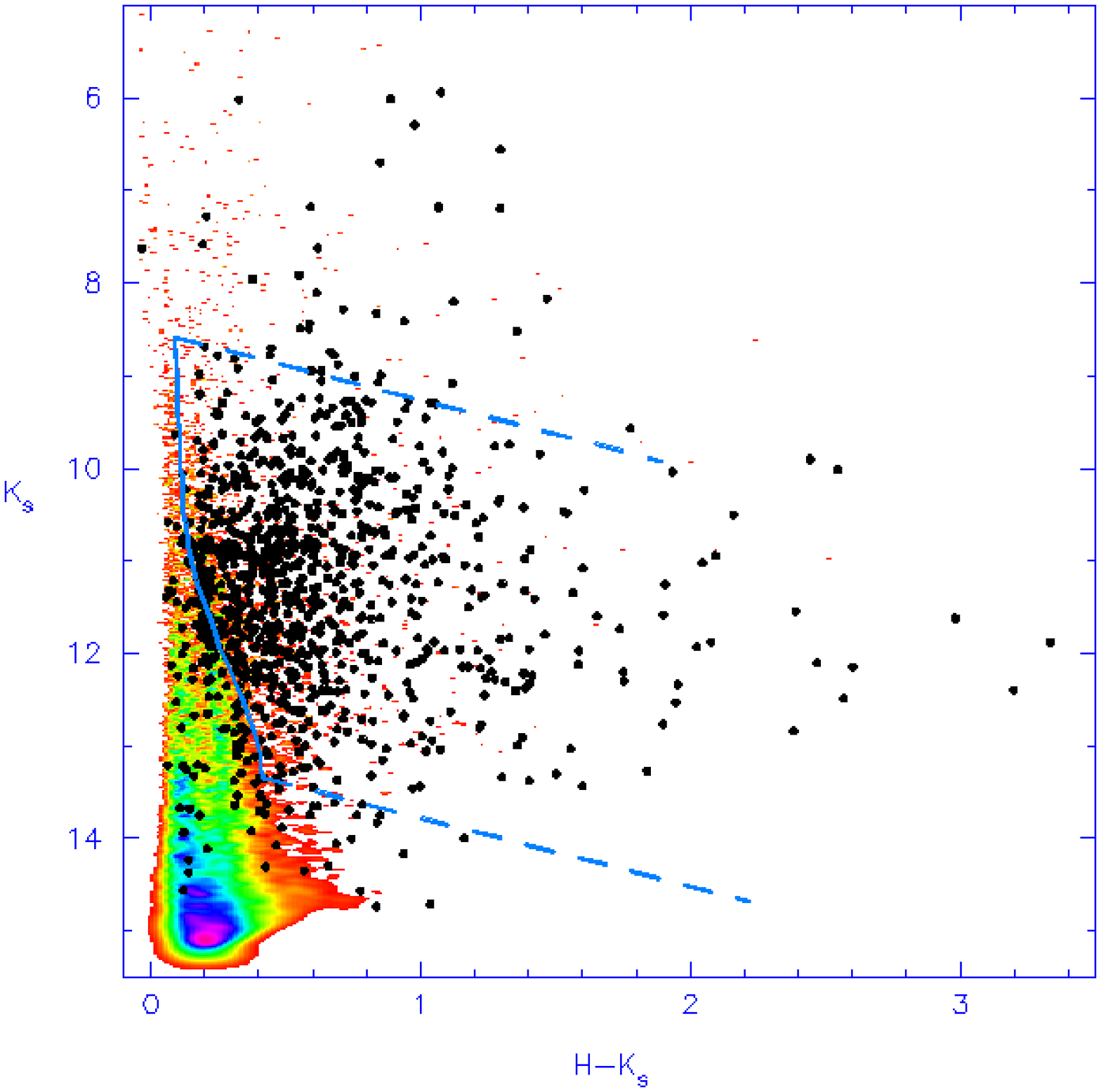}
\epsscale{1.00}
\caption{
  \KB\ vs. \HK\ color-magnitude diagram for all stars (color scale) and 
  the variable stars (black circles) identified from the 16 nights in which
  all tiles were observed. For reference, the solid blue curve shows 
  the 1~Myr pre-main-sequence isochrone from  \citet{DM97} for stellar masses 
  between 0.08\msun\ and 3.0\msun. The dashed lines indicate the 
  reddening vector for 10 magnitudes of visual extinction from \citet{Cohen81} 
  transformed into the 2MASS photometric system \citep{Carp01}. The 
  lowest halftone is 1\% of the peak density. This figure shows that the 
  observed magnitudes and colors for the majority of the variable population 
  is consistent with reddened pre-main-sequence stars with masses 
  \aboutless 3\msun.
  \label{fig:khk}
}
\end{figure}

The \KB\ vs. \HK\ color-magnitude diagram shown in Figure~\ref{fig:khk} 
further supports the conclusion that most of the variable stars are young 
(\aboutless 1~Myr), low mass (\aboutless 1\msun), moderately reddened members 
of the Orion~A molecular cloud.  The color scale in this figure represents all 
detected stars, \about 85\% of which are likely field 
stars (see the Appendix). The non-variable stars (predominantly field stars) 
are located primarily to the left of the 1~Myr isochrone, while most of the 
variable stars (filled symbols) are located to the right and are consistent 
with reddened pre-main-sequence objects.

\begin{figure}
\insertplot{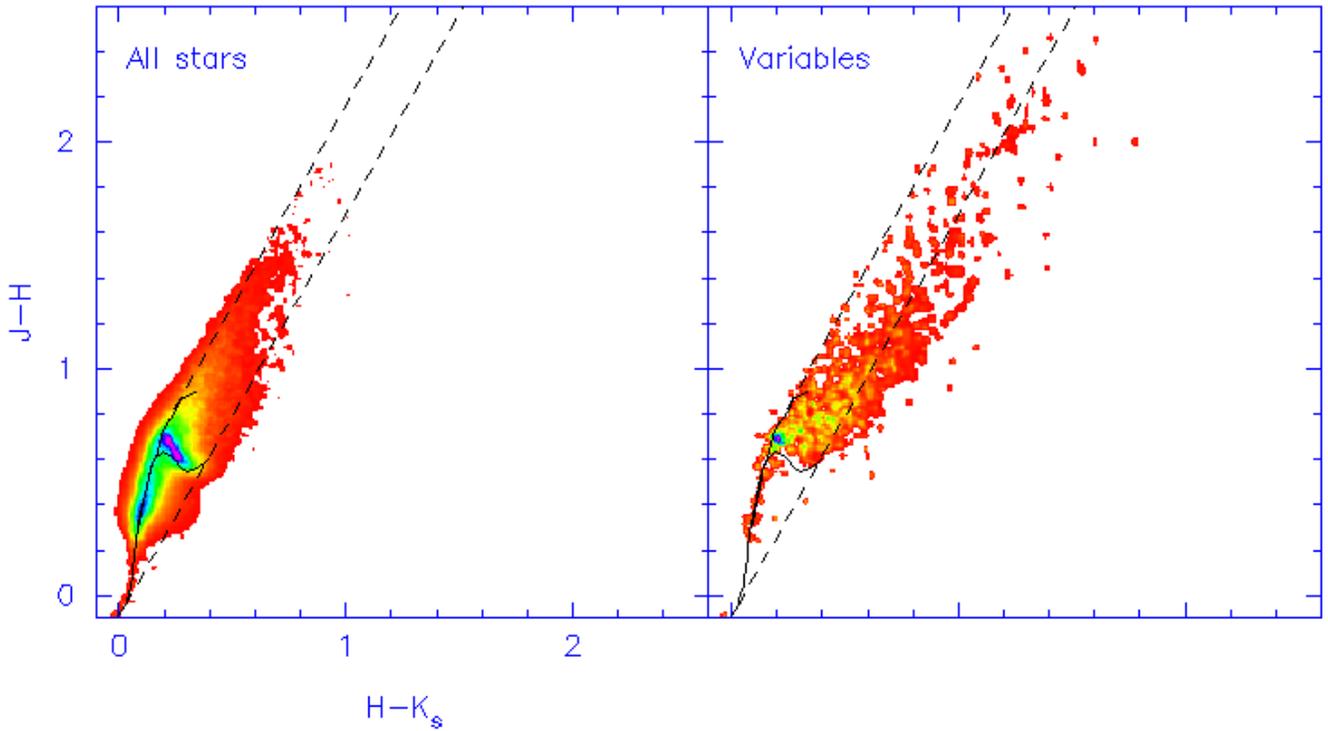}{7.8}{8.4}{0.0}{1.0}{0.7}{1}
\caption{
  $J-H$ vs. \HK\ color-color diagram for all stars (left panel) and the 
  variable stars (right panel) identified from the 16 nights in which
  all tiles were observed. The lowest halftone in each panel begin at 1\%
  of the peak density. The stars represented in the left panel are dominated 
  by field stars unrelated to the Orion~A molecular cloud. The variable stars 
  are on average redder than the field star population, and \about 30\% have 
  colors indicating the presence of a near-infrared excess.
  \label{fig:jhhk}
}
\end{figure}

To further examine the properties of the variable star population, 
Figure~\ref{fig:jhhk} shows the $J-H$ vs. \HK\ color-color diagram for all 
stars in our sample (left panel) compared to the variable stellar population 
(right panel). This figure shows that the majority of non-variables have 
colors consistent with unreddened main sequence stars either in front of 
the cloud or at a variety of distances but along lines of site exterior to the 
cloud boundaries, or reddened main sequence and giant stars seen through the 
cloud. The variable stars are systematically redder compared to the 
non-variable stars, many with colors that place them in a region of the 
diagram that cannot be explained by interstellar extinction. These 
near-infrared colors are characteristic of young, low mass Classical T Tauri 
stars (CTTS's) surrounded by optically thick accretion disks 
\citep{Lada92,Meyer97} and moderately reddened. Approximately 30\% of the 
variable stars have near-infrared colors consistent with CTTS's as identified 
in the $J-H$ vs. \HK\ color-color diagram, suggesting that up to 50\% of the 
variable star population may be CTTS's if the efficiency factor for this 
diagram as described in the Appendix of \cite{Lynne98a} is adopted. Many of 
the variable stars, however, do not have such distinctive near-infrared 
colors. The variable stars without substantial near-infrared excesses are 
either CTTS's with small excesses or weak-line T~Tauri stars (WTTS's) which 
as a class do not have strong signatures of optically thick circumstellar 
disks. Near-infrared variability therefore appears to be a characteristic of 
both CTTS's and WTTS's.

\section{Characteristics of the Variability}
\label{characteristics}

After establishing that most of the near-infrared variable stars are young, 
low mass stars associated with the Orion~A molecular cloud,
we now investigate in more detail the variability characteristics exhibited 
by these stars.  Our goal is to use the statistics inherent in our large sample
to categorize the predominant types of near-infrared photometric fluctuations. 
To include as complete a sample as possible in 
characterizing the photometric variations, we incorporate all of the 
March/April 2000 time series data for the stars listed in 
Table~\ref{tbl:variables} (unless otherwise stated). Strictly speaking, this 
results in non-uniform time coverage for stars as a function of their location 
in the survey region and also incorporates some stars that were included as 
variables based on somewhat subjective criteria. In practice, however, 
\about 77\% of the variables are located in the center tiles 3-5 (see
Table~\ref{tbl:log}) and were observed on at least 25 of the possible 29 
nights. Further, the number of stars included in the variable star list for 
reasons other than high Stetson index amount to less than 10\% of the entire 
sample. Therefore, the following results should closely reflect those obtain 
from a more uniformly selected sample.

\subsection{Amplitude of Magnitude and Color Changes}
\label{amplitudes}

The amplitude of the variability was characterized by computing the observed 
RMS and peak-to-peak fluctuations in the magnitudes and the colors for 
individual stars. Since the observed photometric fluctuations reflect both 
actual astrophysical variability and photometric noise, a statistical 
correction needs to be applied to recover the intrinsic amplitude. For the RMS 
values, the actual variability amplitude ($\sigma_{\rm var}$) was estimated 
by subtracting in quadrature the expected RMS due to photometric noise 
($\sigma_{\rm noise}$; see Eq.~\ref{eq:sigma_noise}) from the observed RMS 
($\sigma_{\rm obs}$; see Eq.~\ref{eq:sigma_obs}) as
\begin{equation}
\label{eq:sigma_var}
   \sigma_{\rm var} = \sqrt{\sigma_{\rm obs}^2 - \sigma_{\rm noise}^2}.
\end{equation}
The noise correction for the peak-to-peak amplitudes is not as straight forward
since it depends on the noise in the individual observations (which vary in 
the time series data) and the number of samples. A Monte Carlo simulation was 
run 500 times for each star to estimate the average expected peak-to-peak 
amplitude due to noise given the number of measurements and the photometric 
uncertainties. This expected peak-to-peak amplitude was subtracted linearly 
from the observed value to estimate the intrinsic peak-to-peak fluctuations.

\begin{figure}
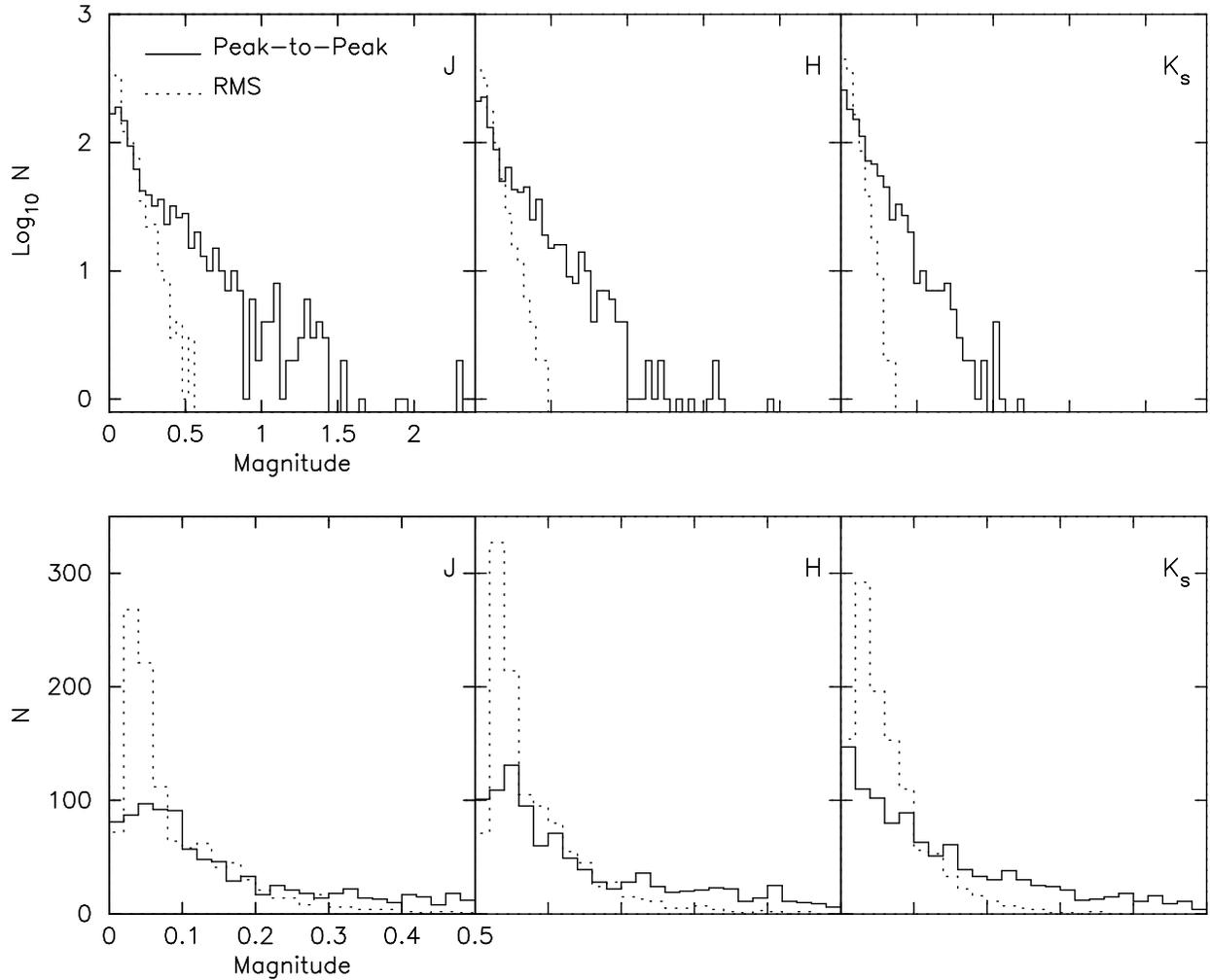

\insertplot{figure18.ps}{7.8}{8.4}{0.0}{0.8}{0.75}{1}
\caption{
  Histograms of the variable star peak-to-peak (solid histogram) and RMS
  (dotted histogram) amplitudes in the $J$, $H$, and \KB\ data after 
  correcting the observed amplitudes for photometric noise (see text). The top 
  panels show the histograms over the full dynamic range, and the bottom 
  panels show in more detail the distribution at low amplitudes which contain 
  most of the variables. The amplitudes were computed using all measurements 
  in the March/April 2000 time series.
  \label{fig:amp_mag}
}
\end{figure}

\begin{figure}
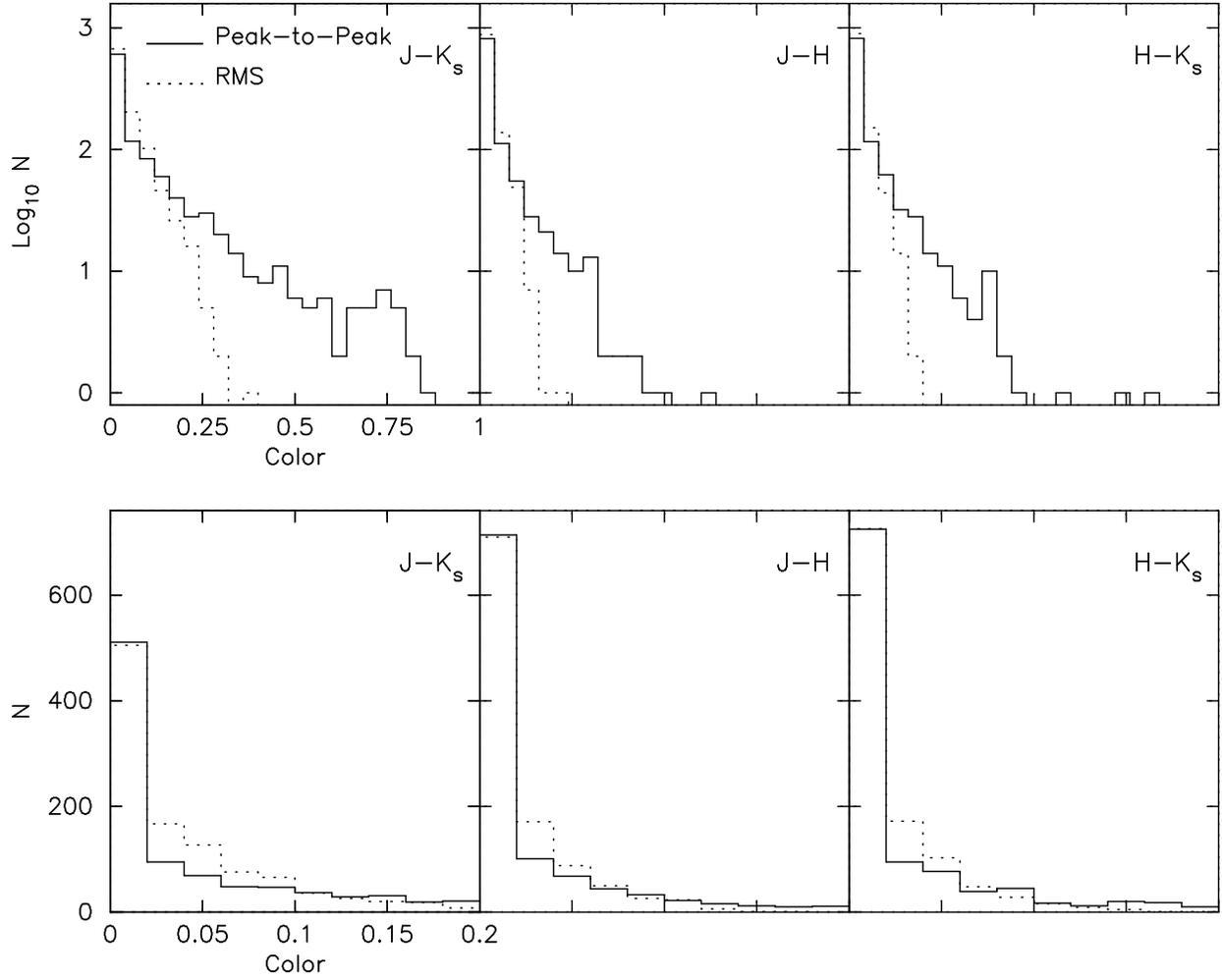

\insertplot{figure19.ps}{7.8}{8.4}{0.0}{0.8}{0.75}{1}
\caption{
  Similar to Figure~\ref{fig:amp_mag}, except for the \JK, $J-H$, and 
  \HK\ colors.
  \label{fig:amp_color}
}
\end{figure}

Histograms of the peak-to-peak and RMS amplitudes for the magnitudes and 
colors, after correcting the observed values for the photometric noise, are 
shown in Figure~\ref{fig:amp_mag} and Figure~\ref{fig:amp_color}.
The top panels in each figure show the histogram over the full dynamic range 
of the amplitudes, and the bottom panels emphasize the distribution at low 
amplitudes where most of the variable stars in fact reside. Statistics on the 
maximum, mean, median, and dispersion in the amplitudes are summarized in 
Table~\ref{tbl:amp}. The peak-to-peak fluctuations in the magnitudes are a 
couple tenths on average, but can be as large as 2.3\M\ at $J$-band and
1.2\M\ at \KB-band. Peak-to-peak fluctuations as large as 1.7\M\ at \KB-band
are observed in three stars when the March~1998 are included. Fluctuations in 
the colors are less pronounced, as most of the photometric variations are 
essentially colorless within the photometric noise of the data. Thus stars 
that exhibit large magnitude {\it and} color variations (see, e.g., 
Figures~\ref{fig:ex_blue}-\ref{fig:ex_red3}) are among the more extreme cases 
of near-infrared variability in our sample.

\subsection{Correlation of Magnitude and Color Changes}
\label{slopes}

For stars with significant color variations, any correlation
between the color-magnitude and color-color changes can serve as a clue 
to the origin of the variability. Cases of stars becoming bluer as they fade 
(e.g. Fig.~\ref{fig:ex_blue}) and others becoming redder as they fade 
(e.g. Fig.~\ref{fig:ex_red1}-\ref{fig:ex_red3}) exist in our data, with
the photometric changes within a star usually well correlated along narrow 
vectors in the various 
color-color and color-magnitude diagrams.  However, not all possible vector
orientations are found in the time series data.  To quantify the observed
correlations, the slopes of the photometric variations in the $J$ vs. $J-H$, 
\KB\ vs. \HK\ and $J-H$ vs. \HK\ diagrams were computed for individual stars. 
For each correlation, only stars in which the observed RMS in the colors
exceeded the expected photometric uncertainties by 50\% were included so that 
the derived slopes would not be dominated by noise in the data. As only \about 
10\% of the variable stars satisfied this criterion, the analysis that follows 
is most appropriate for those objects with relatively large-amplitude color 
variations and may not be applicable to those with largely colorless 
variability. The routine FITEXY \citep{Press92}, which incorporates 
uncertainties in both axes in computing the best fit linear model to the data, 
was used to derive the slopes. A slope angle of 0\arcdeg\ is defined as 
positive color change along the X-axis with no magnitude or color change along 
the Y-axis. The slope angle increases counter clockwise in the color-color 
diagram and clockwise in the color-magnitude diagram (since magnitudes are 
plotted with decreasing values towards the top). 

Histograms of the derived slopes in various color-magnitude and color-color
diagrams are shown in Figure~\ref{fig:slopes}. The open histogram represents 
all stars for which slopes were derived. The hatched histogram are stars
where the slopes have been determined to better than 20\% accuracy, with a 
typical 1$\sigma$ uncertainty of \aboutless 5\arcdeg. In the $J-H$ vs. 
\HK\ diagram, all but 5 of the stars have positive slopes between \about 
30\arcdeg\ and 60\arcdeg. The photometric correlations for the 5 stars with 
negative slopes do not vary along a well-defined vector in the color-color 
diagram and the slope is not meaningful. Thus the dominant type of photometric 
variability in the near-infrared color-color diagram has both colors becoming 
redder together. In the $J$ vs. $J-H$ diagram, the predominant trend is that 
stars become fainter as the colors get redder with a slope between \about 
50\arcdeg\ and 80\arcdeg. Two stars have negative slopes with uncertainties 
less than 20\%; both are long term variables with large magnitude and color 
variations revealed by including the March~1998 data. In the \KB\ vs. \HK\ 
diagram the derived slopes show two distinct trends. In addition to stars with 
colors becoming redder as they fade (positive slopes; e.g. 
Figs.~\ref{fig:ex_red1}-\ref{fig:ex_red3}), a number of variables have colors 
becoming bluer as the stars become fainter (negative slopes; e.g. 
Fig.~\ref{fig:ex_blue}); \about 1\% of the total number of variables are of 
this kind.

\begin{figure}
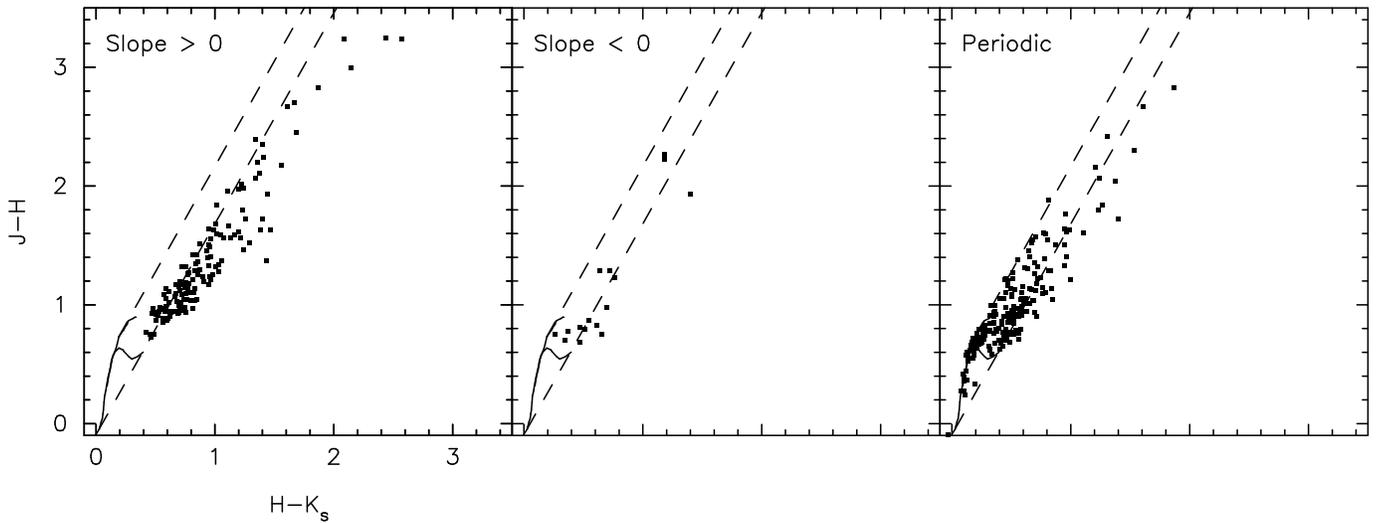

\insertplot{figure21.ps}{6.8}{8.1}{-1}{0.8}{0.8}{1}
\caption{
  $J-H$ vs. \HK\ diagrams for three groups of variable stars.
  The left panel shows variable stars that have significant fluctuations in 
  both the observed \KB\ magnitudes and \HK\ colors, and the photometric 
  fluctuations are such that the colors become redder as the star gets fainter
  (i.e. slope in the \KB\ vs \HK\ diagram is positive). The middle panel 
  shows stars that have negative slopes such that the colors become bluer
  as the stars gets fainter. The right panel is the color-color diagram for 
  stars that are identified as periodic variables in the near-infrared. Stars
  with significant color and magnitude variations are generally redder than 
  the periodic stars, and in particular, stars with positive slope variations 
  tend to have near-infrared excesses more so than periodic variables.
  \label{fig:jhhk_types}
}
\end{figure}

Figure~\ref{fig:jhhk_types} compares $J-H$ vs \HK\ color-color diagrams 
for stars with positive and negative slope in the \KB\ vs. \HK\ 
color-magnitude diagram. (This figure also includes a panel for periodic stars,
which are discussed in Section~\ref{periodicity}.) 
The majority (\about 90\%) of variable stars do not 
have large color variations and do not appear in this figure. On average, 
the stars with large color variations tend to have redder near-infrared 
colors compared to the variable star population as a whole (see 
Fig.~\ref{fig:jhhk}). Further, \about 76\% of the stars with positive slope 
variations have near-infrared excesses characteristic of CTTS's, 
compared to \about 53\% of the small number of stars with negative slope 
variations and 30\% for the uniformly select variable star population of 
\about 1000 stars (Section~\ref{mean}). 

\subsection{Temporal Properties}

\subsubsection{Time scales for Variability}
\label{timescales}

The temporal variability characteristics can be evaluated on time periods of 1 
day to \about 1 month using the March/April 2000 observations, on \about 2 
month time scales by incorporating the February~2000 data, and on \about 2 
year time scales with addition of the March 1998 data. The temporal properties 
of the March/April 2000 observations were evaluated using the autocorrelation 
function (ACF), which measures the similarity of photometric measurements over 
different time samples. While photometry is nominally available in our data set 
on a daily basis, the time series can contain gaps of up to 4 days. To account 
for this non-uniform sampling, the Fourier transform of the observed 
measurements was computed using the \citet{Scargle89} algorithm for unevenly 
sampled data. The power spectrum was then computed from the Fourier transform, 
and the inverse transform of the power spectra yielded the ACF. The resulting 
ACF was normalized by the ACF of the sampling function for each star
\citep{Scargle89} and was sampled every 1 day in accordance with the nominal 
separation between our observations. A positive value of the ACF at a given 
time indicates that the photometry is correlated on that time scale, while 
negative value indicates the photometry is uncorrelated. The time scale of the 
variability can then be characterized by computing the largest time lag before 
the ACF first becomes negative. Since the mean value of the photometry was 
subtracted from the data before computing the ACF, the time lag is a crude 
measure of the number of consecutive days a star remains brighter or fainter 
than the mean magnitude over the time series. The longest variability time
scale that can be estimated from the data then is approximately half of the 
total time period of the observations. The maximum time scale will vary 
between 9-18 days depending on the spatial location of the star.
For variable stars locate in the center three tiles (\about 77\%
of the total number of variable stars), the maximum variability time scale that 
can be inferred is \about 14 days.

For random noise with an infinite number of samples, the distribution of time 
lags is a delta function at $\Delta t=0$; a finite number of samples though
broadens the ACF. To estimate the expected distribution of time lags for 
random noise, the ACF was computed for each star as described above but 
replacing the observed magnitudes by a random number with a gaussian 
probability distribution with dispersion given by the photometric 
uncertainties. The time sampling in these simulated data are identical to that 
in the real observations.

\begin{figure}
\plotone{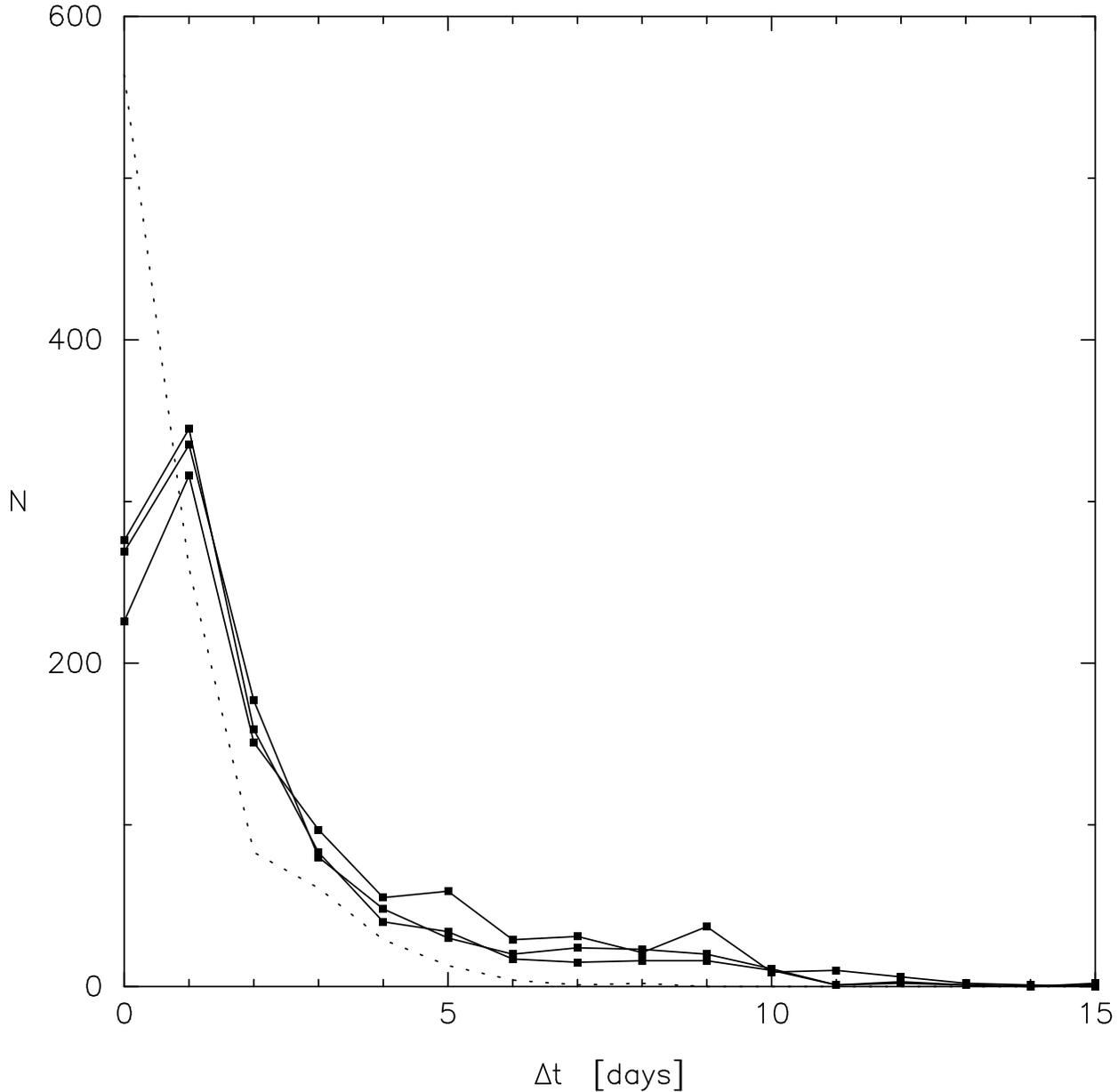}
\caption{
  Distribution of time lags inferred from the autocorrelation function for
  the variables stars in the March/April 2000 time series data. The three 
  solid lines represent the time lags for the $J$, $H$, and \KB\ data.
  The dotted line shows the time lags for a simulated data set with the 
  same random noise characteristics and time sampling as the observations. 
  The maximum time lag possible in this ACF analysis is approximately half
  the time period of the observations. For 77\% of the variable stars, the 
  maximum possible time lag is \about 14 days.
  This figure shows that in the \about 1 month time series observations,
  most of the variability occurs of time scales of a few days, but can be as 
  long as 10-15 days in the more extreme cases.
  \label{fig:acf}
}
\end{figure}

Figure~\ref{fig:acf} shows the inferred time lags in the three
bands for all identified variables (solid lines) and the simulated $J$-band 
data (dotted line). The $H$ and \KB\ band simulated data are similar 
to the $J$-band simulation and for clarity are not shown. This figure 
demonstrates that the random noise data has a peak near zero time lag as 
expected since each data point is independent. By contrast, the distribution 
of time lags for the variable stars peaks at 1 day and has a more extended 
tail toward larger time lags than the simulated observations. The shape of 
these distributions suggests then that most of the variability occurs on time 
scales of less than a few days.

The near-infrared variability characteristics on \about 2~month and 2~year time 
scales were assessed using the February 2000 and March 1998. (Note that the 
February~2000 data are available only for sources north of 
$\delta=-6$\arcdeg.) A long term variable was identified as a star in which 
the $J$, $H$, or \KB\ magnitudes from these earlier measurements differed 
from the average March/April 2000 data by more than four times the observed 
RMS scatter in that time period. A 4$\sigma$ deviation was chosen since only 
\about 1 star in the entire source list should exhibit a fluctuation 
that large due to random noise. Given typical photometric uncertainties for 
the variable stars of 0.02-0.03\M, the 4$\sigma$ criteria imposes a 
minimum amplitude change of \about 0.1\M. By this definition, a total of 14
stars exhibited variability in the February~2000 data, but only 5 of these 
were variables not previously identified based on the Stetson index in 
Sample~1 or Sample~2 (see Table~\ref{tbl:samples}). A total of 72 stars 
exhibited significant photometric fluctuations on \about 2 year time scales, 
with 26 of these new variables not previously identified from the 
March/April 2000 data based on the Stetson statistic. The fact that few 
new long term variables were identified indicates that most of the short term
variations are not part of larger amplitude variations occurring long time
periods.

\subsubsection{Periodic Behavior}
\label{periodicity}

The properties of variable stars can be further characterized by determining 
if the photometric variations are typically aperiodic or periodic. 
The Lomb periodogram algorithm in \citet{Press92} was used to compute the power
spectrum of all 17,808 stars in our sample and to determine the ``false alarm 
probability'' (FAP) that the highest peak in the power spectrum could result 
by chance. The highest frequency searched was 0.5~days$^{-1}$, and the 
shortest independent frequency was the reciprocal of the time period of the 
March/April~2000 data sequence for that star. So that at least 2 full cycles 
are present in the time series data, only periods less than half the total 
time span of the observations were considered reliable. Stars in the overlap 
regions of adjacent tiles have more than one photometric measurement recorded 
per night. Such clumping in the time sampling can invalidate the FAPs, and 
therefore only one of the overlap measurements was considered. The power 
spectrum and FAP were computed in this manner for all stars and in each band 
that contain $\ge$ 8 reliable photometric measurements.

Due to the multi-band nature of our time series data, the periodicity derived 
for a star can have high significance under several different scenarios.
In generating a list of periodic stars we nominally required that the 
false-alarm-probability be $\le 10^{-4}$ in any one of the following 
combinations: (1) a single band, (2) the product of any two bands, or (3) the 
product in all three bands. In practice, only one star had a FAP less than 
$10^{-4}$ in a single band. The effective single-band FAP through this 
procedure is $\sqrt{10^{-4}} \approx 0.01$ for two bands and $\root 3 \of 
{10^{-4}} \approx 0.05$ for three bands. When the product of the FAP in two or 
three bands was used, it was further required that the periods agree to within 
20\%. 

Table~\ref{tbl:periodic} summarizes the periods and false-alarm-probabilities 
for the 233 stars identified as periodic variables in this study. Only 8 
(3.4\%) of these periodic sources were not identified as variable stars from 
the Stetson statistic.  Conversely, the fraction of the near-infrared variable 
star population that is periodic is \about 18\%, however, this is a lower 
limit to the actual percentage given the conservative FAP limits used to 
establish periodicity.\footnote{For example, the BN object discussed by 
\cite{Lynne01} is not identified as periodic using the adopted FAP criterion.} 
If a FAP of 10$^{-3}$ is used, an additional \about 100 stars are identified 
with consistent periods in all three bands. Furthermore, optical monitoring 
studies have identified \about 760 periodic stars within a \about 60\arcmin\ 
radius of the Trapezium region of the ONC \citep{Stassun98,Herbst00b,Rebull01}, 
only \about 100 of which are in common to our periodic stars. The $J-H$ vs.
\HK\ color-color diagram for the periodic stars identified in our sample is
shown in Figure~\ref{fig:jhhk_types}, where it is shown that most of the 
periodic stars have colors consistent with WTTS's.

To verify that our criteria produced a reliable sample of periodic
stars, robust estimation with replacement \citep{Press92} was used to find
the number of false periods one would expect to detect among the 17,808 
stars in our sample. For each time sample, the observed magnitude was replaced
by a magnitude chosen randomly from the time series photometry for that star.
The $J$, $H$, and \KB\ photometry were re-distributed in parallel since 
the coincidence of periodic behavior in these bands were used to select 
periodic stars from the real data. The power spectra of the redistributed data
were computed and the number of significant periodic stars assessed using the 
same criteria described above. Of the 17,808 stars in the Monte Carlo 
simulation, only 2 were identified as periodic.

Finally, we examined the accuracy of our derived periods by comparing them
with those found from optical monitoring observations 
(Stassun \etal~1998; Herbst \etal~2000; Rebull~2001 and references therein).
Figure~\ref{fig:optical} compares the periods for 109 stars identified as 
periodic at both optical ($I_C$) and near-infrared ($J$ and/or $H$ and/or \KB) 
wavelengths. The periods agree to within 10\% for \about 80\% of the stars, 
suggesting that for these stars the origin of the periodicity is the same at 
optical and near-infrared wavelengths. The biggest discrepancy in the derived 
periods occurs when our reported period is actually an alias of a more 
significant period at $<$2 days found from optical monitoring at higher time 
resolution than our time series. Figure~\ref{fig:optical} suggests that this 
occurs for \about 10\% of the stars in our periodic sample. Three of the stars 
with optical periods greater than 10 days have near-infrared periods that 
differ by a factor of two or more. In a few cases, the near-infrared 
period is approximately half that of the optical period. These may be examples 
of ``period doubling'', where the presence of two star spots on opposite sides
of the star causes the periodogram analysis to derive a period half that of 
the actual value \citep{Herbst00b}. The other discrepant source (with an 
optical period of \about 60~days and a near-infrared period of \about 14 days) 
has been noted to have uncertain optical period \citep{Rebull01}.

A histogram of our derived periods is shown in Figure~\ref{fig:periods}. The 
shaded region indicates periods that are suspected aliases of sub-2 day 
periods based on comparison with optical monitoring data; only \about 1/2 of 
our periodic sample has the information necessary to make this comparison due
largely to our larger spatial coverage compared to optical period searches.
The frequency distribution is characterized by a peak at 2-3 days 
and a slow decline towards longer periods. The range of amplitudes and periods 
to which our data are sensitive was established by replacing the time
series data for each variable star with a sinusoidal signal plus gaussian 
random noise that has a dispersion equal to the photometric noise of the 
actual data. Periodic stars in these simulated data were then identified 
as described above using an maximum effective FAP of 10$^{-4}$. The results 
indicate 
that our data are roughly uniformly sensitive to periods between 2 and 10 
days, although with reduced sensitivity at 2 and 3 days due to the 1 day time 
sampling of the observations. Approximately half of the simulated stars with 
peak-to-peak amplitudes of \about 0.08\M\ and \about 90\% with peak-to-peak 
amplitudes of 0.16\M\ were identified as periodic in the simulations. By 
comparison, 55\% of the total variable star population and 60\% of the 
identified periodic stars have $H$-band peak-to-peak amplitudes $\ge 0.08$\M.

\section{Origins of Near-Infrared Variability}
\label{origin}

We turn now from observational characterization of the near-infrared 
photometric variations to examination 
of possible physical origins of the variability. The primary observational
constraints as established in Section~\ref{characteristics} are
(1) nearly all of the identified variables are young, low mass stars 
    associated with the Orion~A molecular cloud;
(2) the typical time scale for the variability is a few days or less;
(3) the light curves exhibit a variety of features ranging from periodic 
    behavior, to discrete variability episodes superposed on otherwise steady 
    light curves, to smooth photometric variations over several days, to
    month long (or longer) rises and fades;
(4) the amplitudes of the fluctuations are \about 0.2\M\ on average, but can 
    reach \about 2\M\ in the extreme;
and
(5) the photometric fluctuations are nearly colorless in most cases, with 
    \about 77\% of the variable stars having color variations $<$0.05\M.

Both the spatial distribution and the observed colors and magnitudes of the
variable stars are consistent with what is expected for the Orion 
pre-main-sequence population, suggesting that much of the observed 
near-infrared variability is intimately related to the properties of young 
stellar objects. Further, the short time scale for the photometric 
fluctuations suggests that the variability originates either within the 
stellar photosphere or close to it in the inner circumstellar environment. As 
established by previous studies 
\citep{Rydgren83,Liseau92,Kenyon94,Skrutskie96,Hodapp99}, there are several
short term phenomena related to young stellar objects that may contribute to 
near-infrared variability, including rotational modulation by cool and hot 
spots on the stellar surface, changes in the line-of-sight obscuration due to
circumstellar dust, variations in accretion geometry or mass transfer rates 
from circumstellar disks, and gradual declines in brightness from EX~Lup or 
FU~Ori type bursts. Mechanisms not necessarily unique to young stellar objects 
may also be operative, such as eclipses due to binary companions. Many of 
these mechanisms were first suspected from optical monitoring observations of 
young stellar objects (e.g. Herbst \etal~1994; Bouvier \etal~1999). Infrared 
observations, however, uniquely probe variability related to circumstellar 
material which radiates at temperatures too cold (\aboutless 2000~K) to 
contribute substantially to optical emission. 

In the following sections we investigate the amplitudes and the 
time scales expected for photometric fluctuations due to star spots, 
extinction, and accretion disk phenomena. 
We begin by analyzing the contributions from cool and hot spots
since these are known to exist on young stars from optical studies and must 
contribute to near-infrared variability as well. However, as will be shown, 
star spots cannot explain all of the observed near-infrared variability 
characteristics and other mechanisms must be present, with extinction and 
accretion-related phenomenon as strong candidates. While throughout this 
discussion we consider the effects of these mechanisms separately, in reality,
a number of mechanisms may be operating simultaneously for any individual
star.

\subsection{Star Spots}

\subsubsection{Models}

Star spots, which can be either cooler or hotter than the photospheric 
temperature, modulate the brightness of a star as stellar rotation alters 
the fractional spot coverage visible to the observer. Cool spots are thought to 
arise from magnetic activity on the stellar surface, whereas hot 
spots are interpreted as regions where material accreting along magnetic field 
lines (e.g. Hartmann, Hewett, \& Calvet 1994) impacts the star. Instead of 
discrete hot spots, a more realistic model may be one in which the accretion 
is confined to a high latitude ring on the stellar surface, where the 
inclination of the ring with respect to the stellar rotation axis depends on 
the orientation of the dipole magnetic field (see, e.g., Mahdavi \& 
Kenyon~1998). In this scenario, variability may result either from simple 
rotation of the inclined accretion ring around the star, or from non-steady 
accretion.

The photometric amplitudes expected from both cool and hot star spots were 
calculated 
assuming that spots can be characterized by a single temperature blackbody, 
$T_{\rm spot}$, that covers a fraction, $f$, of the stellar photosphere with 
an effective temperature $T_*$. The amplitude of the photometric variations, 
relative to pure photospheric emission, can then be expressed as (see, e.g., 
Vrba \etal~1986)
\begin{equation}
\label{eq:spots}
    \Delta m(\lambda) = -2.5\: {\rm log}_{10}\{1 - 
        f[1.0 - B_\lambda(T_{\rm spot})/B_\lambda(T_*)]\},
\end{equation}
where $B_\lambda(T)$ is the Planck function. This star spot model ignores
limb darkening, inclination effects, and opacity differences as a function
of wavelength between the spot and the stellar photosphere. The models shown 
in Figure~\ref{fig:models} assume a stellar effective temperature of 4000~K, 
corresponding to a 0.5\msun\ star at age 1~Myr \citep{DM97}, and cool and hot 
spot temperatures of 2000~K and 8000~K respectively. Results are presented for 
spots that cover 1, 2, 5, 10\%, 20\%, and 30\% of the stellar surface. These 
models encompass the more extreme spot parameters inferred from optically 
selected sample of low mass pre-main-sequence stars 
\citep{Bouvier89,Bouvier93,FE96}. 

Figure~\ref{fig:models} shows that cool and hot spots can be distinguished 
observationally in the near-infrared based on the amplitude of the photometric
fluctuations. While hot and cool spots with small fractional coverages 
both produce low amplitude, nearly colorless fluctuations, the maximum
amplitude from cool spots is \about 0.4\M\ at $J$-band, while hot spots can 
produce photometric fluctuations as large as 1\M\ for sufficiently hot spot 
temperatures and/or large fractional coverages. Further, the maximum changes 
in the $J-H$ and \HK\ colors that can be produced with the spot parameters 
considered here is \about 0.03\M\ for cool spots but 0.1-0.2\M\ for hot spots.

The time scale for variability caused by cool or hot star spot modulation is 
governed by the stellar rotation period.\footnote{\citet{Smith95} suggest 
though that the period measured from hot spots may represent the beat 
frequency between the stellar rotation frequency and the orbital frequency at 
the magnetospheric boundary.}
The frequency distribution of periods in T Tauri stars derived from optical 
monitoring observations imply rotation periods of \aboutless 10 days for 
\about 90\% of the known periodic stars in Orion 
\citep{Stassun98,Herbst00b,Rebull01}, 
assuming that the observed periodicity is a result of rotational modulation by 
star spots. Similar rotation periods have been derived from Doppler imaging of 
T Tauri stars \citep{Joncour94b,Rice96,JKH97,N98}. Rotational velocities 
derived from high resolution spectroscopy ($v\ {\rm sin}\:i$) also imply that 
young stellar objects have rotational periods on the order of a few days 
\citep{Bouvier86,Hartmann86}. Time scales of a few days are consistent with
the results of our ACF analysis (Section~\ref{timescales}).

Cool and hot spots are perhaps distinguishable based not on the time scale of 
the photometric variability, but on the time scale over which it persists.  
Cool spots are thought to be relatively stable features that can last 
several years or more. Hence observed periodicity is often a repeatable result,
especially in WTTS's which do not have additional variability components 
related to accretion phenomena. Hot spots, however, generally last only a few 
days or weeks as evidenced by period changes and even 
disappearance/reappearance in a few cases where hot spot periods have been 
detected (Vrba \etal~1989,1993). Thus hot spots tend to produce irregular 
variability, especially if the accretion of material onto the star is unsteady 
as has been inferred in some stars \citep{Gullbring96,Basri97,Smith99}, or if 
the geometry is complicated by misaligned rotation and magnetic dipole axes 
(e.g. Mahdavi \& Kenyon~1998; Bouvier et al. 1999). Based on these tendencies, 
\citet{Herbst94} introduced a classification scheme in which periodic 
fluctuations from cool spots (mainly in WTTS's, but in some CTTS's as well) 
are Type I variables, irregular fluctuations in CTTS's from hot spots are 
Type~II variables, and periodic fluctuations in CTTS's from hot spots are 
Type~IIp variables. (Type III variability in this classification scheme is 
discussed in Section~\ref{disks}).

\subsubsection{Comparison to Observations: Cool Spots}

For cool spot parameters typically inferred from optical observations 
($T_*$-$T_{\rm spot}$ \aboutless 2000~K, $f$ \aboutless 30\%; see
Bouvier \& Bertout~1989, Bouvier \etal~1993, Fern\'andez \& Eiroa~1996), the
expected peak-to-peak amplitudes should be \aboutless 0.4\M\ in the $J$, $H$, 
and \KB\ bands and \aboutless 0.03\M\ in the $J-H$ and \HK\ colors 
(Figure~\ref{fig:models}).  These amplitudes are only approximate given the 
simplicity of the model with the essential predictions that the color 
variations from cool spots should be relatively low and appear periodic
(moreso in WTTS's than CTTS's) by analogy with Type I optical variability 
\citep[see also Herbst, Maley, \& Williams~2000]{Herbst94}. In our data, 65\% 
of the periodic variables stars have amplitudes $\le$ 0.4\M\ at $J$-band and 
$\le$ 0.03\M\ in the $J-H$ color, 77\% have color amplitudes $\le$ 0.05\M, and 
80\% have near-infrared colors consistent with WTTS's. Thus rotational 
modulation by cool spots can account for the variability characteristics in 
the majority of the periodic stars. 

\begin{figure}
\plotone{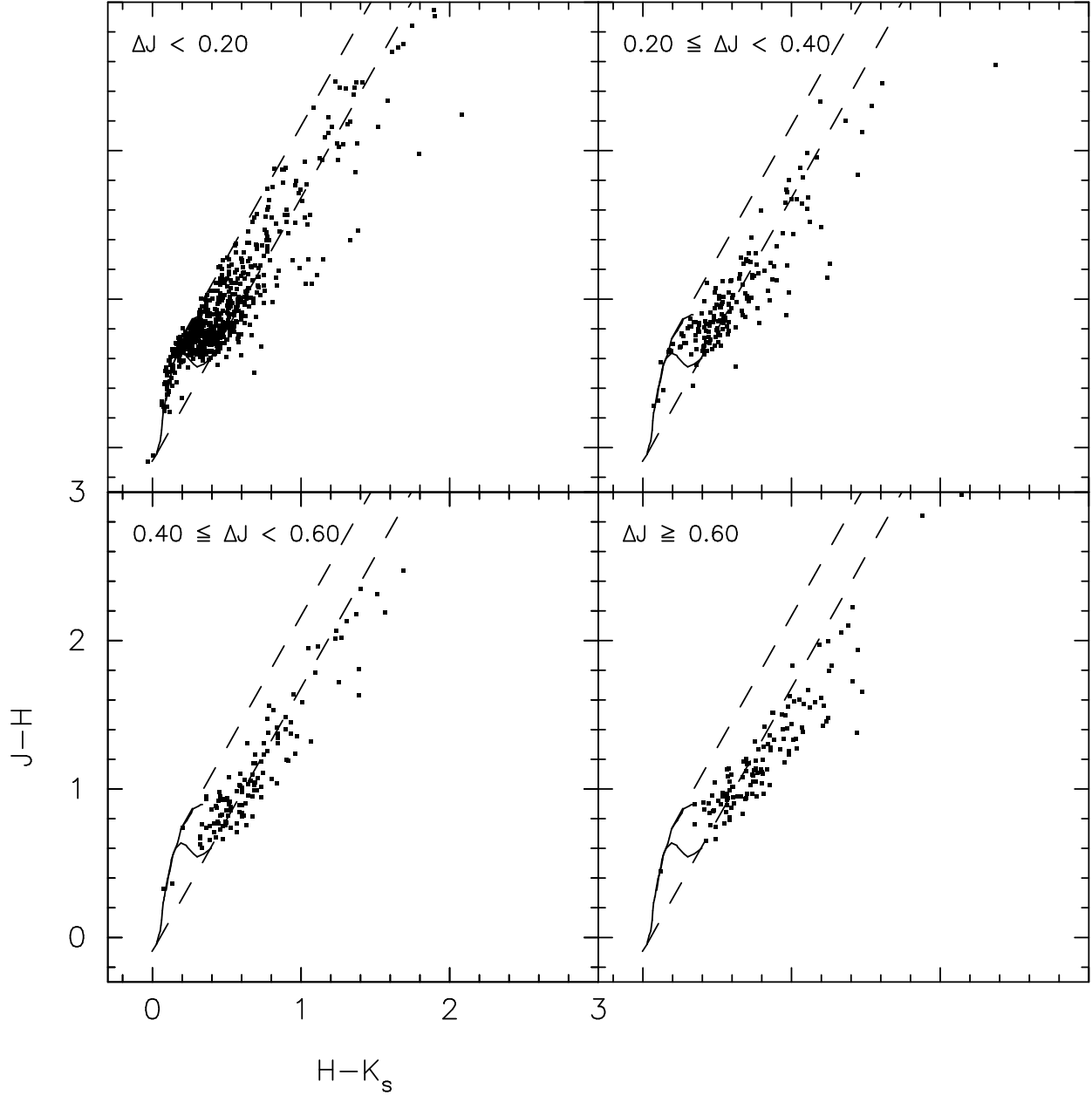}
\caption{
  $J-H$ vs. \HK\ diagrams for variable stars as a function of the 
  $J$-band peak-to-peak amplitude. This figure shows that stars with larger
  $J$-band amplitudes tend to have redder colors and larger near-infrared 
  excesses than stars with smaller fluctuations.
  \label{fig:jhhk_grid}
}
\end{figure}

Periodic stars, however, account for only \about 18\% of the total variable 
population, and an additional \about 600 stars (\about 50\% of the identified 
variables) also have low amplitude, nearly colorless photometric fluctuations 
but have not been identified as periodic. While arbitrarily low photometric 
fluctuations can be produced by many of the physical mechanisms discussed in 
this section, Figure~\ref{fig:jhhk_grid} suggests that rotational modulation 
by cool spots may cause much of the low amplitude variability, independent of 
whether a period is actually detected. This figure shows the 
$J-H$ vs. \HK\ color-color diagram as a function of the peak-to-peak $J$-band 
amplitude. Most stars with low amplitudes ($\le$ 0.2\M) have near-infrared 
colors consistent with WTTS's or stars with small infrared excesses. As the 
$J$-band amplitude increases, the near-infrared colors tend to become redder 
and an increasing fraction contain a near-infrared excess characteristic of 
the optically thick circumstellar disks of CTTS's. This trend suggests that 
different mechanisms related to the absence or presence of an accretion disk 
may be producing the low and high amplitude variability. Of the variability 
mechanisms examined here, only cool spots and eclipsing binaries do not 
require the presence of a circumstellar disk. Since eclipsing systems cannot 
account for the large number of stars with low amplitude variability, cool 
spots appear as a more likely explanation.

Cool spot modulation produces larger amplitude fluctuations at optical 
wavelengths, and so optical surveys should provide a more complete census of 
the periodic variables. As mentioned in Section~\ref{periodicity}, optical 
studies have found \about 750 periodic stars in the Orion Nebula region 
\citep{Stassun98,Herbst00b,Rebull01}, of which \about 330 are identified as
near-infrared variables and \about 100 as periodic in this study. Of those 
optical periodic stars also identified as near-infrared variables (but not 
necessarily periodic near-infrared variables), \about 70\% have low 
peak-to-peak amplitudes in the $J$-band magnitudes ($\le$ 0.4\M) and $J-H$ 
colors ($\le$ 0.03\M). Further, \about 88\% of the \about 650 optical periodic 
variables not identified as periodic in the near-infrared have near-infrared 
colors consistent with reddened WTTS's. 

Based on the observational evidence just described we speculate that the 
photometric fluctuations observed in many of the low amplitude variable stars 
are due to cool spot modulation. As a lower limit to the star that may have 
variability due to cool spot modulation, we find that 57\% of the variables 
have $J$-band amplitudes $\le$ 0.4\M, $J-H$ amplitudes $\le$ 0.03\M, and no 
near-infrared excess as expected for cool spot parameters inferred from 
optical observations. Given the simplicity of the star spot model, if we 
crudely assume that the maximum $J-H$ color change that can be produced by 
cool spots is 0.05\M, we derive a upper limit of 77\% as the percentage of the
variability that may be attributed to cool spots.

\subsubsection{Comparison to Observations: Hot Spots}

\begin{figure}
\epsscale{0.85}
\plotone{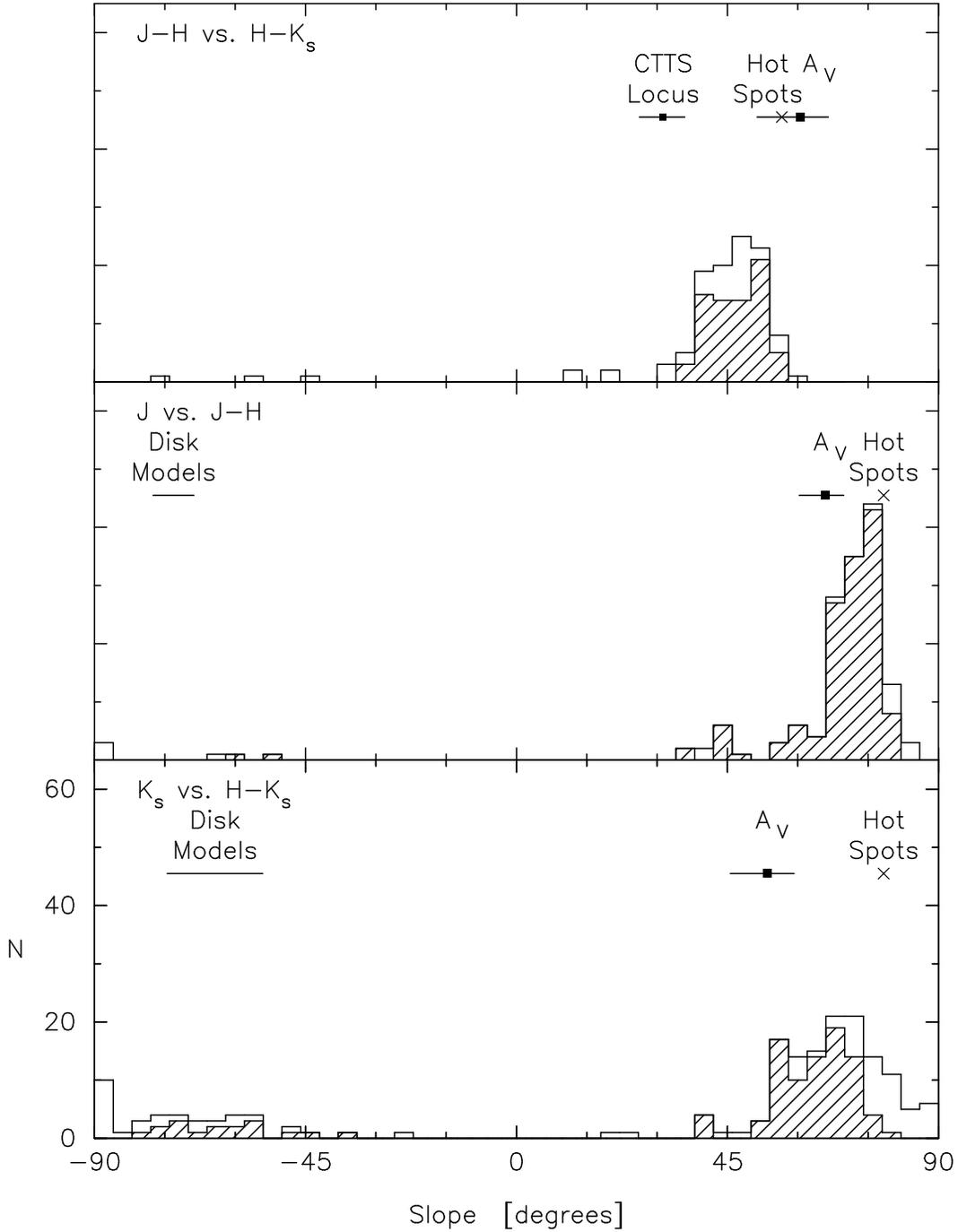}
\epsscale{1.00}
\caption{
  Histograms of the derived slopes in the $J-H$ vs. \HK, $J$ vs. $J-H$, 
  and \KB\ vs. \HK\ diagrams. Only variable stars in which the observed
  RMS of the appropriate colors/magnitudes exceeded the expected RMS by a 
  factor of 1.5 are shown. The open histogram is for all stars that meet
  these criteria, and the hatched histogram is those stars in which the
  slope have been determined to an accuracy of better than 20\%. The
  predicted slopes based on hot spot models, extinction variations, and
  circumstellar disks models (see text) are indicated. The predicted slopes 
  from cool spot models are not shown since they cannot account for the 
  amplitude of the color variability observed in most of these objects. This 
  figure shows that while each of these models can account for some aspect 
  of the variability amongst these stars, none alone can account for all of 
  the observed trends.
  \label{fig:slopes}
}
\end{figure}

While cool spots can plausibly account for the low amplitude variables with 
small color variations, they cannot explain the 23\% of the variable stars 
that have color variation exceeding \about 0.05\M\ (see 
Fig.~\ref{fig:models}). Stars 
with significant color variations were analyzed in Section~\ref{slopes} where 
it was shown that these objects tend to have colors consistent with CTTS's 
(see Fig.~\ref{fig:slopes}). Therefore, again by analogy with optical 
variability characteristics, we investigate whether hot spots, which cause 
Type~II optical variability in CTTS's in the \citet{Herbst94} classification 
scheme, can account for the color amplitudes observed in some of the 
near-infrared variable stars.

For hot spot parameters typically inferred from optical observations 
($T_{\rm spot}$ \aboutless 8000~K, $f$ \aboutless 10\%), hot spots can cause 
peak-to-peak amplitudes at near-infrared wavelengths as large as \about 
0.2-0.4\M\ in the magnitudes and \aboutless 0.06-0.12\M\ in the colors (see 
Fig.~\ref{fig:models}).  About 10\% of the near-infrared variable stars have 
magnitude and color amplitudes larger than these hot-spot predictions
(see Section~\ref{slopes}). A combination of hotter spot temperatures 
(\aboutmore 10,000~K) and higher coverages (\aboutmore 20\%) are needed to 
explain these stars from hot spots alone. These spots parameters are evidently 
rare, but may be possible and simply not identified previously due to small 
number statistics that are overcome by the extensive near-infrared variable 
sample obtained here.

Because of the substantial color variability expected from hot spots, a more 
quantitative comparison between the observations and the hot spot model can be 
made by assessing correlated color and magnitude changes in individual stars. 
Figure~\ref{fig:slopes} showed the observed slopes in various color-color and 
color-magnitude diagrams as discussed in Section~\ref{slopes}. The predicted 
slope from the hot spot model is also indicated on this figure, which varies 
only by a few degrees for stellar effective temperatures between 3000 and 
6000~K and spot temperatures up to 40,000~K. Figure~\ref{fig:slopes} shows 
that many of the observed slopes in the $J$ vs. $J-H$ diagram can be 
accounted for quantitatively by hot spots (and extinction variations as 
discussed in the following section). However, in the $J-H$ vs. \HK\ and \KB\ 
vs. \HK\ diagrams, while there is an approximate correspondence between the 
observed and predicted slopes, the observed slopes are systematically 
shallower than expected if the variability is due solely to hot spots. These
differences may not be significant though given the simplicity of the hot spot 
model. In addition, the \KB\ vs. \HK\ diagram contains a number of stars with 
negative slopes in that the colors become bluer as the star gets fainter. Such 
variations are completely inconsistent with the hot spot model, and suggest 
that additional mechanism(s) are present that contribute to the near-infrared 
variability, especially at \KB-band.

\subsection{Extinction}

\subsubsection{Models}

Variability due to extinction can result from inhomogeneities in either the 
inner circumstellar environment or the ambient molecular cloud that move 
across the line of site. Photometric changes induced by visual extinction 
variations of $\Delta {\rm A_V} = \pm 2$\M\ are illustrated in the middle 
column in Figure~\ref{fig:models}, although the extinction changes in practice 
can be of arbitrary value. The extinction vectors were calculated from the 
interstellar reddening law measured by \citet{Cohen81} and transformed into 
the 2MASS photometric system \citep{Carp01}. If the extinction variations 
originate from a circumstellar disk and the dust grains have conglomerated 
into larger particles to produce a more grey extinction law, the amplitude of 
the color variations will be less than indicated. For the interstellar 
reddening law, Figure~\ref{fig:models} shows that extinction variations 
produce photometric slopes at near-infrared wavelengths similar to those 
expected from hot spots except in the \KB\ vs. \HK\ diagram, where the 
extinction slope is shallower by \about 25\arcdeg. 

The time scale for extinction variations caused by the ambient molecular cloud
is set by the size scale of the structural inhomogeneities and the velocity
field of the molecular gas. The typical line of sight full-width-at-maximum 
line width observed in the Orion molecular cloud from single dish \thcoj\ 
observations is \about 2\kms\ \citep{Bally87}. Thus the time scale for one
point in the cloud to transit a 1.5\rsun\ radius star (representative of a 
\about 0.5\msun\ at an age of 1~Myr) is \about 12 days. The line wings in 
molecular line profiles can extend to several kilometers per second, which may 
shorten the appropriate time scales to a few days.

If structural inhomogeneities in a circumstellar disk occult the star to cause 
extinction variations, a variety of time scales are possible. Possible 
inhomogeneities include azimuthal asymmetries in the plane of the disk, a 
warped or flared outer disk \citep{Bertout00}, or a warped/precessing inner 
disk \citep{Lai99,Bouvier99,TP00}. Azimuthal asymmetries or warped features 
will produce variability on time scales dictated by the rotational velocity of 
the disk and the size scales of the inhomogeneous features. The time for one
position in the disk to transit the star ranges from hours within a few 
stellar radii to a week or more for radii \aboutmore 30~AU. (If a warp 
extends for a significant azimuthal distance around the star, however,
the relevant time scale will be given by the rotation period at that radius.)
The fraction of the stars that show extinction variability will depend on the 
probability that these inhomogeneous features transit the star. These 
predictions are heavily model dependent, but \citet{Bertout00} suggests that 
as many as 20\% of the stars with circumstellar disks may be occulted by the 
flared outer disk. Azimuthal asymmetries require viewing the disk nearly edge 
on, and are not likely to contribute substantially to the total number of 
variables. The magnetospheric accretion columns and the inner disk wind are 
likely optically thin and would not contribute to extinction variations 
(Kenyon, Yi, \& Hartmann~1994 and references therein). 

\subsubsection{Comparison to Observations}

If the ambient molecular cloud produces extinction variations, then the
variable star population exhibiting significant color variations should
contain a mixture of all (non-foreground) populations to which our survey
is sensitive.  This would include background field stars and Orion population
stars with and without near-infrared excesses (i.e. CTTS's and WTTS's). Since 
field stars and WTTS's outnumber CTTS's, CTTS's should be the least represented 
group in the variable star population. However, as shown in the left panel of 
Figure~\ref{fig:jhhk_types}, stars with large color variations preferentially 
have near-infrared excesses characteristic of circumstellar disks and CTTS's. 
This, combined with the longer time scale (\about 12 days) expected for cloud
transit events compared to the typical observed variability time scale
(\aboutless 3 days; see Figure~\ref{fig:acf}) suggests that if extinction 
changes do contribute to near-infrared variability, the source of the 
extinction is in the inner circumstellar environment and not the ambient 
cloud.  

Visual extinction variations of \about 10\M\ are need to account for the 
extreme magnitude and color variations indicated in Table~\ref{tbl:amp}.
The typical observed variability time scale of \aboutless 3 days indicates
that any inhomogeneities in the disk structure must transit the star within a
similar interval. For a stellar radius of 1.5\rsun, the implied velocity of 
the inhomogeneous region is \aboutmore 10\kms. Assuming Keplerian rotation, 
the maximum radial distance from a 0.5\msun\ star that produces such 
velocities is \about 7 AU. If the size of the warp is significantly larger the
stellar diameter, the warp would need to be located at smaller radial distances
to be consistent with the observed variability size scale. This suggests that 
the outer disk flares described by \citet{Bertout00} are not likely 
responsible for the short term photometric variability observed here, but may 
still be contribute to the longer term fluctuations. 

The slope of the photometric correlations shown in Figure~\ref{fig:slopes} can
also be used to test if the observed variability may be due to extinction
variations. The expected slope of the interstellar extinction law is indicated 
by the filled squares in Figure~\ref{fig:slopes}. This figure shows that
extinction variations (along with hot spots) can quantitatively explain most 
of the observed slopes in the $J$ vs. $J-H$ diagram. However, as with the 
hot spot model, extinction cannot in detail account for the observed slopes in 
the $J-H$ vs. \HK\ and \KB\ vs. \HK\ diagrams. In the color-color diagram, the 
observed slopes are not as steep as expected from extinction. In the \KB\ vs. 
\HK\ diagram, while extinction can account for many of the observed slopes,
the average observed slope is steeper than expected for interstellar extinction 
and shallower than predicted by hot spots. Invoking a more grey extinction 
(although not completely grey, as color variability is observed) would provide 
better agreement with the observed slopes in the $J-H$ vs. \HK\ and \KB\ vs. 
\HK\ diagrams, but then could not simultaneously account for the already good 
agreement with the $J$ vs. $J-H$ slopes. Therefore, extinction cannot simply 
account for the observed variability characteristics. Further, similar to hot 
spots, extinction cannot explain any of the stars that have negative slopes in 
the \KB\ vs. \HK\ diagram.

\subsection{Accretion Disks}
\label{disks}

The above discussion indicates that the simple star spot and extinction 
models cannot explain all of the variability characteristics of stars that 
exhibit color variations. The two main discrepancies are the slopes
of the positive photometric correlations in the $J-H$ vs. \HK\ and \KB\ vs. 
\HK\ diagrams (but not the $J$ vs. $J-H$ diagram), and any of the stars with 
negative slopes in the \KB\ vs. \HK\ diagram. This result suggests that an 
additional mechanism may be present that affects the emission at longer 
wavelengths more so than at shorter wavelengths. Since the stars that cannot 
be explained by the star spot and extinction models tend to have near-infrared 
excesses, we consider emission from a circumstellar disk as a source of 
near-infrared variability.

\subsubsection{Models}

Disks contribute to the near-infrared emission through two means. First, they 
absorb radiation from the star and re-radiate it at longer wavelengths, 
producing the infrared excess that is commonly observed in young stellar 
objects. The shortest wavelength at which a near-infrared excess is produced 
depends on how close the disk extends to the stellar surface and how hot the 
inner disk material becomes \citep{Meyer97}. A second source of near-infrared
emission from circumstellar disks occurs when energy is released as material 
is radially transported through a viscous, optically thick, accretion disk 
\citep{Lynden74}. Photometric variability due to a circumstellar disk can 
result then from changes in the mass accretion rate and consequently the 
accretion luminosity\footnote{\citet{Herbst99} suggested that changes in the 
accretion luminosity produce the irregular optical photometric variability 
sometimes observed in stars more massive that typical CTTS's, or Type~III 
variability in the \citet{Herbst94} classification scheme. \citet{Grinin98}, 
however, attribute Type~III variability to variable circumstellar extinction.},
or changes in the inner disk 
structure that alters the amount of absorbed and re-processed stellar 
radiation. Changes in the re-processed radiation may result, for example, from 
physical variations in the disk inner hole radius, or variations in the 
thickness of the inner edge of a warped disk. Such complicated inner disk 
geometries are predicted to occur when the dipole magnetic field is misaligned 
with the stellar rotation axis \citep{MK98,Lai99,TP00}. 

Figure~\ref{fig:models} shows the expected photometric variations for changes
in the mass accretion rate and inner hole radius of a circumstellar disk
around a star with an effective temperature of 4000~K. These model results are
from the same models used by \citet{Meyer97}, and were kindly provided by 
N. Calvet. The open and filled triangles in Figure~\ref{fig:models}
represent mass accretion rates of \mdot = 10$^{-8.5}$ and 10$^{-7.0}$\myear, 
respectively, and for each accretion rate, the triangles represent the 
photometric emission for an inner hole radius of 1, 2, and 4\rsun, all for
a disk inclination angle of 45\arcdeg. As the inner hole radius increases, 
the star/disk system gets fainter in the near-infrared with less emission 
absorbed and reprocessed by the circumstellar disk. However, unlike the hot 
spot and extinction models, the system colors become bluer (approaching the
stellar color) as it becomes fainter since the disk radiates more strongly at 
\KB-band than at $J$ or $H$. 
For hole sizes \aboutmore 4~\rsun\ and a mass accretion rate of \about 
10$^{-8.5}$\myear, no near-infrared excess emission is produced.
The photometric variations expected from these disks models
can be as large as 1\M\ at $K$-band for the hole sizes and accretion rates 
illustrated in Figure~\ref{fig:models}. Photometric variations of several 
magnitudes can be achieved if the accretion rates increase to 
\mdot \about $10^{-5}$-$10^{-4}$\myear. These accretion disk models also 
predict a shallower slope in the $J-H$ vs. \HK\ diagram than expected from 
extinction or hot spot variability, which is consistent with the observed 
locus of CTTS's in this color-color diagram \citep{Meyer97}.

To examine the time scales associated with 
viscous accretion disks, we considered an accretion disk around a 0.5\msun\ 
star using the $\alpha$ viscosity parameterization \citep{SS73} and assuming 
$\alpha = 0.01$ \citep{Hartmann98}. The accretion rate can change globally 
within the disk or in a localized region. The time scale to transport material 
from the outer to the inner disk is \aboutmore 1000 years for \mdot \about 
10$^{-8}$\myear\ and radii greater than 1 AU \citep{Frank85}. Accretion rates 
may also vary if the disk equilibrium is perturbed. Thermal instabilities in 
the disk can be on the order of \about 1 day in the inner disk region given 
the uncertainty in $\alpha$ \citep{Frank85}. Another possible mechanism that 
may lead to variable mass accretion rates in the inner disk is a time-variable 
magnetic field. In the current paradigm of magnetospheric accretion, the size 
of the inner hole of a circumstellar disk in steady state is a function of the 
mass accretion rate and the stellar magnetic field strength \citep{Konigl91}. 
If the magnetic field strength varies in time, the magnetosphere will 
intersect the disk at a radius other than the corotation radius, which will 
modulate the accretion flow in the inner disk. \citet{Armitage95} and 
\citet{Clarke95} (see also Kenyon, Yi, \& Hartmann~1996) invoked this model to 
explain the variability characteristics in T Tauri stars over many years, but 
since the viscous time scale in the inner disk approaches \about 10 days for 
large values of $\alpha$, this may contribute to shorter term variability as 
well. Finally, the growth time of a warping instability in the inner disk, 
which may lead to variations in the amount of re-processed emission, is \about 
6 days \citep{Lai99}.

\subsubsection{Comparison to Observations}

The observed time scales of the near-infrared fluctuations rule out any global
changes in the disk mass accretion rate (expected on time scales \aboutmore 
$10^3$ years, if indeed occurring) as the origin of the near-infrared 
variability observed here. Short term variations in the mass accretion rate or 
geometry in the inner disk are somewhat speculative and it is not clear that 
they can in fact produce variability on time scales of days. Nonetheless, the 
following discussion considers short term variations in the disk properties as 
possible origins of the observed near-infrared photometric fluctuations. 

The accretion disk models plotted in Figure~\ref{fig:models} reproduce the 
observed locus of CTTS's in the $J-H$ vs. \HK\ color-color diagram 
\citep{Meyer97}, and this observed CTTS's locus slope is indicated in 
Figure~\ref{fig:slopes}. In the $J-H$ vs. \HK\ diagram, the variable stars 
follow an observed slope that is steeper on average than expected from a 
time-variable disk (although extinction and hot spots are no more successful 
in uniquely explaining the observed slopes). In the color-magnitude diagrams, 
the observed slopes are positive, while the predicted slopes from the disk 
models are distinctly negative. Thus the majority of the stars exhibiting 
color variability can not have their variability explained solely by these 
disk models. However, a small number of stars (\about 17) do have negative 
observed slopes in the \KB\ vs. \HK\ diagram (see, e.g., 
Fig.~\ref{fig:ex_blue}) that agree quantitatively with that expected from the 
disk models. In fact, 7 of the 17 stars with significant negative 
slopes may have a near-infrared excess indicative of a circumstellar
disk. About half of the stars with negative slopes do not show a near-infrared 
excess, although it should be noted that not all stars with \KB\ excesses will 
be identified in the $J-H$ vs. \HK\ diagram.

Finally, we briefly discuss the limits set by our data on the frequency of
FU~Ori type bursts in which a star brightens by several magnitudes in
\aboutless 1 year, presumably due to a dramatic increase in accretion driven by 
large-scale disk instabilities \citep{Hartmann93}. All stars brighter than
\KB=14\M\ in the March/April~2000 data are apparent in the March~1998 
images, and the largest observed change in the \KB-band magnitudes is 1.7\M,
with an approximately equal number of stars decreasing in brightness
by $>$1\M\ as increasing over this \about 2 year period. Thus no convincing 
evidence exists for a FU~Ori type burst in these monitoring observations. A 
lower limit on the time scale for such bursts is given then by the number of 
stars in the Orion~A molecular cloud (\about 2700; see Appendix) and the time 
spanned by the observations (\about 2 years), or \aboutmore 5400 years. This 
is consistent with the time scale between FU~Ori bursts commonly assumed 
(\about 10$^4$ years; Herbig~1977, Hartmann, Kenyon, \& Hartigan~1993).

\subsection{Eclipsing Systems}
\label{eclipses}

Although the above mechanisms are the leading candidates to account for most 
of the observed near-infrared variability, a small number of systems have 
variability properties consistent with eclipsing binary systems. This type of 
variability is distinguished by simultaneous drops in the $J$, $H$, and \KB\ 
amplitudes on a single observation for the 1 day time interval of our
observations (see, e.g., BM~Ori in Fig.~\ref{fig:ex_eclipse}). As an initial 
screening for potential eclipsing systems, the March/April 2000 data was 
searched for stars in which at least 2 of 
the 3 bands decreased in brightness by more than 5$\sigma$ from the mean 
magnitude, where $\sigma$ is the observed photometric rms after excluding the 
faintest 20\% of the measurements (as to exclude the actual eclipses from the 
RMS measurements). These criteria yielded 73 stars, all of which were 
previously identified as variable based on the Stetson index or other 
criteria. The light curves for these stars were then visually inspected to 
identify possible eclipsing systems. The 22 candidate eclipsing systems in our 
sample are listed in Table~\ref{tbl:variables}. All but 2 of these stars
are located within the spatial boundaries of the stellar density enhancement
identified here (see the Appendix) and are likely 
pre-main-sequence stars. If these stars can be verified as eclipsing systems
and followed up spectroscopically to determine the orbital elements, they
can provide valuable means to measure the stellar masses and test 
pre-main-sequence evolutionary tracks. Only one candidate in our list, 
BM~Ori, is known to be an eclipsing system from previous studies (see, e.g., 
Antokhina, Ismailov, \& Cherepashchuk 1989). For the remaining sources, the 
drop in the $J$, $H$, and \KB\ amplitudes either does not repeat within the 
time range covered by these observations, or the star exhibits additional
variability characteristics such that the amplitude drops may be attributed 
to other phenomena. The eclipsing nature of these systems then remain to be 
confirmed.

\section{Discussion}
\label{discussion}

Although the current data are not sufficient to identify uniquely the 
predominant variability mechanisms on a star-by-star basis, we can make
a few general statements about the origins of the observed 
near-infrared variability. Rotational modulation by cool 
spots is the leading candidate for the low amplitude variables since many of 
these stars have photometric characteristics analogous to Type~I optical 
variability \citep{Herbst94}. Approximately 56-77\% of the variable stars
may have variability attributed to cool spot modulation. The origin of the 
larger amplitude variable stars that also exhibit color variations is not 
clear. Hot spots, extinction, and accretion disk phenomenon can explain the 
coarse properties of these stars, but the detailed correlation between the 
colors and magnitudes cannot be accounted for by any single mechanism. The 
largest discrepancy is among photometric correlations involving the \KB-band 
photometry ($J-H$ vs.  \HK\ and \KB\ vs. \HK), indicating that relatively cool 
variability phenomenon are present in many of the stars. This suggests that
either 
(1) the hot spot and extinction models used here are too simplistic, 
(2) a process not identified in this study is contributing to the variability, 
or 
(3) that several of these mechanisms are operative in any individual star. 
It would not be surprising, and possibly even expected, that more than one 
variability mechanism is present since several of the identified phenomena 
(hot spots, extinction, and accretion variability) are all related to the 
presence of a circumstellar disk. Since extinction tends to make stars 
fainter, while hot spots and accretion tend to make stars brighter, the net
result on the photometric properties of any individual star is not easy to
predict. 

Evidence for multiple variability mechanisms within a single star is in fact 
available for a few objects. YY~Ori, shown in Figure~\ref{fig:ex_blue}, 
is the namesake for a class of objects displaying broad, inverse P Cygni 
absorption lines that indicate matter infalling onto the star at nearly 
free-fall velocities \citep{Walker72}. YY Orionis stars generally have 
ultraviolet excesses \citep{Walker83}, presumably a result of the accretion 
shock that develops when infalling material impacts the star. YY~Ori is 
suspected to contain a hot spot from optical observations \citep{Bertout96} 
which will also contribute to near-infrared variability. However, the 
near-infrared observations are best explained by variations in the accretion 
disk properties (see Fig.~\ref{fig:ex_blue}).\footnote{The light curves of 
other YY~Orionis stars within our survey region \citep{Walker72,Harder98} were 
examined to see if they also show negative slopes in the \KB\ vs. \HK\ 
diagram. One (NS~Ori) shows similar, but more complex, photometric trends as 
YY~Ori, and three others (CE~Ori, SU~Ori, and XX~Ori), while detected as 
variable stars in our data, do not have negative slopes.} Other stars in our 
survey show evidence for multiple variability mechanisms within the 
near-infrared data itself. The star shown in Figure~\ref{fig:ex_red2}, 
for example, initially shows photometric fluctuations most consistent with 
extinction, but then shows a clear near-infrared excess which subsequently 
disappears. As another example, the long term variable shown in 
Figure~\ref{fig:ex_long2} faded at $J$-band over a 2 year time period while 
simultaneously brighten at \KB-band. This behavior cannot be produced by any 
single variability mechanism considered here.

Finally, we briefly discuss the implications that near-infrared variability
has for determining the stellar mass and age distributions in young star 
forming regions. These distributions are frequently established either by 
modeling the observed colors and magnitudes of the young stars (e.g. 
Comer\'on, Rieke, \& Neuh\"{a}user~1999; Muench, Lada, \& Lada~2000; 
Hillenbrand \& Carpenter~2000), or by combining photometric and spectroscopic 
observations (e.g. Greene \& Meyer~1995; Meyer~1996; Luhman \etal~1998). 
Variability can influence the results since it sets a limit to 
how accurately the luminosity of the star can be measured. Assuming that the 
near-infrared variability characteristics established here are typical for 
most star forming regions, our results suggest variability will have a 
relatively minor effect in analyzing the mass and age distributions for the
majority of the stellar population. For a 1~Myr, 0.5\msun\ star, the average 
$J$-band dispersion of 0.09\M\ translates into a mass and age uncertainty of 
\about 10\% \citep{DM97}, which are minor relative to current uncertainties in 
pre-main-sequence evolutionary tracks. Given that most stars in the Orion~A 
molecular cloud are not detected as variable in this study, the mass and 
age uncertainties for the typical star will be even less. Even for stars that 
exhibit large amplitude variability, the color variations follow roughly
the interstellar reddening law, suggesting that the simultaneous observations
can reasonably estimate the extinction from the observed colors and spectral
type (assuming the spectral variability is negligible). However, if the 
broad-band photometric observations are taken on separate days, or if the 
stars exhibit spectral as well as photometric variability, photometric errors
of up to \about 2\M\ may be introduced in estimating the stellar luminosities,
which can lead to mass errors of a factor of 2-3 and age errors of up to 
an order of magnitude for a 1~Myr, 0.5\msun\ star. Only a small percentage of
the stars should be affected in this manner, indicating that one must consider
variability in interpreting the tails of mass and age distributions
(see also Hartmann~2001).

\section{Summary}
\label{summary}

We have investigated the $J$, $H$, and \KB\ near-infrared variability 
characteristics of stars in a \about $0.84^\circ\times6^\circ$ region centered 
on the Trapezium region of the Orion Nebula Cluster using the southern 2MASS 
telescope. The primary set observations analyzed here were conducted on nearly 
a nightly basis for a one month time period in March/April 2000, but also 
include data from March~1998 and February~2000. These data are used to 
establish the near-infrared variability characteristics of young stars on time 
scales of \about 1-36~days, \about 2 months, and \about 2 years, and to 
investigate possible mechanisms that may cause near-infrared variability in 
pre-main-sequence objects.

Variable stars were identified primarily by quantifying the correlated 
photometric fluctuations in the $J$, $H$, and \KB\ bands using the 
\citet{Stetson96} variability index. The light curves of the identified stars 
exhibit a diversity of features including periodic fluctuations, continuous 
aperiodic variations over day time scales, long term variability over \about 
2 years, steady rises or fades, photometric fluctuations on discrete days 
imposed on otherwise constant brightness (including eclipses), colorless 
variability, stars that get redder as they fade, and stars that get bluer as 
they fade. Examples of many of these phenomenon can be found in 
Figures~\ref{fig:ex_limit}-\ref{fig:ex_long2}, and light curves for all 
identified variable stars can be found at 
http://www.astro.caltech.edu/$\sim$jmc/variables/orion as well as in the
electronic version of this article.

In total, 1235 variable stars were identified, compared
to the estimated \about 2700 stars with $K_s \le 14$\M\ that are associated 
with the Orion~A molecular cloud. The observed spatial distribution, 
magnitude, and colors suggest that \about 93\% of the variables are 
pre-main-sequence stars associated with Orion~A. This sample of 
variable stars were analyzed as an ensemble to establish the characteristics 
of the amplitude, time scales, and any correlated magnitude-color fluctuations 
for pre-main-sequence stars. The mean peak-to-peak
photometric fluctuation is \about 0.2\M\ in each band, although the
more extreme variable stars have amplitudes as large as \about 2\M. Most of
the variability is essentially colorless within the photometric noise, with
77\% of the variables stars having peak-to-peak $J-H$ color variations 
less than 0.05\M. The time scale of the photometric fluctuations as 
established from the autocorrelation function are typically less than a few 
days. Only \about 2\% of the variable stars were identified based solely on 
their long term variability with amplitude fluctuations \aboutmore 0.1\M, 
suggesting that most near-infrared variability is a result of short term 
processes.

The variability characteristics established from these observations 
constrain the mechanisms that lead to near-infrared photometric variability
in individual sources. While any variety of mechanism can produce the 
observed low amplitude fluctuations, stars with low amplitudes typically show
colorless photometric variations and near-infrared colors consistent with
reddened photospheres (i.e. lack near-infrared excesses), and often exhibit 
periodic photometric fluctuations. These characteristics are analogous to that 
observed in Type~I optical variability in the \citep{Herbst94} classification 
scheme, and we suggest that the variability in 56-77\% of the stars is a
result of rotational modulation by cool spots. Cool spots cannot explain the 
23\% of the stars that have color variations \aboutmore 0.05\M, however, and 
alternate mechanisms were investigated to explain these variability 
characteristics, including hot spots, extinction and accretion disks. 
Rotational modulation by hot spots and extinction variations can account for 
the amplitude of the photometric variations and the observed correlation 
between the $J$ and $J-H$ variations. However, the more extreme
variable stars would require spot temperatures and spot coverages larger
than inferred from optical observations. In addition, neither hot spots nor
extinction can quantitatively explain the correlations between the magnitude
and color variations in the $J-H$ vs. \HK\ and \KB\ vs. \HK\ diagrams or 
account for stars that become bluer as they fade. Accretion disk models
were investigated as another source of near-infrared variability. By invoking
changes in the mass accretion rate or changes in the inner hole radius of the
circumstellar disk (whether by dynamical, opacity, or temperature variations),
accretion disk models can account for the \about 17 stars that become bluer as 
they decrease in brightness. Like hot spots and extinction, however, 
accretion disk models cannot explain the detailed distribution of stars in the 
color-color diagram. This suggests that either another variability mechanism
not considered here may be operative, or that the observed variability 
represents the net results of several of these phenomena. In particular,
hot spots, extinction variations, and accretion variations all relate to the
properties of the circumstellar disk, and it is plausible, if not likely, that 
more than one of these mechanisms is simultaneously contributing to the
near-infrared variability in many of these stars.

\acknowledgements

We thank Nuria Calvet for allowing us to use the accretion disk model results
shown in Figure~\ref{fig:models}, Bill Herbst for discussions about 
photometric periodicity in young stellar objects, and Luisa Rebull for sharing
her optical periodicity results with us prior to publication. We would also 
like to thank the 2MASS Observatory Staff and Data Management Team for 
acquiring and pipeline processing the special survey observations used in this 
investigation. This publication makes use of data products from the Two Micron 
All Sky Survey, which is a joint project of the University of Massachusetts 
and the Infrared Processing and Analysis Center, funded by the National 
Aeronautics and Space Administration and the National Science Foundation. 
2MASS science data and information services were provided by the InfraRed 
Science Archive (IRSA) at IPAC. This research has made use of the SIMBAD 
database, operated at CDS, Strasbourg, France. JMC acknowledges support from 
Long Term Space Astrophysics Grant NAG5-8217 and the Owens Valley Radio 
Observatory, which is supported by the National Science Foundation through 
grant AST-9981546.

\appendix

\section{The Stellar Population in the Orion~A Molecular Cloud}

\citet{Lynne98b} demonstrated that the Trapezium cluster, a \about 
$3'\times5'$ region centered near $\theta^1$~C Ori \citep{Herbig86},
constitutes the core of the Orion Nebula Cluster. They further showed that 
the Orion Nebula Cluster contains upwards of a few thousand stars extended
over at least a 18\arcmin\ (2.5~pc) region and is elongated along the 
direction of the Orion~A molecular cloud with an aspect ratio of \about 2:1. 
\citet{Ali95} indicated though that the observed $K$-band stellar surface 
density is higher than the expected field star density over much of the 
$39'\times39'$ region they surveyed. \citet{Carp00} further found from 
analysis of the 2MASS Second Incremental Release data that the enhanced 
\KB-band star counts extend for \about 2\arcdeg\ in declination. The 
observations obtained here are spatially more complete than the data analyzed 
in that study and allow for a more thorough investigation of the spatial 
extent and membership of the northern portion of the Orion~A molecular cloud 
as now described.

\begin{figure}
\insertplot{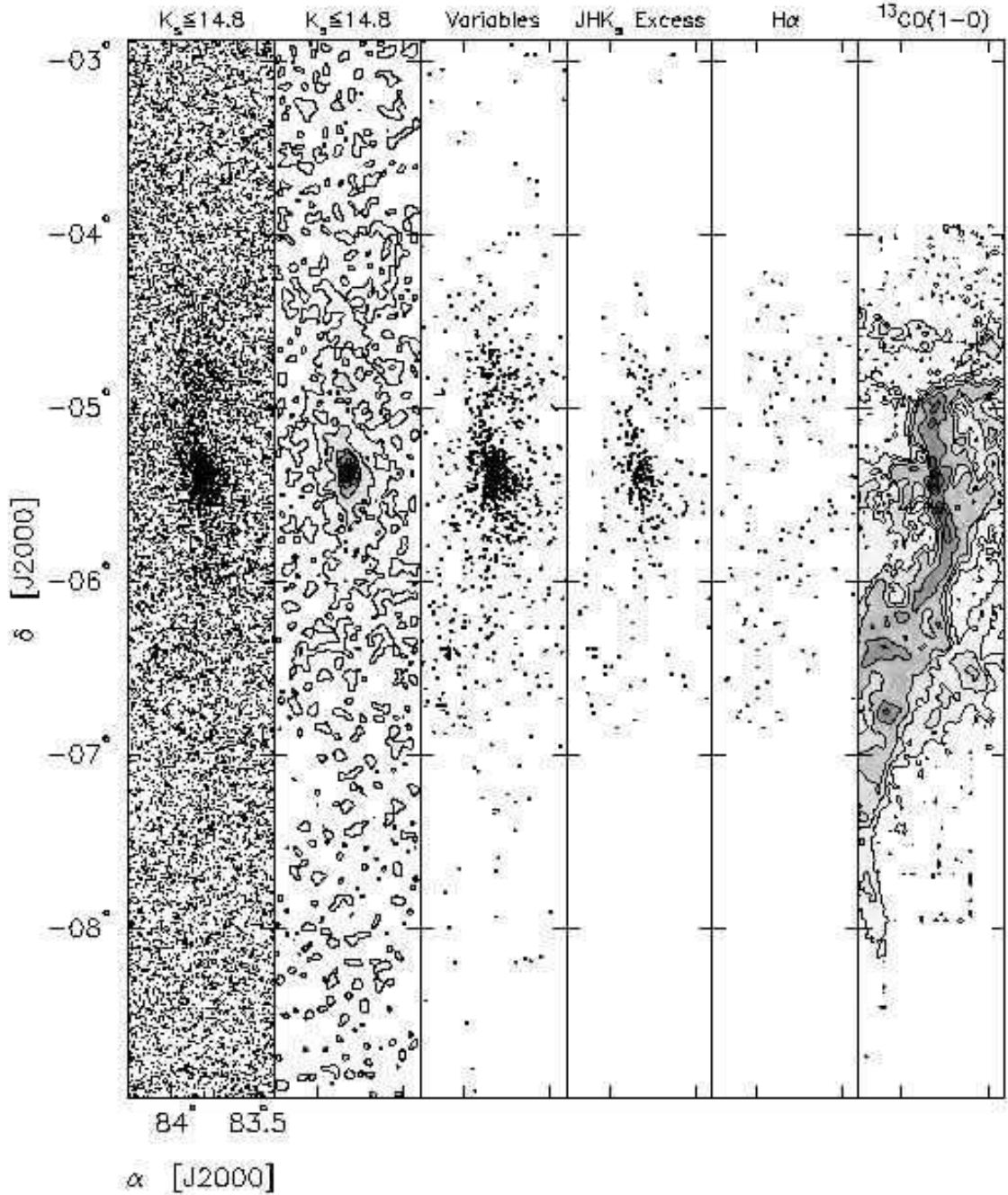}{6.8}{8.4}{0.3}{1.8}{0.90}{0}
\caption{
 Spatial distribution of stars and molecular gas toward the Orion Nebula 
 Cluster. Starting with the left-most panel, these figures show the
 ({\it a}) spatial distribution of stars with \KB $\le 14.8$\M;
 ({\it b}) surface density of stars with \KB $\le 14.8$\M, where the surface 
           density map was created by convolving the stellar spatial 
           distribution shown in ({\it a}) with a gaussian kernel of size 
           $\sigma=60$\arcsec;
 ({\it c}) spatial distribution of variable stars from Sample~1 (see 
           Table~\ref{tbl:samples});
 ({\it d}) distribution of variable stars shown in ({\it c}) that have a 
           near-infrared excess in the $J-H$ vs. \HK\ color-color diagram;
 ({\it e}) distribution of H$\alpha$ emitting stars from the Kiso H$\alpha$
           survey with a Kiso class of 3, 4, or 5 (Wiramihardja 
           \etal~1991,1993);
 ({\it f}) contour map of the integrated \thcoj\ emission ($\int T_R dv$) from 
           \citet{Bally87}. The contours levels are 1, 5, 10, 20, 30, 40, and 
           50\kkms.
 These panels indicate that the variable stellar population follow the large
 scale spatial distribution of stars in the Orion~A molecular cloud as traced 
 by the total \KB-band star counts and $H\alpha$ emitting objects. 
 \label{fig:radec}
}
\end{figure}

In general it is not possible to identify the individual stars associated with
the Orion~A molecular cloud from $J$, $H$, and \KB\ photometry alone unless a
star has a near-infrared excess. Since not all stars have such distinctive
colors, we identified the spatial distribution of stars statistically by 
subtracting the expected field star population from the observed star counts 
using the semi-empirical procedure described by \citet{Carp00}. Briefly, the 
field star contamination in the absence of extinction from the Orion~A 
molecular cloud itself was assessed using the regions ($\alpha < 84$\arcdeg, 
$\delta < -7$\arcdeg) and ($\delta > -4$\arcdeg) that are off of the molecular 
cloud and away from the stellar density peaks (see Fig.~\ref{fig:radec}). The 
mean \KB-band surface density in these regions is 0.66\sqamin\ for stars 
brighter than \KB=14.8\M\ and 0.006\sqamin\ for the variable stars. These 
results indicate that the field star contamination to the variable population
is only \about 7\% and is therefore neglected throughout this paper.
The field star surface density for the $K_s\le14.8$\M\ star counts decreases 
by \about 10\% from $\delta > -4.2$\arcdeg\ to $\delta < 
-7$\arcdeg, which is consistent with the expected variation predicted by 
Galactic star count models due to variations in the Galactic latitude across 
the region surveyed \citep{Wainscoat92}. The variation with Galactic longitude 
is predicted to be \about 1\%. The expected surface density of unreddened 
field stars was estimated by making a linear fit to the stellar surface 
density map in the off-cloud regions as a function of Galactic latitude. 
Extinction from the 
Orion~A molecular cloud, however, will decrease the number of background field 
stars predicted by this linear fit. The reddened field star density as a 
function of position in the cloud was estimated using the \thcoj\ map from 
\citep{Bally87} as a tracer of the cloud extinction and the 
\citet{Wainscoat92} star count model to estimate the fraction of field stars
that are background to the molecular cloud (see Carpenter~2000 for complete 
details). This reddened field star model was subtracted from the observed 
surface density map to yield the stellar population associated with the 
Orion~A molecular cloud.

\begin{figure}
\plotone{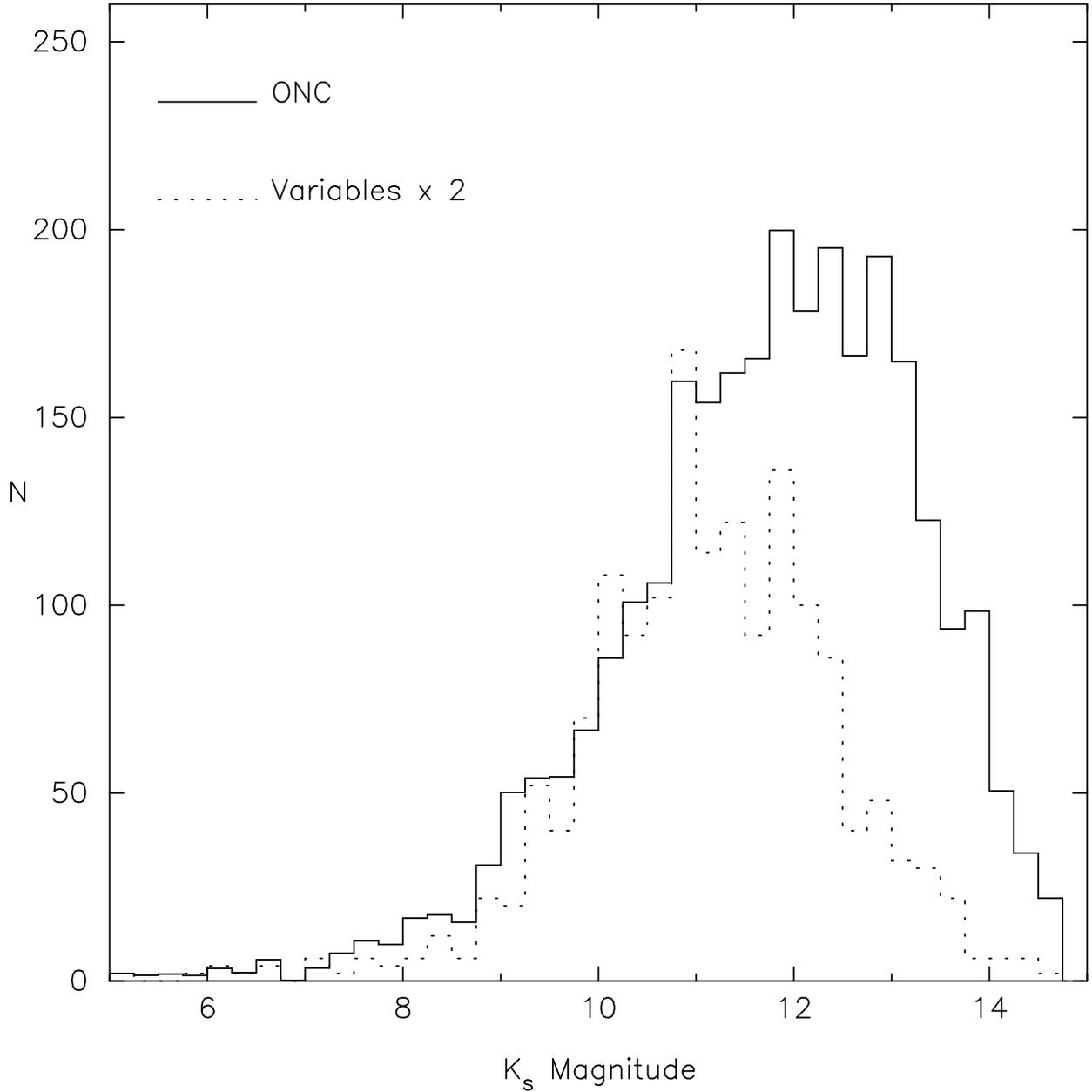}
\caption{
  Histogram of the \KB-band magnitudes for stars in the Orion~A molecular cloud
  and the variable star population. The field star contribution to the cloud 
  population has been subtracted
  from the observed star counts using the procedure described in the Appendix
  applied to differential magnitude intervals. The completeness limit of the 
  observations is \KB=14.8\M. This figure indicates that the variable star 
  population identified with these observations tend to be the brighter stars 
  in the cluster. The lack of faint variable stars is likely a result of
  increased photometric noise at these magnitudes that masks any low amplitude
  photometric fluctuations.
  \label{fig:khist}
}
\end{figure}

\begin{figure}
\insertplot{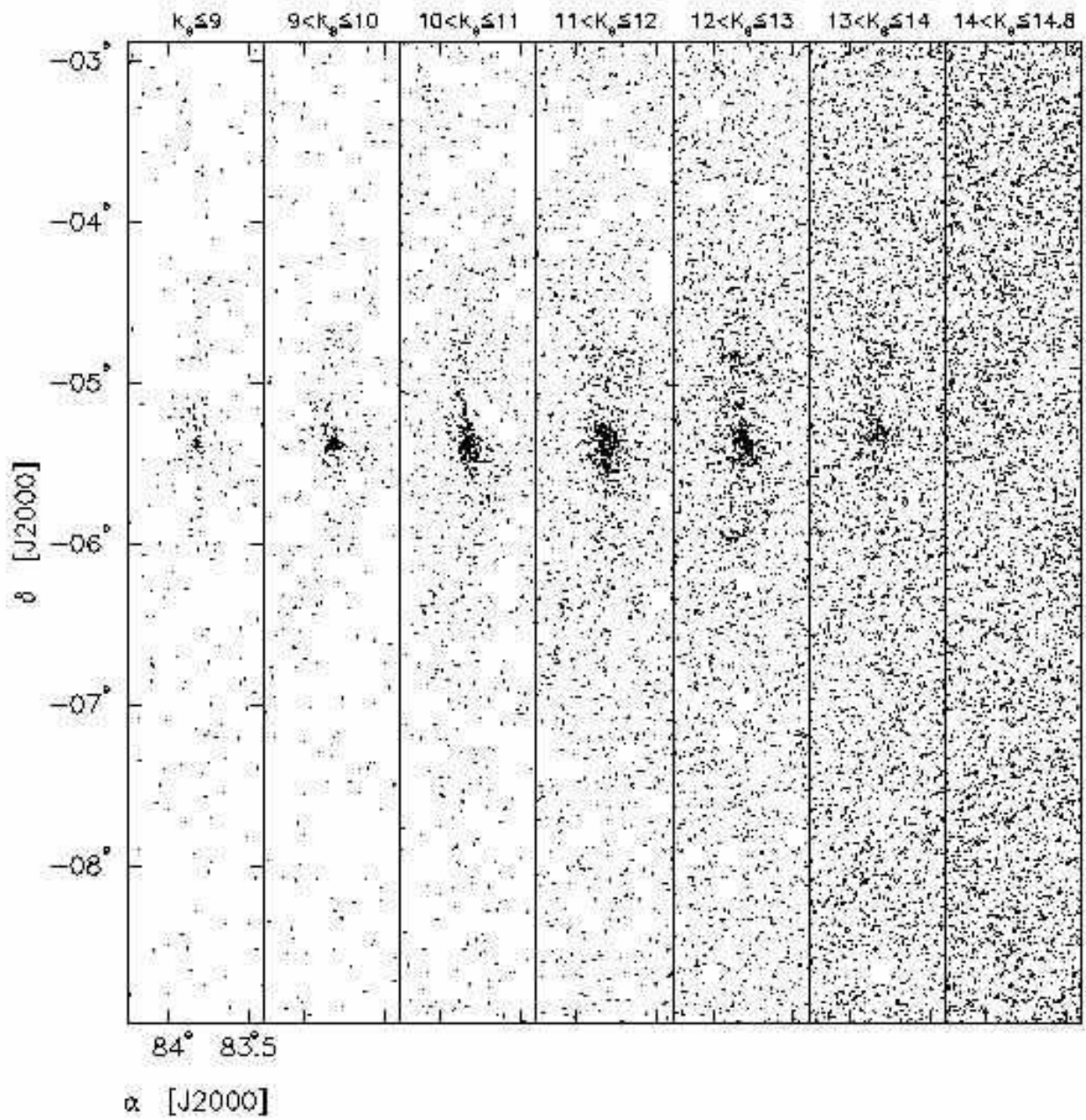}{7.8}{8.4}{0.0}{1.8}{1.00}{0}
\caption{
  Spatial distribution of stars in differential \KB\ magnitude intervals.
  This figure demonstrates that the stellar population associated with the 
  Orion~A molecular cloud is prominent only for \KB\ \aboutless 14\M. At 
  fainter magnitudes, the star counts are dominated by the field star 
  population.
  \label{fig:kmag}
}
\end{figure}

Generally the star count analysis should be performed over all \KB-band
magnitudes to obtain the most complete estimate of the stellar population. In 
practice, a spatial enhancement in the stellar surface density is only
prominent at brighter magnitudes as shown in Figure~\ref{fig:kmag}, which 
shows the spatial distribution of stars in differential magnitude 
intervals. An excess of stars is discerned against the field star population 
for magnitudes \KB\ \aboutless 13\M, but is no longer prominent at \KB\ 
\aboutmore 14\M\ (see also Fig.~\ref{fig:khist}). Including the star counts 
fainter than \KB=14\M\ will add noise to the statistical analysis, and 
therefore, only stars with magnitudes \KB\ $\le$ 14\M\ were used in this
analysis.

The extent and number of stars in the Orion~A molecular cloud were estimated 
by forming closed contours in the field star subtracted surface density map
at various density levels. For each density, we computed the number of 
observed stars in excess of the field star model, the number of variable 
stars, and the spatial extent assuming a distance of 480~pc \citep{Genzel81}.
The spatial extent is characterized by a major and minor axis, defined as the 
largest angular extent in declination at the defined surface density level, 
and the perpendicular angular dimension needed in order to reproduce the total 
contour area. The lowest surface density considered begins at \about 1$\sigma$ 
(0.25\sqamin) above the mean field star density, where $\sigma$ is the Poisson 
noise in the field star contamination in the absence of extinction from the 
molecular cloud. Table~\ref{tbl:cluster} summarizes the angular extent and 
number of stars as a function of the surface density. At the lowest surface 
density level identified here, the closed contour extends over a 
$0.4^\circ\times2.4^\circ$ region ($3.4\:\rm{pc}\times20\:\rm{pc}$) and 
contains \about 2700 stars brighter than \KB=14\M. These source counts 
represent a lower limit to the total stellar population in the northern portion
of the Orion~A molecular cloud due to unresolved binaries and the difficulty 
in identifying point sources in the Trapezium region with its bright nebular
background and high overall source density. A more global star count analysis 
indicates that these observations cover the full area of the stellar density
enhancement even though the Orion~A molecular cloud extends further to the 
southeast \citep{Carp00}. The angular extent of the density enhancement is
similar to that seen in OB stars and H$\alpha$ emitting objects 
\citep{Gomez98}.

The lowest surface density contour encompasses several regions that have been 
previously labeled as individual clusters, including the ONC, the Trapezium 
(which is now considered part of the ONC; Hillenbrand \& Hartmann 1998), 
L1641N \citep{Strom93,Hodapp93}, OMC2 \citep{Jones94}, smaller groups of 
stars associated with low luminosity IRAS sources \citep{Chen94}, and a more 
dispersed grouping of stars near \NGC1977. While each of these regions 
represent a local stellar density peak, the results obtained here suggest that 
they are all connected as part of an extended ridge of stars that have formed 
within a localized region of the Orion~A molecular cloud. This assertion is 
supported by the spatial continuity of the enhanced star counts, their close 
correspondence with the shape of Orion~A molecular cloud (see 
Fig.~\ref{fig:radec}), and observational signatures of young stellar ages 
(\about 1~Myr) over the entire region, either from direct spectroscopic 
evidence \citep{Hodapp93,Lynne97} or from the presence of outflows and 
near-infrared excesses common in pre-main-sequence stars 
\citep{Strom93,Chen94,Jones94}. 

Whether the enhanced ridge of star counts should be identified as a single
object or several clusters is somewhat ambiguous. Historically clusters have 
been defined as gravitationally bound groups of stars. The elongated shape of 
the stellar density enhancement observed in Orion~A suggests that it is 
unlikely it will persist as a bound group once the molecular gas is dispersed. 
Localized regions within the stellar density enhancement may remain
gravitationally bound, though, and specifically, the ONC as defined by 
\citet{Lynne98b} {\it may} emerge as a cluster if a factor of \about 2 more 
stars are formed in the 2.5~pc region centered on the Trapezium. Near-infrared 
star count analyses in the past decade though have loosely identified clusters 
as groups of stars forming within a localized region of a molecular cloud that 
may not necessarily emerge as gravitationally bound entities. By this 
definition, the boundaries of the previously identified ONC, L1641N, and OMC2 
clusters are not easily distinguished, and it is not clear how, or if, the
stellar density enhancement should be divided into separate clusters.



\newcommand{\h}{$^{\rm h}$}
\newcommand{\m}{$^{\rm m}$}
\newcommand{\s}{$^{\rm s}$}


\begin{deluxetable}{cccc}
\tablewidth{275pt}
\tablecaption{Coordinates of Observed Tiles\label{tbl:coords}}
\tablehead{
  \colhead{Tile}                       & 
  \colhead{Right Ascension}            &
  \multicolumn{2}{c}{Number of Nights Observed} \\
  \cline{3-4} \\
  \colhead{}         & 
  \colhead{(J2000)}  &
  \colhead{Complete} &
  \colhead{Partial}
}
\startdata
1 &  05\h33\m47\s & 18 & 3\\
2 &  05\h34\m16\s & 22 & 3\\
3 &  05\h34\m45\s & 26 & 3\\
4 &  05\h35\m14\s & 28 & 3\\
5 &  05\h35\m42\s & 26 & 3\\
6 &  05\h36\m11\s & 21 & 3\\
7 &  05\h36\m40\s & 16 & 3\\
\enddata
\end{deluxetable}
\clearpage





\begin{deluxetable}{rcccccccl}
\tablewidth{460pt}
\tablecaption{Observing Log\label{tbl:log}}
\tablehead{
\colhead{UT Date}  & 
\colhead{Tile 7}  & 
\colhead{Tile 6}  & 
\colhead{Tile 5}  & 
\colhead{Tile 4}  & 
\colhead{Tile 3}  & 
\colhead{Tile 2}  & 
\colhead{Tile 1}  &
\multicolumn{1}{l}{Comments}
%
}
\startdata
March 19, 1998   & x & x & x & x & x & x & x & $\delta > -6^\circ$\\
March 22, 1998   & x & x & x & x & x & x & x & $\delta < -6^\circ$\\
February 6, 2000 & x & x & x & x & x & x & x & $\delta > -6^\circ$\\
March 4, 2000    & x & x & x & x & x & x & x & \\
March 5, 2000    & x & x & x & x & x & x & x & \\
March 6, 2000    & x & x & x & x & x & x & x & \\
March 7, 2000    & x & x & x & x & x & x & x & \\
March 9, 2000    &   &   & x & x & x & x &   & \\
March 11, 2000   & x & x & x & x & x & x & x & \\
March 12, 2000   & x & x & x & x & x & x & x & \\
March 13, 2000   & x & x & x & x & x & x & x & \\
March 14, 2000   & x & x & x & x & x & x & x & \\
March 15, 2000   & x & x & x & x & x & x & x & \\
March 16, 2000   & x & x & x & x & x & x & x & \\
March 18, 2000   & x & x & x & x & x & x & x & \\
March 19, 2000   & x & x & x & x & x & x & x & \\
March 20, 2000   & x & x & x & x & x & x & x & \\
March 21, 2000   & x & x & x & x & x & x & x & \\
March 22, 2000   & x & x & x & x & x & x & x & \\
March 23, 2000   & x & x & x & x & x & x & x & \\
March 24, 2000   &   & x & x & x & x & x & x & \\
March 25, 2000   &   & x & x & x & x & x & x & \\
March 26, 2000   &   & x & x & x & x & x &   & \\
March 27, 2000   &   & x & x & x & x & x &   & \\
March 28, 2000   &   & x & x & x & x & x &   & \\
March 30, 2000   &   &   & x & x & x &   &   & \\
March 31, 2000   &   &   & x & x & x &   &   & \\
April  1, 2000   &   &   & x & x & x &   &   & \\
April  5, 2000   &   &   &   & x & x &   &   & \\
April  6, 2000   &   &   &   & x &   &   &   & \\
April  7, 2000   &   &   &   & x &   &   &   & \\
April  8, 2000   &   &   & x &   &   &   &   & \\
\enddata
\end{deluxetable}
\clearpage






\begin{deluxetable}{l@{\extracolsep{50pt}}r}
\tablewidth{250pt}
\tablecaption{Stellar Samples\label{tbl:samples}}
\tablehead{
  \colhead{Sample}         & 
  \colhead{N$_{\rm{stars}}$}
}
\startdata
 All stars\tablenotemark{a}             & 17808 \\
 All variables                          &  1236 \\
 Sample 1 variables\tablenotemark{b}    &  1006 \\
 Sample 2 variables\tablenotemark{c}    &  1054 \\
 2 Month variables\tablenotemark{d}     &    14 \\
 2 Year variables\tablenotemark{e}      &    72 \\
 Periodic                               &   233 \\
 Eclipsing Candidates                   &    22 \\
 Subjective Additions                   &    60 \\
\enddata
\tablenotetext{a}{$J\le16.0^{\rm m}$, $H\le15.4^{\rm m}$, or $K_s\le14.8^{\rm m}$}
\tablenotetext{b}{Identified from 16 nights common to all tiles}
\tablenotetext{c}{Identified from March/April 2000 observations}
\tablenotetext{d}{Between February 2000 and March/April~2000}
\tablenotetext{e}{Between March 1998 and March/April~2000}
\end{deluxetable}
\clearpage






\begin{deluxetable}{rcc@{\extracolsep{10pt}}
 r@{\extracolsep{10pt}}r@{\extracolsep{10pt}}r@{\extracolsep{10pt}}
 c@{\extracolsep{10pt}}c@{\extracolsep{10pt}}c@{\extracolsep{10pt}}
 c@{\extracolsep{0pt}}c@{\extracolsep{0pt}}c@{\extracolsep{0pt}}
 @{\extracolsep{10pt}}c@{\extracolsep{0pt}}c}
\tabletypesize{\scriptsize}
\tablewidth{480pt}
\tablecaption{Near-Infrared Variable Stars (INCOMPLETE LIST)\label{tbl:variables}}
\tablehead{
  \colhead{}  &
  \multicolumn{2}{c}{Equatorial (J2000)}  &
  \multicolumn{3}{c}{Magnitudes\tablenotemark{a}}   &
  \multicolumn{3}{c}{Observed RMS\tablenotemark{a}} &
  \multicolumn{3}{c}{N\tablenotemark{a}}        &
  \multicolumn{2}{c}{Variability\tablenotemark{a}}  \\
  \cline{2-3}
  \cline{4-6}
  \cline{7-9}
  \cline{10-12}
  \cline{13-14}\\
  \colhead{ID}        & 
  \colhead{$\alpha$}  &
  \colhead{$\delta$}  &
  \colhead{$J$}       &
  \colhead{$H$}       &
  \colhead{$K_s$}     &
  \colhead{$J$}       &
  \colhead{$H$}       &
  \colhead{$K_s$}     &
  \colhead{$J$}       &
  \colhead{$H$}       &
  \colhead{$K_s$}     &
  \colhead{Index}     &
  \colhead{Flags\tablenotemark{b}}
}
\startdata
    91 &   83.410336 & $  -6.000067$ & 13.724 & 13.086 & 12.814 &  0.034 &  0.031 &  0.042 &  22 &  21 &  21 & $  0.46$ & 110000\\
   123 &   83.412085 & $  -7.502039$ & 11.428 & 10.809 & 10.573 &  0.054 &  0.062 &  0.054 &  19 &  19 &  19 & $  0.89$ & 000001\\
   237 &   83.418483 & $  -3.056776$ & 14.113 & 13.511 & 13.126 &  0.052 &  0.056 &  0.084 &  21 &  21 &  21 & $  0.79$ & 110000\\
   292 &   83.421295 & $  -3.224478$ & 14.872 & 14.484 & 14.351 &  0.111 &  0.117 &  0.142 &  20 &  20 &  20 & $  1.09$ & 110000\\
   307 &   83.421983 & $  -4.823802$ & 12.192 & 11.525 & 11.292 &  0.035 &  0.038 &  0.036 &  20 &  20 &  20 & $  0.87$ & 110000\\
   341 &   83.423342 & $  -6.102054$ & 14.510 & 13.573 & 12.854 &  0.218 &  0.185 &  0.128 &  19 &  19 &  19 & $  3.55$ & 110000\\
   376 &   83.424761 & $  -6.263577$ & 14.036 & 13.334 & 12.895 &  0.111 &  0.089 &  0.072 &  19 &  19 &  19 & $  1.95$ & 110000\\
   517 &   83.432022 & $  -6.408023$ & 11.194 & 10.568 & 10.386 &  0.040 &  0.041 &  0.037 &  19 &  19 &  19 & $  1.00$ & 110100\\
   609 &   83.436659 & $  -5.236045$ & 10.770 & 10.041 &  9.654 &  0.037 &  0.034 &  0.034 &  20 &  20 &  20 & $  0.45$ & 100000\\
   621 &   83.437227 & $  -5.519041$ & 11.242 & 10.464 & 10.160 &  0.042 &  0.046 &  0.034 &  20 &  20 &  20 & $  0.98$ & 110100\\
   654 &   83.438991 & $  -6.676367$ & 14.855 & 14.229 & 13.634 &  0.080 &  0.103 &  0.103 &  19 &  19 &  19 & $  1.39$ & 110000\\
   662 &   83.439227 & $  -4.591010$ & 16.249 & 15.359 & 14.595 &  0.227 &  0.209 &  0.186 &  17 &  20 &  20 & $  0.90$ & 110000\\
   665 &   83.439355 & $  -6.951656$ & 12.719 & 12.027 & 11.816 &  0.050 &  0.042 &  0.045 &  19 &  19 &  19 & $  1.09$ & 110000\\
   666 &   83.439399 & $  -5.609014$ & 12.041 & 11.244 & 10.732 &  0.042 &  0.049 &  0.099 &  20 &  20 &  20 & $  0.80$ & 000001\\
   703 &   83.441158 & $  -5.549489$ & 16.445 & 15.077 & 13.972 &  0.287 &  0.427 &  0.296 &  18 &  20 &  20 & $  1.98$ & 001000\\
   724 &   83.442073 & $  -5.574040$ & 11.789 & 11.133 & 10.955 &  0.038 &  0.029 &  0.024 &  20 &  20 &  20 & $  0.48$ & 110000\\
   766 &   83.444522 & $  -5.390434$ & 13.681 & 12.270 & 11.059 &  0.229 &  0.245 &  0.258 &  20 &  20 &  20 & $  6.96$ & 110000\\
   839 &   83.448550 & $  -5.430148$ & 12.552 & 11.878 & 11.560 &  0.038 &  0.034 &  0.025 &  20 &  20 &  20 & $  0.67$ & 100000\\
   897 &   83.450932 & $  -5.223924$ & 13.007 & 12.208 & 11.726 &  0.059 &  0.046 &  0.047 &  20 &  20 &  20 & $  0.97$ & 110000\\
   996 &   83.456430 & $  -5.605810$ & 12.775 & 11.706 & 10.975 &  0.376 &  0.307 &  0.207 &  20 &  20 &  20 & $  9.56$ & 110000\\
  1048 &   83.459572 & $  -6.363610$ & 13.973 & 13.004 & 12.448 &  0.363 &  0.315 &  0.242 &  19 &  19 &  19 & $  7.54$ & 110000\\
  1079 &   83.461424 & $  -5.010936$ & 13.118 & 12.387 & 12.080 &  0.034 &  0.039 &  0.035 &  20 &  20 &  20 & $  0.83$ & 110000\\
  1114 &   83.463838 & $  -5.387879$ & 13.146 & 11.102 &  9.771 &  0.218 &  0.173 &  0.128 &  20 &  20 &  20 & $  3.89$ & 110000\\
  1117 &   83.463960 & $  -4.806136$ & 12.483 & 11.671 & 11.288 &  0.125 &  0.106 &  0.070 &  20 &  20 &  20 & $  2.77$ & 110000\\
  1150 &   83.465796 & $  -5.551001$ & 15.361 & 13.466 & 12.252 &  0.329 &  0.264 &  0.185 &  20 &  20 &  20 & $  3.97$ & 110000\\
  1185 &   83.467603 & $  -6.984333$ & 13.165 & 12.501 & 12.267 &  0.049 &  0.024 &  0.025 &  19 &  19 &  19 & $ -0.16$ & 000001\\
  1186 &   83.467627 & $  -7.924482$ & 14.406 & 14.029 & 13.910 &  0.072 &  0.049 &  0.067 &  19 &  19 &  19 & $ -0.04$ & 000001\\
  1189 &   83.467694 & $  -8.469353$ & 14.283 & 13.777 & 13.685 &  0.063 &  0.052 &  0.045 &  19 &  19 &  19 & $ -0.01$ & 000001\\
  1199 &   83.468187 & $  -5.697293$ & 11.704 & 10.978 & 10.685 &  0.063 &  0.039 &  0.043 &  20 &  20 &  20 & $  0.92$ & 110000\\
  1340 &   83.476564 & $  -5.753533$ & 13.690 & 12.918 & 12.473 &  0.131 &  0.104 &  0.083 &  20 &  20 &  20 & $  2.44$ & 110000\\
  1426 &   83.481348 & $  -5.432632$ & 13.342 & 12.044 & 11.262 &  0.250 &  0.175 &  0.125 &  20 &  20 &  20 & $  5.62$ & 110000\\
\enddata
\tablenotetext{a}{Photometric parameters computed using all available data}
\tablenotetext{b}{Each digit represents a different variability characteristic.}
\tablenotetext{\ }{First digit : Variable on 16 nights common to entire area}
\tablenotetext{\ }{Second digit: Variable in March/April 2000 data}
\tablenotetext{\ }{Third digit : Long term variability in February 2000 and/or March 1998}
\tablenotetext{\ }{Fourth digit : Periodic variable}
\tablenotetext{\ }{Fifth digit : Eclipsing candidate}
\tablenotetext{\ }{Sixth digit : Selected as variable from subjective inspection
 of light curves}
\end{deluxetable}
\clearpage






\begin{deluxetable}{ccrrc}
\tablewidth{320pt}
\tablecaption{Stellar Population Associated with Orion~A\label{tbl:cluster}}
\tablehead{
  \colhead{Surface Density} & 
  \colhead{Size}            &
  \colhead{N($K_s\le$14)}   &
  \colhead{N$_{var}$}       &
  \colhead{$f_{var}$}      \\
  \colhead{(arcmin$^{-2}$)}    &
  \colhead{($'$ $\times$ $'$)} & 
}
\startdata
 0.25 & $24\times144$  & 2704 & 786 & 0.29 $\pm$ 0.010 \\
 0.50 & $20\times87$   & 2148 & 627 & 0.29 $\pm$ 0.011 \\
 0.75 & $14\times86$   & 1881 & 554 & 0.29 $\pm$ 0.012 \\
 1.00 & $12\times57$   & 1488 & 445 & 0.30 $\pm$ 0.014 \\
 1.25 & $12\times40$   & 1262 & 386 & 0.31 $\pm$ 0.016 \\
 2.50 & $9.3\times20$  &  895 & 258 & 0.29 $\pm$ 0.018 \\
 3.75 & $7.9\times17$  &  752 & 202 & 0.27 $\pm$ 0.019 \\
 5.00 & $7.0\times14$  &  621 & 158 & 0.25 $\pm$ 0.020 \\
 7.50 & $4.9\times9.5$ &  397 &  94 & 0.24 $\pm$ 0.025 \\
10.00 & $4.2\times6.0$ &  253 &  52 & 0.21 $\pm$ 0.029 \\
\enddata
\end{deluxetable}
\clearpage





\begin{deluxetable}{cc@{\extracolsep{-10pt}}ccc@{\extracolsep{10pt}}c@{\extracolsep{-10pt}}ccc}
\tablewidth{446pt}
\tablecaption{Variable Star Amplitudes in March/April 2000\label{tbl:amp}}
\tablehead{
  \colhead{Band}       & 
  \multicolumn{4}{c}{Peak-to-Peak (mag)} &
  \multicolumn{4}{c}{RMS (mag)}\\
\cline{2-5}
\cline{6-9}
  \colhead{}           & 
  \colhead{Maximum}    & 
  \colhead{Mean}       &
  \colhead{Median}     &
  \colhead{Dispersion} &
  \colhead{Maximum}    & 
  \colhead{Mean}       &
  \colhead{Median}     &
  \colhead{Dispersion}
}
\startdata
 $J$      & 2.31 & 0.26 & 0.15 & 0.31 & 0.54 & 0.09 & 0.06 & 0.09\\
 $H$      & 1.95 & 0.22 & 0.12 & 0.25 & 0.45 & 0.08 & 0.05 & 0.07\\
 $K_s$    & 1.16 & 0.17 & 0.11 & 0.18 & 0.36 & 0.07 & 0.05 & 0.05\\
 $J-K_s$  & 1.22 & 0.10 & 0.02 & 0.16 & 0.38 & 0.04 & 0.02 & 0.05\\
 $J-H$    & 0.61 & 0.03 & 0.00 & 0.07 & 0.21 & 0.02 & 0.00 & 0.03\\
 $H-K_s$  & 0.82 & 0.04 & 0.02 & 0.08 & 0.19 & 0.02 & 0.00 & 0.03\\
\enddata
\end{deluxetable}
\clearpage






\begin{deluxetable}{rrrr@{\extracolsep{20pt}}
 r@{\extracolsep{10pt}}r@{\extracolsep{10pt}}r@{\extracolsep{10pt}}}
\tabletypesize{\scriptsize}
\tablewidth{250pt}
\tablecaption{Periodic Stars (INCOMPLETE LIST)\label{tbl:periodic}}
\tablehead{
  \colhead{ID}  &
  \multicolumn{3}{c}{Periods (days)}   &
  \multicolumn{3}{c}{FAP}  \\
  \cline{2-4}
  \cline{5-7}\\
  \colhead{}       & 
  \colhead{$J$}    &
  \colhead{$H$}    &
  \colhead{$K_s$}  &
  \colhead{$J$}    &
  \colhead{$H$}    &
  \colhead{$K_s$}
}
\startdata
   517 &   6.52 &   6.16 &   6.25 &     0.03 &     0.03 &     0.04\\
   621 &   2.66 &   2.60 &   2.65 &     0.07 &     0.05 &     0.03\\
  1721 &   5.06 &   4.95 &   5.06 & $<$0.01 &     0.02 &     0.01\\
  1936 &   4.21 &   4.17 &   4.21 &     0.02 &     0.01 &     0.02\\
  1994 &   2.65 &   2.68 &   2.66 &     0.02 &     0.06 &     0.05\\
  2276 &   6.11 &   5.87 &   6.03 & $<$0.01 &     0.01 &     0.02\\
  2342 &   4.58 &   4.54 &   4.67 &     0.02 &     0.01 &     0.03\\
  2718 &   9.96 &  10.18 &  10.41 &     0.01 &     0.05 &     0.08\\
  3179 &   8.81 &   8.81 &   8.64 &     0.02 &     0.02 &     0.04\\
  3238 &   2.25 &   2.26 &   2.26 &     0.07 &     0.01 &     0.05\\
  3285 &   5.52 &   5.39 &   5.33 &     0.01 &     0.01 &     0.08\\
  3644 &  10.65 &  10.41 &  10.65 &     0.01 &     0.01 &     0.02\\
  3730 &   8.18 &   7.90 &   7.76 & $<$0.01 & $<$0.01 & $<$0.01\\
  3960 &   5.09 &   5.03 &   4.98 &     0.01 &     0.02 & $<$0.01\\
  3993 &   9.54 &   9.35 &   9.16 & $<$0.01 & $<$0.01 & $<$0.01\\
  4034 &   7.05 &   7.05 &   6.94 & $<$0.01 & $<$0.01 & $<$0.01\\
  4094 &   4.36 &   4.36 &   4.40 &     0.06 &     0.06 &     0.02\\
  4097 &   2.06 &   2.07 &   2.04 & $<$0.01 & $<$0.01 &     0.01\\
  4127 &   9.16 &   8.98 &   9.16 &     0.06 &     0.01 &     0.02\\
  4156 &  10.65 &  10.18 &   9.96 &     0.05 &     0.02 &     0.04\\
\enddata
\end{deluxetable}
\clearpage


\clearpage

\clearpage

\clearpage

\clearpage

\clearpage

\clearpage

\clearpage

\clearpage

\clearpage

\clearpage

\clearpage

\clearpage

\clearpage

\begin{figure}
\insertplot{figure13.ps}{7.8}{8.4}{0.0}{0.8}{0.75}{1}
\caption{
  Photometric data for star 5841 (also known has JW~101 and V1314~Ori), a
  second example of a star that exhibits long term photometric variability
  relative to the March/April 2000 time series data (see also 
  Fig.~\ref{fig:ex_long1}). In this case, the star is identified as a 
  variable in the March/April 2000 data, but exhibits even larger fluctuations
  in the March~1998 data. The long term variability is such that the star
  got fainter at $J$-band while simultaneously getting brighter at \KB-band.
  \label{fig:ex_long2}
}
\end{figure}
\clearpage

\clearpage

\clearpage

\clearpage

\clearpage

\clearpage

\clearpage

\clearpage

\clearpage

\clearpage

\begin{figure}
\plotone{figure23.ps}
\caption{
  Comparison of the optical, $I$-band periods 
  \citep{Stassun98,Herbst00b,Rebull01} with the $H$-band
  periods derived in this study. For \about 80\% of the stars, the optical
  and near-infrared periods agree to better than 10\%. Stars with periods less 
  than \about 2 days as indicated by the optical data are aliased to longer 
  periods in this study due to the 1 day time sampling of the near-infrared
  observations. Stars that have optical periods roughly twice that of the 
  near-infrared period may be examples of ``period doubling'' which results
  when multiple star spots are present \citep{Herbst00b}. One star with an 
  optical period of \about 60~days and a near-infrared period of \about 
  14~days is not shown in this figure since the optical period is 
  uncertain \citep{Rebull01}.
  \label{fig:optical}
}
\end{figure}
\clearpage

\begin{figure}
\plotone{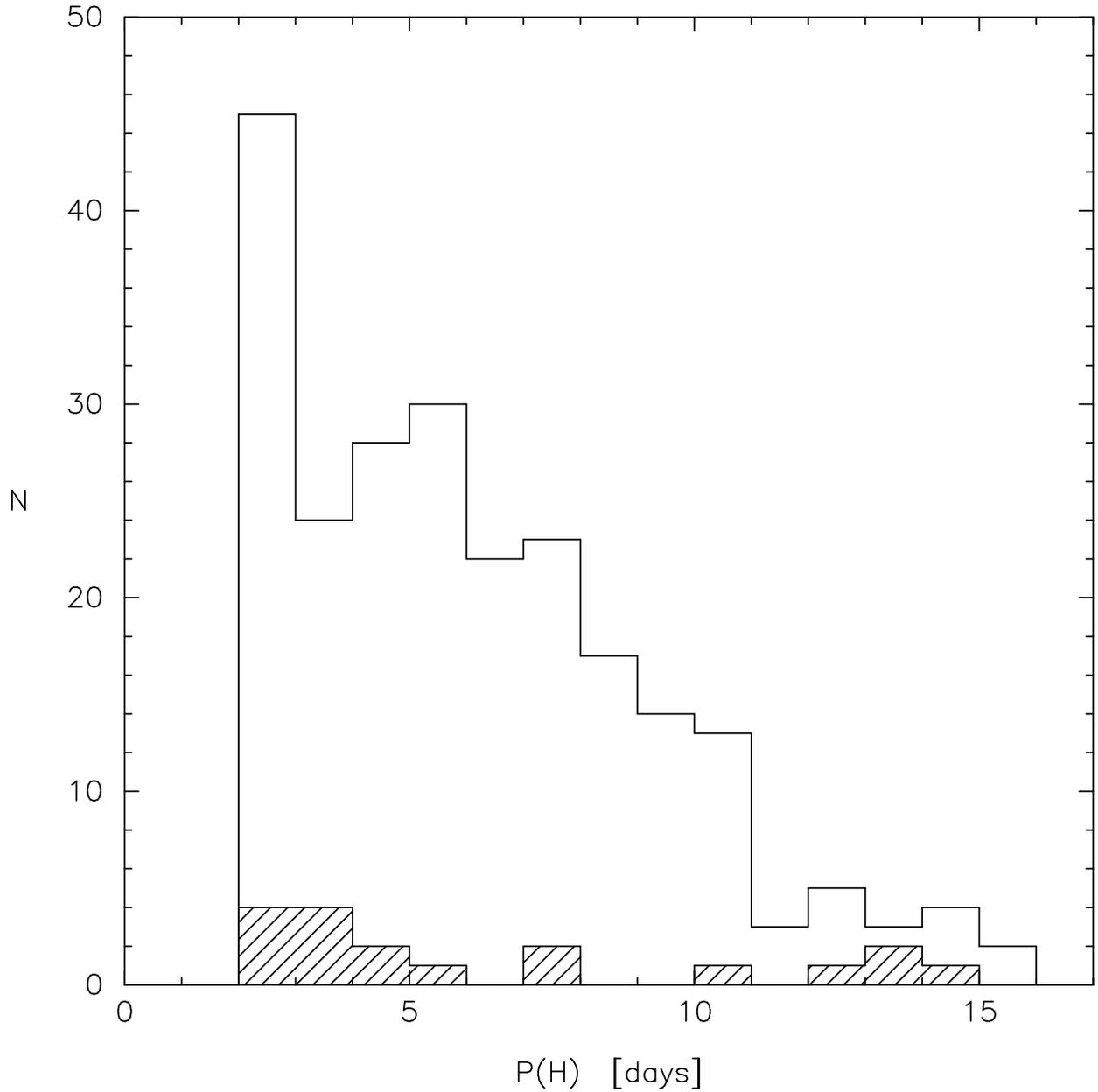}
\caption{
  Frequency distribution of periods for stars in this study that have 
  false-alarm-probabilities less than 10$^{-4}$. The open histogram is for
  all the stars, and the hatched histogram are stars which are suspected to
  be aliased with a sub-2 day period based on comparison to the optical 
  derived periods (see Fig.~\ref{fig:optical}). Only about half the stars 
  represented by the open histogram have the necessary information to 
  establish if the inferred period is aliased in this manner.
  \label{fig:periods}
}
\end{figure}
\clearpage

\begin{figure}
\epsscale{0.90}
\plotone{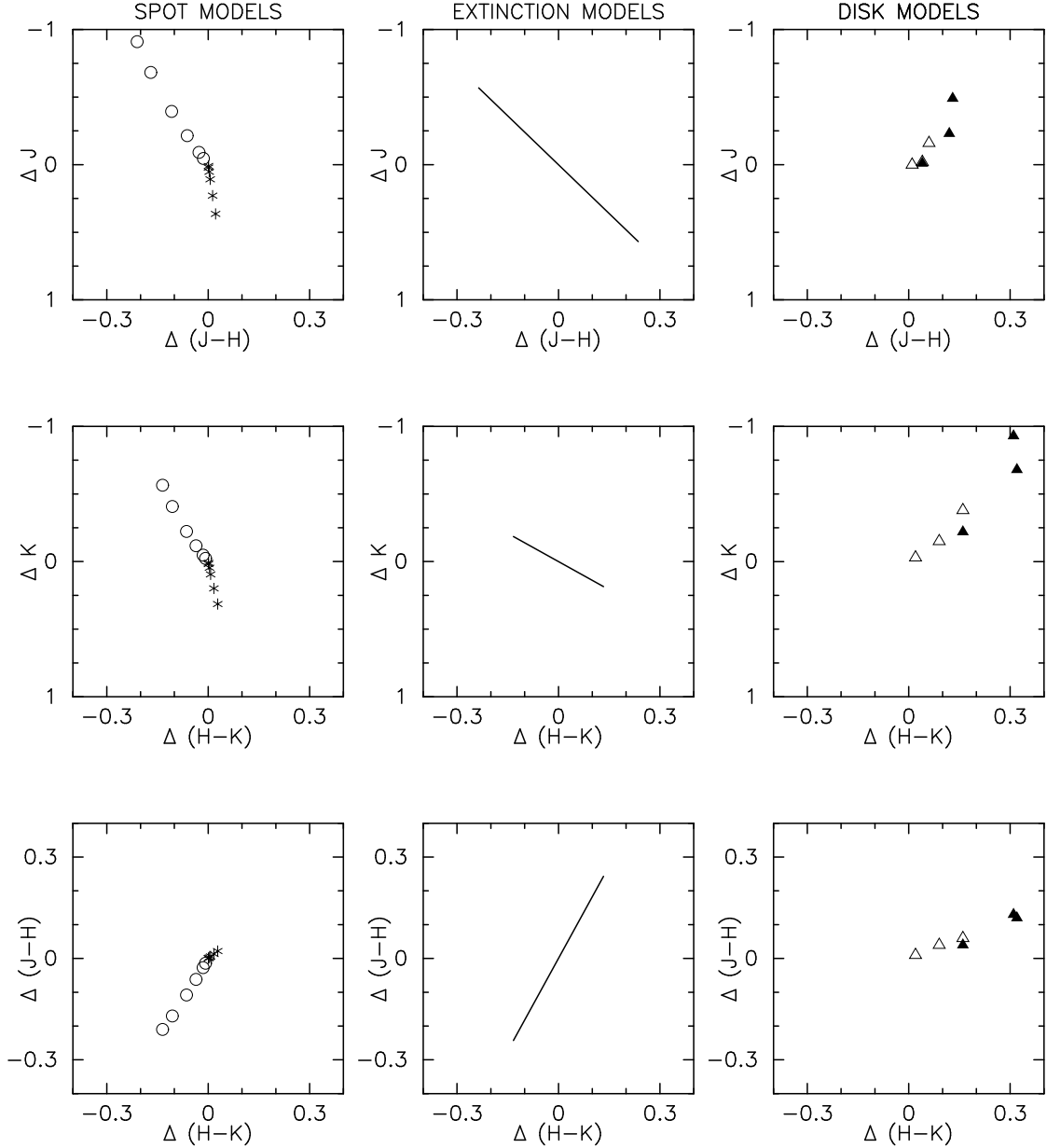}
\epsscale{1.00}
\caption{
  Model $J$ vs. $J-H$, \KB\ vs. \HK, and $J-H$ vs. \HK\ diagrams for cool and
  hot spots, extinction, and accretion disk variations. These models can be
  compared with the observed diagrams in 
  Figures~\ref{fig:ex_limit}-\ref{fig:ex_long2} to investigate the origin of
  the near-infrared variability. The models assume an photospheric
  temperature of 4000~K, appropriate for a 1~Myr, \about 0.5\msun\ star 
  \citep{DM97}, spot temperatures of 2000~K for cool spots (asterisk symbols)
  or 8000~K for hot spots (open circles), and spot coverages of 1, 2, 5, 10,
  20, and 30\% (see Eq.~\ref{eq:spots}). The extinction vectors were calculated 
  from the interstellar reddening law from \citet{Cohen81} transformed into 
  the 2MASS color system \citep{Carp00}. The length of the vectors 
  correspond to $\Delta\rm{A_V} = \pm 2$\M.  The disk models have been 
  provided courtesy of N. Calvet, and represent the effects of varying the 
  mass accretion rate and inner hole size of the accretion disks. Results are 
  presented for mass accretion rates of 10$^{-8.5}$\myear\ (open triangles) 
  and 10$^{-7.0}$\myear\ (filled triangles), and for each accretion rate, 
  inner hole sizes of 1, 2, and 4\rsun. The larger hole sizes correspond to 
  small infrared excesses.
  \label{fig:models}
}
\end{figure}
\clearpage

\clearpage

\end{document}